\documentclass[preprint,notoc]{JHEP3}

\usepackage{axodraw}

\def\PlusBreak#1{+ \nonumber \\          
          &&  \hphantom{#1} \! \null   +}
\def\MinusBreak#1{- \nonumber \\         
          &&  \hphantom{#1} \!  \null  -}

\def\qb{{\bar{q}}}

\def\Tr{\mathop{\rm tr}\nolimits}

\def\BigglBl{\Biggl[}

\def\bigglB{\biggl[}
\def\biggrB{\biggr]}
\def\bigglP{\biggl(}

\def\BiglP{\Bigl(}

\def\ksl{\not{\hbox{\kern-2.3pt $k$}}}

\def\e{\epsilon}

\def\Ord{{\cal O}}
\def\cm{{\cal M}}

\def\Re{{\rm Re}}
\def\Nf{{N_{\! f}}}
\def\Nfsq{{N_{\! f}^2}}
\def\la{\langle}
\def\ra{\rangle}
\def\RS{{\scriptscriptstyle\rm R\!.S\!.}}

\def\bom#1{{\mbox{\boldmath $#1$}}}
\def\MSbar{\overline{\rm MS}}
\def\DRbar{\overline{\rm DR}}

\def\FDH{{\rm FDH}}
\def\lr{\leftrightarrow}
\def\li#1{{\mathop{\rm Li}\nolimits}_#1}
\def\Li{\mathop{\rm Li}\nolimits}
\def\ggtogg{{gg \to gg}}
\def\qqtogg{{q\bar{q} \to gg}}
\def\qgtoqg{{qg \to qg}}
\def\qgtogq{{qg \to gq}}

\def\tg{\tilde{g}}

\def\alphas{\alpha_s}

\def\Susy{{\scriptscriptstyle\rm SYM}}

\def\eqn#1{eq.~(\ref{#1})}
\def\eqns#1#2{eqs.~(\ref{#1}) and~(\ref{#2})}

\def\rg{r_\Gamma}

\def\Boxsix{{\rm Box}^{(6)}}

\def\Trifour{{\rm Tri}^{(4)}}

\def\spa#1.#2{\left\langle#1\,#2\right\rangle}
\def\spb#1.#2{\left[#1\,#2\right]}
\def\lor#1.#2{\left(#1\,#2\right)}
\def\trc{{\rm Tr}}

\def\tS{{\tt S}}
\def\tT{{\tt T}}
\def\tU{{\tt U}}
\def\ibar{\bar{\imath}}

\def\tab#1{table~\ref{#1}}
\def\Tab#1{Table~\ref{#1}}

\def\AL#1#2#3#4#5#6#7#8{A^L_{ {#1^#5} {#2^#6} {#3^#7} {#4^#8}}}
\def\AR#1#2#3#4#5#6#7#8{A^R_{ {#1^#5} {#2^#6} {#3^#7} {#4^#8}}}
\def\ALh#1#2#3#4#5#6#7#8{A^{L, [1/2]}_{ {#1^#5} {#2^#6} {#3^#7} {#4^#8}}}

\preprint{
  hep-ph/0304168\\
  SLAC--PUB--9675\\
  UCLA/03/TEP/10\\
  DESY 03--034\\
  April, 2003}

\title{
Two-Loop Helicity Amplitudes for Quark-Gluon Scattering in QCD and 
Gluino-Gluon Scattering in Supersymmetric Yang-Mills Theory}

\author{Zvi Bern\thanks{Research supported by the US Department of 
Energy under grant DE-FG03-91ER40662.} \\
	Department of Physics and Astronomy\\
	UCLA, Los Angeles, CA 90095-1547, USA\\
	E-mail: \email{bern@physics.ucla.edu}}

\author{Abilio De Freitas \\
	Deutsches Elektronen Synchrotron \\
	DESY, D-15738 Zeuthen, Germany\\
	E-mail: \email{dfreitas@ifh.de}}

\author{Lance Dixon\thanks{Research supported by the US Department of 
Energy under contract DE-AC03-76SF00515.}\\
	Stanford Linear Accelerator Center, Stanford University\\
	Stanford, CA 94309, USA\\
	E-mail: \email{lance@slac.stanford.edu}}

\abstract{ We present the two-loop QCD helicity amplitudes for
quark-gluon scattering, and for quark-antiquark annihilation 
into two gluons.  These amplitudes are relevant for
next-to-next-to-leading order corrections to (polarized) jet
production at hadron colliders.  We give the results in the
`t~Hooft-Veltman and four-dimensional helicity (FDH) variants of
dimensional regularization.  The transition rules for converting
the amplitudes between the different variants are much more 
intricate than for the previously discussed case of gluon-gluon
scattering.  Summing our two-loop expressions over helicities
and colors, and converting to conventional dimensional regularization,
gives results in complete agreement with those of Anastasiou, Glover,
Oleari and Tejeda-Yeomans.  We describe the amplitudes for $2 \to 2$
scattering in pure $N=1$ supersymmetric Yang-Mills theory, obtained
from the QCD amplitudes by modifying the color representation and
multiplicities, and verify supersymmetry Ward identities in the FDH
scheme. }


\dedicated{\centerline{Submitted to JHEP}}

\begin{document}

\section{Introduction}\label{IntroSection}

Recent years have seen rapid progress in our ability to compute
two-loop matrix elements with more than a single kinematic
variable~\cite{BRY,AllPlusTwo,BhabhaTwoLoop,GOTYqqqq,GOTYqqgg,GOTYgggg,
gggamgamPaper,PhotonPaper,BDDgggg,TwoLoopee3Jets,TwoLoopee3JetsHel}.  
The progress has relied in part on new developments in loop
integration~\cite{IBP,PBScalar,NPBScalar,Lorentz,NPBReduction,
PBReduction,IntegralsAGO} and in understanding the infrared
divergences of the theory~\cite{Catani}.  The new two-loop amplitudes
will be essential for reducing theoretical uncertainties in a number
of physical quantities. (For a recent summary describing the various
expected improvements see {\it e.g.} ref.~\cite{GloverReview}.)  
In particular, in jet physics an important source of theoretical 
uncertainty is from
missing higher order corrections to the widely used NLO
calculations~\cite{Aversa,EKS,JETRAD}, which have been crucial for
matching theory to experiment.  For the case of hadron colliders, one
of the essential ingredients to performing next-to-next-to-leading
order (NNLO) calculations are the parton distribution functions.
Recently, global fits to the data for the parton distribution
functions have been performed~\cite{MRSTNNLO} within an approximate
NNLO framework~\cite{NNLOPDFApprox}. There are, however sizable
uncertainties associated with the experimental input to the parton
distribution functions~\cite{PDFuncertainty}.  Nevertheless, an exact
NNLO computation of jet production rates would be very welcome.
Besides reducing the scale uncertainties for jet rates, such a
computation will allow a better understanding of energy flow within
jets, as a jet may consist of up to three partons at this order.  For
very large momentum transfer, improvements can also be made by
resumming threshold logarithms~\cite{Kidonakis}.

Besides the two loop matrix elements, an NNLO calculation of two jet
production at hadron colliders requires the tree amplitudes for six
external partons~\cite{TreeSixPoint,MPReview} and the one-loop
amplitudes for five external
partons~\cite{OneloopFivePoint,Oneloopqqggg}, which have been known
for some time now.  Anastasiou, Glover, Oleari, and Tejeda-Yeomans
have provided the NNLO interferences of the two-loop amplitudes with
the tree amplitudes, for all QCD four-parton processes, summed over
all external helicities and
colors~\cite{GOTYqqqq,GOTYqqgg,GOTYgggg}. The helicity amplitudes for
$g g \rightarrow gg$ have also been computed, using the spinor
helicity formalism~\cite{SpinorHelicity}. In this paper, we present
the $\bar qq \rightarrow gg$ and $q g \rightarrow q g$ two-loop
helicity amplitudes.  The four-quark helicity amplitudes will be
presented elsewhere.

A useful property of helicity amplitudes is that they expose the full
dependence on color and spin.  Many formal properties of scattering
amplitudes are simpler in a helicity basis and/or after color
decomposition.  Such properties include supersymmetry Ward
identities~\cite{SWI,TwoLoopSUSY}, collinear
limits~\cite{MPReview,Neq4,LoopReview}, and high-energy
behavior~\cite{BFKL,TwoloopBFKL}.  The full color dependence is also
useful for understanding the structure of the infrared
divergences~\cite{Catani,BDDgggg,StermanIR}.

This additional formal information is not necessary for the main
phenomenological application, NNLO jet production in collisions of
unpolarized hadrons.  On the other hand, the helicity amplitudes can have
phenomenological applications for jet production in collisions of {\it
polarized} protons, as are being carried out at the relativistic heavy 
ion collider (RHIC) at Brookhaven.  This may, for example, help to 
determine the poorly-known polarized gluon distribution in the
proton~\cite{SofferVirey}.  Theoretical predictions of the relevant
observables require scattering amplitudes for polarized partons.
Currently, predictions are available through NLO~\cite{dFFSV}; the
helicity amplitudes presented here are a prerequisite for improving
the predictions to NNLO accuracy.  Our results also serve as a check
of the results of ref.~\cite{GOTYqqgg} for unpolarized $\qqtogg$
and $\qgtoqg$ scattering.

While preparing our results for publication, we became aware
of a similar computation being completed simultaneously~\cite{GTYqqgg}.
Ref.~\cite{GTYqqgg} computes the one- and two-loop helicity amplitudes
for $\qqtogg$ and $\qgtoqg$ scattering, in the 
't~Hooft-Veltman (HV) scheme~\cite{HV} (see below).
A slightly different method from ours is used to extract the helicity 
amplitudes.  We have compared our results for the two-loop amplitudes 
(and also the one-loop amplitudes), through the finite terms required for 
NNLO cross sections, and we are in complete agreement.

In this paper we also describe the helicity amplitudes for
gluino-gluon scattering in $N=1$ supersymmetric $SU(N)$ gauge theory.
Due to supersymmetry Ward identities~\cite{SWI}, these amplitudes are
simply related to the $N=1$ gluon-gluon scattering amplitudes already
presented in ref.~\cite{BDDgggg}.  The supersymmetric amplitudes are a
close cousin of QCD amplitudes, differing only in fermion
multiplicities and non-abelian charge assignments: The $N=1$
supersymmetric amplitudes are obtained from the QCD amplitudes by
replacing the quarks with a single adjoint color representation
fermion, effectively converting them to a gluino superpartner of the
gluon.

Because the scattering amplitudes possess both infrared and ultraviolet 
divergences, some care is required to ensure that the regularization 
procedure preserves supersymmetry.  Several versions of dimensional 
regularization are commonly used for loop calculations in QCD.  
The widely used conventional dimensional regularization (CDR) 
scheme~\cite{CDR} breaks supersymmetry --- it alters the balance between 
bosonic and fermionic states at order $\e$, where $\e = (4-D)/2$ and
$D$ is the number of dimensions.  The CDR scheme is traditionally 
employed in calculations of amplitude interferences, such as in
refs.~\cite{EllisSexton,BhabhaTwoLoop,GOTYqqqq,GOTYqqgg,GOTYgggg}.  
In the helicity approach, the two commonly used schemes are the
't~Hooft-Veltman (HV) scheme~\cite{HV} and the four-dimensional
helicity (FDH) scheme~\cite{BKgggg,TwoLoopSUSY}.  These schemes differ
in the number of polarization states for unobserved gluons.  The
't~Hooft-Veltman (HV) scheme~\cite{HV} contains $2 - 2\e$ virtual
gluon states (as does the CDR scheme), whereas the four-dimensional 
helicity (FDH) scheme~\cite{BKgggg,TwoLoopSUSY} assigns $2$ states.  
The FDH scheme is related to dimensional reduction (DR)~\cite{DR}, 
but is more compatible with the helicity method, because it allows 
two transverse dimensions in which to define helicity.  
It is also possible to define a scheme, labeled by a parameter
$\delta_R$, which interpolates between the HV ($\delta_R=1$) and
FDH ($\delta_R=0$) schemes.  We shall present the $q\bar{q} \to gg$
and $qg\to qg$ amplitudes in this general $\delta_R$ scheme.
A more detailed description of the differences between schemes, 
as well as a definition of the FDH scheme beyond one loop, has been 
given recently~\cite{TwoLoopSUSY}. 

The supersymmetry preserving properties of the FDH scheme have been
verified explicitly at two loops, for particular helicity
configurations of four-gluon amplitudes that vanish at
tree level~\cite{TwoLoopSUSY,BDDgggg}.  In this paper, we explicitly
verify that the FDH scheme preserves supersymmetry for the case of
gluino-gluon scattering.  Part of this check involves relating
gluino-gluon amplitudes to gluon-gluon amplitudes, for helicity
configurations which are {\it non-vanishing} at tree-level.  Such a
test is somewhat more stringent than previous
ones~\cite{BDDgggg,BGMvdB,TwoLoopSUSY} because of the more intricate infrared
divergences.  The test additionally provides a nontrivial check on the
calculation of the quark-gluon scattering amplitudes, as well as on
the consistency of the FDH scheme.

In general, scattering amplitudes in massless QCD possess strong
infrared (soft and collinear) divergences.  Using dimensional
regularization, the amplitudes generically contain poles in $\e$ up to
$1/\e^4$.  Catani has organized these divergences into a compact form
predicting their structure~\cite{Catani}.  We use Catani's formula and
color space notation to organize the helicity amplitudes into singular
terms (which do contain order $\e^0$ terms in their series expansion in
$\e$), plus finite remainders.  The precise form of the $1/\e$ poles
was not predicted {\it a priori} in ref.~\cite{Catani} for general
processes at two loops.  It is now clear, however, that these terms
have a universal structure depending only on the external legs, based
on explicit calculation~\cite{BhabhaTwoLoop,GOTYqqqq,GOTYqqgg,GOTYgggg,
gggamgamPaper,PhotonPaper,BDDgggg,TwoLoopee3Jets,TwoLoopee3JetsHel} 
and on matching to resummed results~\cite{StermanIR}.

For the quark-gluon scattering amplitudes discussed here,
ref.~\cite{GOTYqqgg} previously computed the interference of the
$1/\e$ pole terms with the tree amplitude, summed over all colors and
helicities.  Here we extract the full color and helicity dependence of
the $1/\e$ pole terms.  For the case of $gg \rightarrow gg$
amplitudes, a term independent of color and helicity was
found~\cite{GOTYgggg}, plus a second ``surprise'' term~\cite{BDDgggg}
with nontrivial color-dependence, which vanishes when the color-summed
interference is performed.  Here we confirm that a similar 
color-dependent term exists for the case of quark-gluon scattering
amplitudes.  We note that a term with similar color structure has
been identified in contributions of one-loop factors for soft
radiation to NNLO processes~\cite{CataniGrazziniSoft}.

The conversion from one variant of dimensional regularization to
another is well understood at one
loop~\cite{BKgggg,KSTfourparton,CST}.  At two loops, the ultraviolet
shift in the coupling constant has been calculated for the commonly
used variants of dimensional regularization~\cite{TwoLoopSUSY}.
However, the infrared aspects have not yet been fully understood for
arbitrary processes.  The scheme dependence of $\ggtogg$ scattering
amplitudes was studied in ref.~\cite{BDDgggg}.   There it was found
that beginning at order $1/\e^2$ the functions appearing in Catani's
infrared decomposition are actually scheme-dependent.  
Here we present the scheme dependence (dependence on $\delta_R$)
for the $\qqtogg$ amplitudes. 
The universal structure of the Catani formula
for infrared divergences suggests that the conversion between schemes
for any two-loop massless QCD amplitude would be controlled by the same
set of functions that we uncover here and in ref.~\cite{BDDgggg}.
However, the $\qqtogg$ scheme-dependence we find here is significantly
more intricate than that found for $\ggtogg$, and we have not yet
identified the general pattern.

The paper is organized as follows.  In section~\ref{IRSection} we
review the infrared and color structure of one- and two-loop QCD
amplitudes used to organize the amplitudes presented in this paper.
In section~\ref{OneloopSection} we describe the one-loop quark-gluon
scattering amplitudes.   Section~\ref{AllOrdersSubsection}
presents them in a form that is valid to all orders in $\e$,
in terms of integrals known through $\Ord(\e^2)$.  Their knowledge at this
order is required for evaluating Catani's formula for the singular parts 
of the two-loop amplitudes through $\Ord(\e^0)$. 
The finite one-loop remainder functions are given in 
section~\ref{OneloopRemainderSubsection}.  
The ``square'' of these functions, summed over colors and helicities,
contributes to the NNLO cross section.  In 
section~\ref{Compare11CDRSection}, we describe how to carry out this sum, 
and compare the result to a similar sum performed in the 
CDR scheme~\cite{GOTYqqgg,AGTYPrivate}.

In section~\ref{TwoLoopFiniteQCDSection} we return to the two-loop
amplitudes.   The finite two-loop remainder functions are
presented in section~\ref{TwoLoopFinRemSubsection} and
appendix~\ref{QCDRemainderAppendix}.  Some auxiliary functions
for describing the shift in the finite remainder functions
are given in appendix~\ref{OrdepsdeltaRemainderAppendix}. 
The interference of the finite two-loop remainder functions with the 
tree amplitudes, summed over colors and helicities, also contributes 
to the NNLO cross section.  In section~\ref{Compare2treeCDRSection}
we describe the computation of this sum, and compare the
results with those obtained in the CDR scheme in ref.~\cite{GOTYqqgg}.

In section~\ref{N=1AmplitudesSection} we discuss the $N=1$
super-Yang-Mills amplitudes obtained by modifying the QCD ones. The
supersymmetry Ward identities are briefly reviewed in
section~\ref{SWISubsection}. In sections~\ref{OneLoopN=1SubSection}
and \ref{TwoLoopFiniteN=1Section} we discuss the results for the one-
and two-loop gluino-gluon scattering amplitudes for pure $N=1$
supersymmetric Yang-Mills theory, after first reviewing the infrared
structure for the theory in section~\ref{IRSusySubsection}.  We verify
that in the FDH scheme the two-loop amplitudes obey the expected
supersymmetry Ward identities, which relate them to the gluon-gluon
scattering amplitudes computed in ref.~\cite{BDDgggg}.  In
section~\ref{ConclusionsSection} we present our conclusions.


\section{Review of infrared and color structure}\label{IRSection}

In this section we review the structure of the infrared singularities
of dimensionally regularized one- and two-loop QCD
amplitudes, using Catani's color space notation~\cite{Catani}, as a 
prelude to presenting the finite remainders of the one- and two-loop 
$\qqtogg$ and $\qgtogq$ amplitudes.

The two processes considered in this paper are
\begin{eqnarray}
    q(p_1,\lambda_1) + \bar{q}(p_2,\lambda_2)
&\to& g(p_3,\lambda_3) + g(p_4,\lambda_4)\,, 
\label{qqgglabel} \\ 
    q(p_1,\lambda_1) + g(p_2,\lambda_2)
&\to& g(p_3,\lambda_3) + q(p_4,\lambda_4)\,, 
\label{qggqlabel} 
\end{eqnarray}
using a ``standard'' ({\it not} ``all-outgoing'') convention for the 
external momentum ($p_i$) and helicity ($\lambda_i$) labeling.  
The Mandelstam variables are $s = (p_1+p_2)^2$, $t = (p_1-p_4)^2$, 
and $u = (p_1-p_3)^2$.  

We use dimensional regularization to handle both ultraviolet and infrared
singularities.  We consider a continuous set of schemes, labeled by a 
parameter $\delta_R$ characterizing the number of virtual gluon degrees 
of freedom circulating in loops.  (Because we are computing helicity 
amplitudes, the number of external gluon states is fixed at two.)
Specifically, when the trace of the Minkowski metric is encountered, we set 
\begin{equation}
\eta^{\mu}{}_{\mu} \equiv D_s \equiv 4 - 2 \e \, \delta_R  \,, 
\label{EtaTrace}
\end{equation}
corresponding to $2(1-\e\,\delta_R)$ gluon states in the loop.
Setting $\delta_R = 1$ corresponds to the HV scheme, which is 
the most closely related to the CDR computation in ref.~\cite{GOTYqqgg}.
Setting $\delta_R = 0$ corresponds to the FDH scheme, which has improved
supersymmetry properties.

The CDR and HV schemes imply the same coupling constant, the 
standard $\MSbar$ coupling, $\bar{\alpha}_s(\mu)$.  The coupling in a general
$\delta_R$ scheme is related to this coupling at NNLO by~\cite{TwoLoopSUSY} 
\begin{eqnarray}
\alpha_s^{\delta_R}(\mu) &=& \bar{\alpha}_s(\mu) \biggl[ 1 
  + {C_A \over 6} (1-\delta_R) {\bar{\alpha}_s(\mu) \over 2\pi} 
  \PlusBreak{ \bar{\alpha}_s(\mu) \bigglB }
    \biggl( {C_A^2 \over 36} (1-\delta_R)^2
        + {7 C_A^2 - 6 C_F T_R \Nf \over 12 } (1-\delta_R) \biggr)   
          \biggl( {\bar{\alpha}_s(\mu) \over 2\pi} \biggr)^2 
  \PlusBreak{ \bar{\alpha}_s(\mu) \bigglB }
    \Ord([\bar{\alpha}_s(\mu)]^3) \biggr] \,.
\label{deltaRMSconv} 
\end{eqnarray}
Henceforth we will suppress the $\delta_R$ index on $\alpha_s(\mu)$.

We work with ultraviolet renormalized amplitudes.  The relation
between the bare coupling $\alphas^u$ and renormalized coupling 
$\alphas(\mu)$, through two-loop order, is~\cite{Catani}
\begin{equation}
\alphas^u \;\mu_0^{2\e} \;S_{\e} = \alphas(\mu) \;\mu^{2\e} 
\left[ 1 - { \alphas(\mu) \over 2\pi } \; { b_0 \over \e } 
         + \biggl( { \alphas(\mu) \over 2\pi} \biggr)^2 
           \left( { b_0^2 \over \e^2 } - { b_1 \over 2\e } \right) 
+ {\cal O}(\alphas^3(\mu)) \right]\,,
\label{TwoloopCoupling}
\end{equation}
where $\mu$ is the renormalization scale, 
$S_\e = \exp[\e (\ln4\pi + \psi(1))]$, and 
$\gamma = -\psi(1) = 0.5772\ldots$ is Euler's constant.
The first two coefficients appearing in the beta function for QCD,
or more generally $SU(N)$ gauge theory with $\Nf$ flavors of
massless fundamental representation quarks, are scheme-independent,
\begin{equation}
b_0 = {11 C_A - 4 T_R \Nf \over 6} \,, \hskip 2 cm 
b_1 = {17 C_A^2 - ( 10 C_A + 6 C_F ) T_R \Nf \over 6} \,,
\label{QCDBetaCoeffs}
\end{equation}
where $C_A = N$, $C_F = (N^2-1)/(2N)$, and $T_R = 1/2$.
(Note that ref.~\cite{Catani} uses the notation 
$\beta_0 = b_0/(2\pi)$, $\beta_1 = b_1/(2\pi)^2$.)

The perturbative expansion of the $\qqtogg$ amplitude is
\begin{eqnarray}
\cm_\qqtogg(\alphas(\mu), \mu;\{p\}) &=&
4\pi\alphas(\mu) \, \bigglB \cm_\qqtogg^{(0)}(\mu;\{p\}) + 
\label{RenExpand} \\
&& \hskip 1.5 cm 
+ { \alphas(\mu) \over 2\pi } \cm_\qqtogg^{(1)}(\mu;\{p\}) +
\nonumber \\
&& \hskip 1.5 cm 
+ \biggl( { \alphas(\mu) \over 2\pi } \biggr)^2 
\cm_\qqtogg^{(2)}(\mu;\{p\}) + \Ord(\alphas^3(\mu)) \biggrB \,, 
\nonumber
\end{eqnarray}
where $\cm_\qqtogg^{(L)}(\mu;\{p\})$ is the $L^{\rm th}$
loop contribution.  The same type of expansion holds for the $\qgtogq$
amplitude.
Equation~(\ref{TwoloopCoupling}) is equivalent to the following
$\MSbar$ renormalization prescriptions at one and two loops,
\begin{eqnarray}
 \cm_\qqtogg^{(1)} 
 &=&  S_\e^{-1} \, \cm_\qqtogg^{(1){\rm unren}}
    - {b_0\over\e} \, \cm_\qqtogg^{(0)} \,,
\label{OneloopCounterterm} \\
 \cm_\qqtogg^{(2)}  
 &=&  S_\e^{-2} \, \cm_\qqtogg^{(2){\rm unren}}   
  -  2 {b_0\over\e} \, S_\e^{-1} 
                    \, \cm_\qqtogg^{(1){\rm unren}}
  + \biggl( { b_0^2\over \e^2 } - {b_1 \over 2 \e} \biggr)
             \, \cm_\qqtogg^{(0)} \,.
\label{TwoloopCounterterm}
\end{eqnarray}

The infrared divergences of renormalized one- and 
two-loop $n$-point amplitudes are given by~\cite{Catani},
\begin{eqnarray}
| \cm_n^{(1)}(\mu; \{p\}) \ra_{\RS} &=& {\bom I}^{(1)}(\e, \mu; \{p\}) 
  \; | \cm_n^{(0)}(\mu; \{p\}) \ra_{\RS} 
  + |\cm_n^{(1){\rm fin}}(\mu; \{p\}) \ra_{\RS} \,,
	\label{OneloopCatani} \\
| \cm_n^{(2)}(\mu; \{p\}) \ra_{\RS} &=& {\bom I}^{(1)}(\e, \mu; \{p\}) 
  \; | \cm_n^{(1)}(\mu; \{p\}) \ra_{\RS} 
	\label{TwoloopCatani} \\ && \null
+ {\bom I}^{(2)}_{\RS}(\e, \mu; \{p\}) \;
         | \cm_n^{(0)}(\mu; \{p\}) \ra_{\RS}
+ |\cm_n^{(2){\rm fin}}(\mu; \{p\}) \ra_{\RS} \,, \hskip .5 cm 
\nonumber
\end{eqnarray}
where the ``ket'' notation $|\cm_n^{(L)}(\mu; \{p\}) \ra_{\RS}$ 
indicates that the $L$-loop amplitude is treated as a vector in 
color space.  The actual amplitude is extracted via
\begin{equation}
\cm_n(1^{a_1},\dots,n^{a_n}) \equiv \la a_1,\dots,a_n \, | \,
\cm_n(p_1,\ldots,p_n)\ra \,,
\label{MnVec}
\end{equation}
where the $a_i$ are color indices.  
The subscript $\RS$ indicates that a quantity depends on the choice of 
regularization and renormalization scheme. The divergences of 
$\cm_n^{(1)}$ are encoded in the color operator ${\bom I}^{(1)}$, 
while those of $\cm_n^{(2)}$ also involve the scheme-dependent operator 
${\bom I}^{(2)}_{\RS}$.

In QCD, the operator ${\bom I}^{(1)}$ is given by
\begin{equation}
{\bom I}^{(1)}(\e,\mu;\{p\}) = \frac{1}{2} {e^{-\e \psi(1)} \over
\Gamma(1-\e)} \sum_{i=1}^n \, \sum_{j \neq i}^n \, {\bom T}_i \cdot
{\bom T}_j \Biggl[ {1 \over \e^2} + {\gamma_i \over {\bom T}_i^2 } \,
{1 \over \e} \Biggr] \Biggl( \frac{\mu^2 e^{-i\lambda_{ij} \pi}}{2
p_i\cdot p_j} \Biggr)^{\e} \,,
\label{CataniGeneral}
\end{equation}
where $\lambda_{ij}=+1$ if $i$ and $j$ are both incoming or outgoing
partons, and $\lambda_{ij}=0$ otherwise. The color charge ${\bom T}_i =
\{T^a_i\}$ is a vector with respect to the generator label $a$, and an
$SU(N)$ matrix with respect to the color indices of the outgoing
parton $i$.  For external gluons $T^a_{cb} = i f^{cab}$, 
so ${\bom T}_i^2 = C_A = N$, and
\begin{equation}
\gamma_g = {11 C_A - 4 T_R \Nf \over 6} \,. 
\label{QCDgluonValues}
\end{equation}
For external fermions, the ratio
\begin{equation}
{ \gamma_q \over {\bom T}_i^2}  = {3\over2} 
\label{QCDfermionValues}
\end{equation}
is independent of the representation.   For quarks, 
${\bom T}_i^2 = C_F = (N^2-1)/(2N)$; for gluinos ${\bom T}_i^2 = C_A = N$.

The operator ${\bom I}^{(2)}_{\RS}$ is given by~\cite{Catani}
\begin{eqnarray}
{\bom I}^{(2)}_{\RS}(\e,\mu;\{p\}) 
& =& - \frac{1}{2} {\bom I}^{(1)}(\e,\mu;\{p\})
\left( {\bom I}^{(1)}(\e,\mu;\{p\}) + {2 b_0 \over \e} \right)
  \PlusBreak{}
 {e^{+\e \psi(1)} \Gamma(1-2\e) \over \Gamma(1-\e)}
\left( {b_0 \over \e} + K_\RS \right) {\bom I}^{(1)}(2\e,\mu;\{p\})
  \PlusBreak{}
  {\bom H}^{(2)}_{\RS}(\e,\mu;\{p\}) \,,
\label{CataniGeneralI2}
\end{eqnarray}
where the coefficient $K_\RS$ depends on $\delta_R$ and 
is given by~\cite{Catani,BDDgggg}
\begin{equation}
K_\RS = \left[ \frac{67}{18} - \frac{\pi^2}{6} 
    - \biggl( {1\over6} + {4\over9} \e \biggr) (1-\delta_R) \right] C_A
 - \frac{10}{9} T_R \Nf \,.    \label{CataniK}
\end{equation}

The function ${\bom H}^{(2)}_{\RS}$ contains only {\em single} poles,
and splits into two types of terms,
\begin{equation}
{\bom H}^{(2)}(\e) = 
{ e^{-\e\psi(1)} \over 4\e \, \Gamma(1-\e) }
 \biggl( { \mu^2 \over -s } \biggr)^{2\e} 
  \Bigl( (2 H_q^{(2)} + 2 H_g^{(2)}) \, {\bom 1} 
       + \hat{\bom H}^{(2)} \Bigr) \,.
\label{OurH}
\end{equation}
From the calculations performed here we find that the term
proportional to the identity matrix in color space ${\bom 1}$ contains
the constants $H_g^{(2)}$ and $H_q^{(2)}$, given by
\begin{eqnarray}
H_q^{(2)} &=& 
\biggl( {13\over2} \zeta_3 - {23\over48} \pi^2 + {245\over216} \biggr) 
 C_A C_F
+ \biggl( - 6 \zeta_3 + {\pi^2\over2} - {3\over8} \biggr) C_F^2
+ \biggl( {\pi^2\over12} - {25\over54} \biggr) C_F T_R \Nf 
\nonumber \\
\hskip0.5cm && 
+ \biggl( - {4\over3} C_A C_F + {1\over2} C_F^2 
           + {1\over6} C_F T_R \Nf \biggr) (1-\delta_R),
\label{Hquark} \\
H_g^{(2)} &=& 
\biggl( {\zeta_3\over2} + {11\over144} \pi^2 + {5\over12} \biggr) C_A^2
- \biggl( {\pi^2\over36} + {58\over27} \biggr) C_A T_R \Nf
+ C_F T_R \Nf + {20\over27} T_R^2 \Nfsq 
\nonumber \\
\hskip0.5cm && 
+ \biggl( - {11\over36} C_A^2 + {1\over9} C_A T_R \Nf \biggr) (1-\delta_R).
\label{Hgluon}
\end{eqnarray}
This term survives the sum over colors, and the expressions for 
$H_g^{(2)}$ and $H_q^{(2)}$ in the HV scheme ($\delta_R=1$) agree, as
expected, with previous color-summed results in the CDR
scheme~\cite{GOTYqqqq,GOTYqqgg,GOTYgggg,AGTYphotons}.
The $H_g^{(2)}$ coefficient agrees for all $\delta_R$ with that extracted
from the $\ggtogg$ helicity amplitudes~\cite{BDDgggg}.

The second term in ${\bom H}^{(2)}(\e)$ has exactly the same type of 
nontrivial color and kinematic dependence found in the $\ggtogg$ 
helicity amplitudes~\cite{BDDgggg}, namely
\begin{equation}
\hat{\bom H}^{(2)} = - 4 \, \ln\biggl( {-s\over-t} \biggr)
                            \ln\biggl( {-t\over-u} \biggr)
                            \ln\biggl( {-u\over-s} \biggr)
      \times \Bigl[ {\bom T}_1 \cdot {\bom T}_2 \,,
                    {\bom T}_2 \cdot {\bom T}_3 \Bigr] \,,
\label{HExtra}
\end{equation}
with $\ln((-s)/(-t)) \to \ln s - \ln(-t) - i\pi$ in the $s$-channel, 
{\it etc.}  (It might seem that the operator 
${\bom T}_1 \cdot {\bom T}_2$ is somewhat ambiguous for the 
$\qqtogg$ process~(\ref{qqgglabel}), since the 
$q\bar{q}$ and $gg$ pairs have different color quantum numbers.
However, the difference between the $q\bar{q}$ and $gg$ 
${\bom T}_1 \cdot {\bom T}_2$ operators is proportional to the 
identity --- it is a difference of Casimirs --- so the commutator
is unambiguous.)

In ref.~\cite{BDDgggg} it was observed for the $\ggtogg$ amplitudes
that the second term in \eqn{OurH} is independent of the helicity 
configuration, and is a nontrivial commutator matrix in color space.  
(The possibility of nontrivial color structure in ${\bom H}^{(2)}(\e)$ 
was pointed out in ref.~\cite{Catani}.)  Now we can see that 
$\hat{\bom H}^{(2)}$ is also independent of whether the external 
lines are quarks or gluons, which buttresses the suggestion~\cite{BDDgggg} 
that it is related to soft, not collinear, virtual contributions.  A similar
color structure emerges in a general analysis of the contributions 
of one-loop factors for soft radiation to NNLO 
processes~\cite{CataniGrazziniSoft}.
Because of the commutator structure, $\hat{\bom H}^{(2)}$ vanishes 
when sandwiched between tree amplitudes, after performing a sum over 
colors; hence it drops out of the color-summed interference of the 
two-loop amplitudes with the tree amplitudes~\cite{BDDgggg}.

To proceed further, we wish to introduce an explicit color basis for the 
amplitudes, and also remove certain overall spinor product factors.
To do the latter, we take the set of independent
helicity configurations $h$ to be
\begin{eqnarray}
h=1: \quad   q(p_1,+) + \bar{q}(p_2,-) &\to& g(p_3,+) + g(p_4,+)\,, 
\label{hel1} \\ 
h=2: \quad   q(p_1,+) + \bar{q}(p_2,-) &\to& g(p_3,-) + g(p_4,+)\,, 
\label{hel2} \\ 
h=3: \quad   q(p_1,+) + g(p_2,-) &\to& g(p_3,+) + q(p_4,+)\,, 
\label{hel3} \\ 
h=4: \quad   q(p_1,+) + g(p_2,-) &\to& g(p_3,-) + q(p_4,+)\,, 
\label{hel4} \\ 
h=5: \quad   q(p_1,+) + g(p_2,+) &\to& g(p_3,+) + q(p_4,+)\,. 
\label{hel5}
\end{eqnarray}
Other configurations are simply related to these by symmetries.
For example, the $q(p_1,-)$ amplitudes are obtained by parity (P);
the $\bar{q}g\to g\bar{q}$ amplitudes are related to 
$\qgtogq$ by charge conjugation (C); 
$q(p_1,+) + \bar{q}(p_2,-) \to g(p_3,-) + g(p_4,-)$ 
is related to process~(\ref{hel1}) by CP; and
$q(p_1,+) + g(p_2,+) \to g(p_3,-) + q(p_4,+)$ 
is related to process~(\ref{hel3}) by time reversal (T). 
Maintaining helicity conservation on the quark line removes half of the
configurations.  Applying parity to the remaining configurations,
in order to let the helicity of $q(p_1)$ be positive, there
are four configurations each for $\qqtogg$ and $\qgtogq$.
\Tab{heltable} relates these eight configurations to the five 
represented by eqs.~(\ref{hel1})--(\ref{hel5}).

\TABULAR[t]
{|c||c|c|}
{\hline
$\hphantom{\;X} \vphantom{\Big|}$ {\bf Amplitude} $\hphantom{\;X}$ 
   & {\bf Value of $h$} 
& {\bf Permutation} \\
\hline\hline $q_1^+ \, \bar{q}_2^- \to g_3^+ \, g_4^+$ 
& $h=1$ & --- \\
\hline $q_1^+ \, \bar{q}_2^- \to g_3^- \, g_4^+$ 
& $h=2$ & --- \\
\hline $q_1^+ \, \bar{q}_2^- \to g_3^+ \, g_4^-$ 
& $h=2$ & $\{ t \lr u \}$ \\
\hline $q_1^+ \, \bar{q}_2^- \to g_3^- \, g_4^-$ 
& $h=1$ & $\{ t \lr u \}$  \\
\hline\hline $q_1^+ \, g_2^+ \to g_3^+ \, q_4^+$ 
& $h=5$ & --- \\
\hline $q_1^+ \, g_2^- \to g_3^+ \, q_4^+$ 
& $h=3$ & --- \\
\hline $q_1^+ \, g_2^+ \to g_3^- \, q_4^+$ 
& $h=3$ & --- \\
\hline $q_1^+ \, g_2^- \to g_3^- \, q_4^+$ 
& $h=4$ & --- \\
\hline}
{\label{heltable}
\small 
Relations between a general helicity configuration containing $q(p_1,+)$
and the five presented in the text, eqs.~(\ref{hel1})--(\ref{hel5}).}

Helicity-dependent, phase-containing factors arise because we
evaluate the amplitudes in the spinor helicity 
formalism~\cite{SpinorHelicity}.  For $h=2,4,5$, we remove
a factor $S_h$ related to the tree amplitude.  For $h=1,3$, the tree 
amplitude vanishes, and we remove a factor related to the one-loop 
amplitude.  We define
\begin{eqnarray}
S_{1} &=& -i {\spa1.3\spb3.4 \over \spa2.3\spa3.4} \,,
   \hskip1.25cm
S_{2}  =  i {{\spa1.3}^3 \spa2.3 \over \spa1.2\spa2.3\spa3.4\spa4.1} \,,
  \nonumber \\
S_{3}  &=&  -i {\spa1.3 \spb3.2 \over \spa2.3\spa3.4} \,,
   \hskip0.4cm
S_{4} 
   = i {{\spa1.3}^3 \spa4.3 \over \spa1.2\spa2.3\spa3.4\spa4.1} \,,
   \hskip0.4cm
S_{5} 
   = i {{\spa1.2}^3 \spa4.2 \over \spa1.2\spa2.3\spa3.4\spa4.1} \,.
\label{SpinorPhases} 
\end{eqnarray}
The spinor inner products~\cite{SpinorHelicity,MPReview} are 
$\spa{i}.j = \langle i^- | j^+\rangle$ and 
$\spb{i}.j = \langle i^+| j^-\rangle$,
where $|i^{\pm}\rangle$ are massless Weyl spinors of momentum $k_i$,
labeled with the sign of the helicity.  They are anti-symmetric, with
norm $|\spa{i}.j| = |\spb{i}.j| = \sqrt{s_{ij}}$, where 
$s_{ij} = 2k_i\cdot k_j$.  The squares of the prefactors $S_h$ enter 
polarized cross sections,
\begin{eqnarray}
|S_{1}|^2 &=& {u\over t} \,,
   \hskip1.25cm
|S_{2}|^2  =  {u^3 \over s^2 t} \,,
  \nonumber \\
|S_{3}|^2  &=&  - {u \over s} \,,
   \hskip1.25cm
|S_{4}|^2 =  - {u^3 \over st^2} \,,
   \hskip1.25cm
|S_{5}|^2 = - {su \over t^2} \,,
\label{SpinorPhasesSquared} 
\end{eqnarray}
where we have included a minus sign for all the $\qgtogq$ cases to
ensure positive tree-level cross sections.

The color decomposition of the amplitudes (or their finite parts,
according to \eqns{OneloopCatani}{TwoloopCatani}) reads
\begin{equation}
 \cm^{(L)}_{h} = S_{h} \times 
 \sum_{c=1}^3 \trc^{[c]} \times M^{(L),[c]}_{h} \,,
 \qquad h=1,2,3,4,5,
\label{RemoveColorPhase}
\end{equation}
where
\begin{eqnarray}
  \trc^{[1]} &=& (T^{a_3} T^{a_4})_{~\ibar_1}^{i_2}\,, \hskip0.5cm
   \trc^{[2]} = (T^{a_4} T^{a_3})_{~\ibar_1}^{i_2}\,, \hskip0.5cm
   \trc^{[3]} = \delta^{a_3a_4} \, \delta_{~\ibar_1}^{i_2}\,,
 \hskip0.7cm h=1,2, \label{TraceBasisqqgg}
\end{eqnarray}
and
\begin{eqnarray}
  \trc^{[1]} &=& (T^{a_3} T^{a_2})_{~\ibar_1}^{i_4}\,, \hskip0.5cm
   \trc^{[2]} = (T^{a_2} T^{a_3})_{~\ibar_1}^{i_4}\,, \hskip0.5cm
   \trc^{[3]} = \delta^{a_2a_3} \, \delta_{~\ibar_1}^{i_4}\,,
 \hskip0.7cm h=3,4,5.
\label{TraceBasisqggq}
\end{eqnarray}
Here $T^a$ are $SU(N)$ generators in the fundamental representation,
normalized according to the convention typically used in helicity
amplitude calculations, $\Tr(T^a T^b) = \delta^{ab}$.  (The $T^a$ used
in this color decomposition should not be confused with the $T_i^a$
appearing in ${\bom I}^{(1)}$, whose representation depends on the
external line;
nor should they be confused with the generators for the quark
representation, which have the more ``standard'' normalization, 
$T_R = 1/2$, as mentioned above.)  
The amplitude components depend on the Mandelstam variables,
$M^{(L),[c]}_{h} \equiv M^{(L),[c]}_{h}(s,t,u)$.   
Often we will suppress the kinematic arguments, unless a permutation 
of them is involved.

Note that we use a charge-conjugated fundamental index for quarks
and anti-quarks in the initial state, {\it i.e.} $\ibar_1$ for $q_1$,
and $i_2$ for $\bar{q}_2$.   Also, amplitudes related to $h=1,2,3,4,5$ 
by charge conjugation (C) will naturally have the fundamental indices 
charge conjugated.

In the basis~(\ref{TraceBasisqqgg}) for $\qqtogg$, the matrix 
${\bom I}^{(1)}$ is
\begin{eqnarray}
&&{\bom I}^{(1)}(\e) = - {e^{-\e\psi(1)} \over \Gamma(1-\e)}
\times
\label{explicitI1} \\ 
&\times&\left(
\begin{array}{ccccccccc}
(-{1\over2N} \xi_q + {N\over2} \xi_g) \tS + N \xi_{qg} \tT    &   0 
   &   \xi_{qg} (\tT - \tU) \\
0 & (-{1\over2N} \xi_q + {N\over2} \xi_g) \tS + N \xi_{qg} \tU 
   &   -\xi_{qg} (\tT - \tU) \\
\xi_{qg} (\tS - \tU)   &  \xi_{qg} (\tS - \tT) 
   & ( {V\over2N} \xi_q + N \xi_g ) \tS 
\end{array}
\right) \nonumber
\end{eqnarray}
where
\begin{eqnarray}
&&\tS = \left({\mu^2\over -s}\right)^\e \,, \qquad
  \tT = \left({\mu^2\over -t}\right)^\e \,, \qquad
  \tU = \left({\mu^2\over -u}\right)^\e \,, \label{STUedef} \\
&& \xi_q = {1\over\e^2} + {3\over 2\e}\,, \qquad
   \xi_g = {1\over\e^2} + {b_0\over N\e}\,, \qquad
 \xi_{qg} = {1\over2} (\xi_q + \xi_g)\,, \label{xidef} \\
&& V = N^2-1.
\label{Vdef}
\end{eqnarray}
The corresponding operator for $\qgtogq$ in the basis~(\ref{TraceBasisqggq})
is obtained by exchanging $\tS$ and $\tT$ in \eqn{explicitI1}.

The tree amplitudes in the color basis~(\ref{TraceBasisqqgg}), 
(\ref{TraceBasisqggq}) are given by $M^{(0),[c]}_h$, where
\begin{eqnarray}
 M^{(0),[c]}_{1} &=& M^{(0),[c]}_{3} = 0, 
 \hskip 4.3 cm c=1,2,3,  \nonumber \\
 M^{(0),[3]}_{h} &=& 0, \hskip 6.0 cm h=1,2,3,4,5, \nonumber \\
 M^{(0),[1]}_{2} &=& 1 \,, \qquad 
 M^{(0),[2]}_{2} = {t\over u} \,, \nonumber \\
 M^{(0),[1]}_{4} &=& 1 \,, \qquad 
 M^{(0),[2]}_{4} = {s\over u} \,, \nonumber \\
 M^{(0),[1]}_{5} &=& 1 \,, \qquad 
 M^{(0),[2]}_{5} = {s\over u} \,.
\label{TreeAmps}
\end{eqnarray}

A typical partonic cross section requires an amplitude interference,
summed over all external colors.  Such interferences are evaluated in
the color bases~(\ref{TraceBasisqqgg}), (\ref{TraceBasisqggq}) as
\begin{equation}
 I^{(L,L')}_{\lambda_1\lambda_2\lambda_3\lambda_4}
 \equiv
\langle \cm_{\lambda_1\lambda_2\lambda_3\lambda_4}^{(L)} 
      | \cm_{\lambda_1\lambda_2\lambda_3\lambda_4}^{(L')} \rangle
 = \sum_{c,c'=1}^3 M^{(L),[c] \, *}_{\lambda_1\lambda_2\lambda_3\lambda_4}
                 {\cal C\!C}_{cc'} 
                  M^{(L'),[c']}_{\lambda_1\lambda_2\lambda_3\lambda_4} \,,
\label{ColorSumGen}
\end{equation}
where the symmetric matrix 
${\cal C\!C}_{cc'} \equiv \sum_{\rm colors} \trc^{[c]\,*} \trc^{[c']}$ 
is
\begin{equation}
{\cal C\!C} = \frac{V}{N}\left(
\begin{array}{ccccccccc}
 V & -1 & N \\
-1 &  V & N \\
 N &  N & N^2
\end{array}
\right) \,.
\label{ColorSumMatrix}
\end{equation}
One can use \tab{heltable} to convert from the 
$\lambda_1\lambda_2\lambda_3\lambda_4$ helicity configuration label
to the label $h \in \{1,2,3,4,5 \}$.
The unpolarized partonic cross section is obtained from the helicity sum
\begin{equation}
 {\bar I}^{(L,L')} \equiv
\sum_{\lambda_i = \pm 1} I^{(L,L')}_{\lambda_1\lambda_2\lambda_3\lambda_4}\,,
\end{equation}
after the usual averaging over initial spins and inclusion of flux
factors.  For example, the helicity sum for the tree-level cross
sections for $\qqtogg$ and $\qgtogq$,
constructed from~\eqns{TreeAmps}{SpinorPhasesSquared} 
in either the HV or FDH scheme, {\it i.e.}, for any $\delta_R$, are
\begin{eqnarray}
 {\bar I}_\qqtogg^{(0,0)} 
&=& 2 V {t^2+u^2 \over tu} 
     \biggl( N {t^2+u^2\over s^2} - {1\over N} \biggr) \,, 
\label{TreeCrossSectionqqgg} \\
 {\bar I}_\qgtogq^{(0,0)} 
&=& - 2 V {s^2+u^2 \over su} 
     \biggl( N {s^2+u^2\over t^2} - {1\over N} \biggr) \,.
\label{TreeCrossSectionqggq}
\end{eqnarray}
%


\section{One-loop amplitudes}\label{OneloopSection}

The one-loop amplitudes for $\qqtogg$ were first evaluated through
$\Ord(\e^0)$ as an interference with the tree amplitude in the 
CDR scheme~\cite{EllisSexton}.  Later they were evaluated as helicity
amplitudes in the HV and FDH (or $\DRbar$) schemes~\cite{KSTfourparton}.

Because ${\bom I}^{(1)}$ contains terms of order $1/\e^2$, the 
${\bom I}^{(1)} | \cm^{(1)} \rangle_\RS$ term in the infrared
decomposition~(\ref{TwoloopCatani}) of the two-loop $\qqtogg$ amplitudes
requires the series expansions of
the one-loop amplitudes through $\Ord(\e^2)$.
In section~\ref{AllOrdersSubsection} we present the all-order 
results in the color bases~(\ref{TraceBasisqqgg}), (\ref{TraceBasisqggq}),
with the normalizations implicit in \eqn{RenExpand},
in terms of integral functions whose series expansions are known to 
the requisite order~\cite{AllPlusTwo,BhabhaTwoLoop}.  

In ref.~\cite{BDDgggg} we showed that $\Ord(\e)$ terms in one-loop
amplitudes such as $\cm_\qqtogg^{(1)}$ are not required for the 
construction of a numerical NNLO program, once such terms have been 
subtracted from $\cm_\qqtogg^{(2)}$ in the framework of 
ref.~\cite{Catani}.  Thus we need only present explicit formulae for
the finite remainders $\cm_\qqtogg^{(1){\rm fin}}$
of the one-loop amplitudes, after ultraviolet 
renormalization~(\ref{OneloopCounterterm})
and subtraction of infrared divergences~(\ref{OneloopCatani}).
We do this in section~\ref{OneloopRemainderSubsection}, for a 
general $\delta_R$ scheme.

In section~\ref{Compare11CDRSection} we give a formula for the
contribution of the finite remainders $\cm_\qqtogg^{(1){\rm fin}}$
to the NNLO cross section.  We compare the result of evaluating the 
formula in the HV scheme to other computations in both the HV and CDR 
schemes~\cite{AGTYPrivate}.

Actually, the formulas~(\ref{Twoloopc12}) and~(\ref{Twoloopc3})
for converting the two-loop finite remainders
$\cm_\qqtogg^{(2){\rm fin}}$ from one scheme to another are 
most compactly presented in terms of the $\delta_R$-dependent parts 
of the one-loop amplitudes at order $\e$; these quantities are 
presented in appendix~\ref{OrdepsdeltaRemainderAppendix}.


\subsection{All orders in $\e$ QCD amplitudes}
\label{AllOrdersSubsection}

We now present the renormalized one-loop $\qqtogg$
amplitudes in the color bases~(\ref{TraceBasisqqgg}), (\ref{TraceBasisqggq}),
with the normalizations implicit in \eqn{RenExpand}, in a form
valid to all orders in $\e$.

At one loop the crossing properties of the amplitudes are relatively 
simple, so we
present the explicit values of the helicity amplitudes for the process
$q \bar q \rightarrow gg$.  The $q g \rightarrow g q$ process may be
obtained from these by crossing the antiquark into the final state,
and gluon 4 into the initial state,
\begin{eqnarray}
&& M^{(1),[c]}_3 (s,t,u) = M^{(1),[c]}_1 (t,s,u)  \,, 
\label{onelooph3cross} \\
&& M^{(1),[c]}_4 (s,t,u) = M^{(1),[c]}_2 (t,s,u)  \,, 
\label{onelooph4cross} \\
&& M^{(1),[c]}_5 (s,t,u) = {s \over u} \, M^{(1),[c']}_2 (t,u,s) \,,
\label{onelooph5cross}
\end{eqnarray}
where $M^{(L),[c]}_h$ is defined in eq.~(\ref{RemoveColorPhase}) with
the color bases (\ref{TraceBasisqqgg}) and (\ref{TraceBasisqggq})
using the helicity configurations $h$ defined in
eqs.~(\ref{hel1})--(\ref{hel5}).  The factor of $s/u$ in the third
relation accounts for the permutation of the $S_2$ prefactor in
eq.~(\ref{SpinorPhases}).  The color label $c'$ needed for the $h=5$
case~(\ref{onelooph5cross}) is given by $c' = 2,1,3$ for $c = 1,2,3$,
respectively.  After crossing, appropriate analytic continuations are
required to bring each function into the physical region; for the
finite parts for convenience we will give the explicit forms in the
different analytic regions.

It is convenient to give the amplitudes in terms of ``primitive''
amplitudes, which are color-stripped building blocks for full
amplitudes in any color representation.  For the case of one-loop
amplitudes with a single external fermion pair, the explicit relations
between the primitive amplitudes and the color decomposed amplitudes
were presented in ref.~\cite{Oneloopqqggg}.  Here we quote the results
for the decomposition and then present the primitive amplitudes.  The
same amplitudes, but in a form valid only through $\Ord(\e^0)$, as
needed in an NLO calculation, may be found in
refs.~\cite{KSTfourparton,Oneloopqqggg}.

The first coefficient in the color basis~(\ref{TraceBasisqqgg}) for
$\qqtogg$ at one loop is expressed in terms of (unrenormalized)
primitive amplitudes as~\cite{Oneloopqqggg}
\begin{eqnarray}
M^{(1),[1]}_1(s,t,u)
& = & N \,   A^L(1_q^{+}, 2_\qb^{-}, 3_g^{+}, 4_g^{+})
    - {1\over N} \,  A^R(1_q^{+}, 2_\qb^{-}, 3_g^{+}, 4_g^{+})
 \PlusBreak{} 
   \Nf \, A^{L,[1/2]}(1_q^{+}, 2_\qb^{-}, 
                             3_g^{+}, 4_g^{+})
\,, \label{oneloopM11} \\
M^{(1),[1]}_2(s,t,u)
& = & N \,  A^L(1_q^{+}, 2_\qb^{-}, 3_g^{-}, 4_g^{+})
    - {1\over N} \,  A^R(1_q^{+}, 2_\qb^{-}, 3_g^{-}, 4_g^{+})
 \PlusBreak{} 
   \Nf \, A^{L,[1/2]}(1_q^{+}, 2_\qb^{-}, 
                             3_g^{-}, 4_g^{+})
     - {b_0 \over \e} M^{(0),[1]}_2 \,, \label{oneloopM21}
\end{eqnarray}
where, as defined in ref.~\cite{Oneloopqqggg},
the $L$ and $R$ superscripts refer to whether the 
external fermion line turns ``left'' or ``right'' upon entering a diagram.
The ``[1/2]'' designation on the last primitive amplitude represents
the subset of contributions with a closed spin 1/2 fermion.
We continue to label the helicities using the ``standard'' convention 
with legs 1 and 2 incoming and legs 3 and 4 outgoing.

The coefficient of the second color factor for $h=1$
in the basis~(\ref{TraceBasisqqgg})
is obtained from the first one by permuting kinematics 
and helicity labels,
\begin{equation}
M^{(1),[2]}_1(s,t,u) = -{t \over u} M^{(1),[1]}_1(s,u,t) 
\,, \label{oneloopM12}
\end{equation}
where the $-t/u$ prefactor accounts for the implicit permutations
of the removed $S_h$ prefactors in eq.~(\ref{SpinorPhases}).
The coefficient of the second color factor for $h=2$ is given by
\begin{eqnarray}
M^{(1),[2]}_2(s,t,u)
&=&  \frac{t}{u} \Bigl( N \,  A^L(1_q^{+}, 2_\qb^{-}, 4_g^{+}, 3_g^{-})
    - {1\over N} \,  A^R(1_q^{+}, 2_\qb^{-}, 4_g^{+}, 3_g^{-})
 \PlusBreak{} 
   \Nf \, A^{L,[1/2]}(1_q^{+}, 2_\qb^{-}, 
                             4_g^{+}, 3_g^{-}) \Bigr)
     - {b_0 \over \e} M^{(0),[2]}_2 \,.
\label{oneloopM22}
\end{eqnarray}

For the third color structure in eq.~(\ref{TraceBasisqqgg}), we have
\begin{eqnarray}
M^{(1),[3]}_1(s,t,u) & = &
   A^L(1_q^{+}, 2_\qb^{-}, 3_g^{+}, 4_g^{+}) + 
   A^R(1_q^{+}, 2_\qb^{-}, 3_g^{+}, 4_g^{+}) 
  - {t \over u} \, A^L(1_q^{+}, 2_\qb^{-}, 4_g^{+}, 3_g^{+}) 
 \MinusBreak{ }
  {t \over u} \, A^R(1_q^{+}, 2_\qb^{-}, 4_g^{+}, 3_g^{+})
\nonumber\\
 & =& 0\,, \label{oneloopM13} \\[1pt plus 4pt]
M^{(1),[3]}_2(s,t,u) & = &
   A^L(1_q^{+}, 2_\qb^{-}, 3_g^{-}, 4_g^{+}) + 
   A^R(1_q^{+}, 2_\qb^{-}, 3_g^{-}, 4_g^{+}) +
  {t\over u} \, A^L(1_q^{+}, 2_\qb^{-}, 4_g^{+}, 3_g^{-}) 
 \PlusBreak{ }
  {t\over u}\, A^R(1_q^{+}, 2_\qb^{-}, 4_g^{+}, 3_g^{-}) + 
  2 {s\over u} \, A^R(1_q^{+}, 4_g^{+}, 2_\qb^{-}, 3_g^{-}) 
      \,,  \label{oneloopM23} 
\end{eqnarray}
where again the ratios of kinematic invariants (including signs) account for
the permutations of the extracted overall prefactors (\ref{SpinorPhases})
from the amplitudes with standard ordering of legs $(1_q,2_\qb,3_g,4_g)$.
The vanishing in \eqn{oneloopM13} follows from a supersymmetry identity.

In order to compress the notation a little, we shall suppress the labels
of legs in the explicit formulas for the primitive amplitudes.
We always take them to be ordered (1,2,3,4) in the following formulas,
and apply permutations as required by 
eqs.~(\ref{oneloopM12})--(\ref{oneloopM23}).
That is, define
\begin{eqnarray}
\AL q \qb g g {{\lambda_1}} {{\lambda_2}} {{\lambda_3}} {{\lambda_4}} 
&\equiv &
A^L (1_q^{\lambda_1},2_\qb^{\lambda_2},3_g^{\lambda_3},4_g^{\lambda_4})
\,, \nonumber \\
\AR q \qb g g {{\lambda_1}} {{\lambda_2}} {{\lambda_3}} {{\lambda_4}} 
&\equiv &
A^R (1_q^{\lambda_1},2_\qb^{\lambda_2},3_g^{\lambda_3},4_g^{\lambda_4})
\,, \nonumber \\
\ALh q \qb g g {{\lambda_1}} {{\lambda_2}} {{\lambda_3}} {{\lambda_4}} 
&\equiv &
A^{L,[1/2]}(1_q^{\lambda_1},2_\qb^{\lambda_2},3_g^{\lambda_3},4_g^{\lambda_4})
\,, \nonumber \\
\AR q g \qb g {{\lambda_1}} {{\lambda_2}} {{\lambda_3}} {{\lambda_4}} &\equiv &
A^{R}(1_q^{\lambda_1},2_g^{\lambda_2},3_\qb^{\lambda_3},4_g^{\lambda_4})
\,. 
\end{eqnarray}
Then, for example, 
$A^L(1_q^+,2_{\bar{q}}^-,4_g^+,3_g^-) = 
\AL q \qb g g + - + - \bigr|_{t\lr u}$.

In this notation, the explicit values of the independent primitive
helicity amplitudes --- in terms of a set of scalar integral functions
--- are:
\begin{eqnarray}
\AL q \qb g g + - + + & = & 
 { \e {} (1- \e \delta_R) \, t \over 4 u} \Biggl[
  {\e \, {} ( (5-2\e) s + 2 t ) \over (1-\e) (1-2\e) (3-2\e) } \, \Trifour(s)
 \PlusBreak{\e (1- \e \delta_R) { t \over 4 u} \BigglBl} 
     {\e \, s \over (1-\e) (1-2\e) }  \, \Trifour(t)
    - s \, \Boxsix(s,t) \Biggr]
\,, \label{ALmpppqqgg}\\[1pt plus 4pt]
%
\AR q \qb g g + - + + & = &   
- \AL q \qb g g + - + +        
  - (1 - \e \delta_R) \ALh q \qb g g + - + + 
\,, \label{ARmpppqqgg}\\[1pt plus 4pt]
%
\ALh q \qb g g + - + + & = &          
 {\e^2 t \over  2 (1-\e) (1-2\e) (3-2\e) } \, \Trifour(s) 
\,, \label{ALhmpppqqgg} \\[1pt plus 4pt]
%
%
%
\AL q \qb g g + - - + & = &
 { (1 - \e \delta_R) s t \over 4 u^2 } \Biggl[
  { \e  \over (1-\e) (1-2 \e) }  
        \Bigl( ((3-\e) s + t) \, \Trifour(s)
 \PlusBreak{ { (1 - \e \delta_R) s \over 4 u^2 } \BigglBl }
              ((1-\e) s - t) \, \Trifour(t) \Bigr)
     - (2-\e) s  \, \Boxsix(s,t) \Biggr]
 \PlusBreak{ }
  { 1 \over 2 u }  \Biggl[ 
   { 1 \over 1 - 2 \e }  \Bigl( s {} ((1-2 \e) s+t) \, \Trifour(s)
 \PlusBreak{ { 1 \over 2 u }  \BigglBl { 1 \over 1 - 2 \e } \BiglP }
               2 t {} ((1-\e) s+(1-2 \e) t) \, \Trifour(t) \Bigr)
 \PlusBreak{ { 1 \over 2 u }  \BigglBl }
  ( (1-2\e) (s^2 + 2 t^2) + (1-4\e) s t ) \, \Boxsix(s,t) \Biggr]
\,, \label{ALmpmpqqgg}\\[1pt plus 4pt]
%
%
\AR  q \qb g g + - - + & = &
 - { (1 - \e \delta_R) s t \over 4 u^2 } \Biggl[ 
  { \e \over (1-\e) (1-2 \e) } \Bigl( ((1+\e) s - (1-2\e) t) \, \Trifour(s)
 \MinusBreak{ { (1 - \e \delta_R) s t \over 4 u^2 } \BigglBl 
              { \e \over (1-\e) (1-2 \e) } \BiglP }
              ((1-\e) s + (3-2 \e) t) \, \Trifour(t) \Bigr)
 \MinusBreak{ { (1 - \e \delta_R) s t \over 4 u^2 } \BigglBl  }
     (\e s - 2 (1-\e) t) \, \Boxsix(s,t) \Biggr]
 \PlusBreak{ }
 { s \over 2 u } \Biggl[ 
   { 1 \over 1-2\e } \Bigl( ((1-2\e) s + t) \, \Trifour(s)
               + 2 \e t \, \Trifour(t) \Bigr)
 \PlusBreak{ { s \over 2 u } \BigglBl  }
     ((1 - 2\e) s - t) \, \Boxsix(s,t) \Biggr]
\,, \label{ARmpmpqqgg}\\[1pt plus 4pt]
%
%
\ALh q \qb g g + - - + & = &
0 
\,, \label{ALhmpmpqqgg} \\[1pt plus 4pt]
%
%
%
\AL q \qb g g + - + -  & = &
 - { \e {} (1-\e \delta_R) s \over 4 u } \Biggl[
   { 1 \over (1-\e) (1-2\e) } \Bigl( ((1-\e) s + t) \, \Trifour(s)
                      + \e t \, \Trifour(t) \Bigr)
 \PlusBreak{ - { \e (1-\e \delta_R) s \over 4 u } \BigglBl  }
      s \, \Boxsix(s,t) \Biggr]
 \MinusBreak{   }
  { s \over 2 (1-2\e) } \Trifour(s) - t \, \Trifour(t)
 - { 1 \over 2 } (s + 2 (1-2\e) t) \Boxsix(s,t) 
\,, \label{ALmppmqqgg}\\[1pt plus 4pt]
%
%
\AR q \qb g g + - + - & = &
 - { \e {} (1-\e \delta_R) s \over 4 u } \Biggl[
   { 1 \over (1-\e) (1-2 \e) } \Bigl( ((1-\e) s + (1-2\e) t) \, \Trifour(s)
 \MinusBreak{ - { \e (1-\e \delta_R) s \over 4 u } \BigglBl XXX }
                      \e t \, \Trifour(t) \Bigr)
    + (s + 2 t) \, \Boxsix(s,t) \Biggr]
 \MinusBreak{ }
   { s \over 2 (1-2 \e) } \, \Trifour(s)
 - { s \over 2 } \, \Boxsix(s,t) 
\,, \label{ARmppmqqgg}\\[1pt plus 4pt]
%
\ALh q \qb g g + - + -  & = &
0 
\,, \label{ALhmppmqqgg} \\[1pt plus 4pt]
%
%
\AR q g \qb g + + - + & = &
 {1\over 4} \,  \e \, t {} (1- \e \delta_R) \Biggl[
   {\e \over (1-\e) (1-2\e)} \Bigl( \Trifour(s)
                                 +  \Trifour(t) \Bigr)
 - \Boxsix(s,t) \Biggr]
\,, \qquad~ \label{ARmppmqgqg}\\[1pt plus 4pt]
%
\AR q g \qb g + - - +  & = &
  { 1 \over 4 } \e {} (1-\e \delta_R) t{} \biggl[
  { 1 \over 1-2\e } \, \Trifour(t) 
     + \Boxsix(s,t) \biggr]
 -  { s \over 2 } \, \Trifour(s)
 \MinusBreak{ }
  { t \over 2 (1-2\e) } \Trifour(t)
- { 1 \over 2 } ((1-2\e) s + t) \, \Boxsix(s,t)
\,. \label{ARmmppqgqg}
\end{eqnarray}
In assigning helicity labels to the above primitive amplitudes 
we take the quark legs to be incoming.

Here $\Trifour(s)$ is the scalar triangle integral in 
$4-2\e$ dimensions with one external massive leg, 
and $\Boxsix(s,t)$ is the all-massless scalar box 
integral in $6-2\e$ dimensions.  The explicit value of the triangle 
integral is 
\begin{equation}
\Trifour(s) = -{\rg \over \e^2} \, (-s)^{-1-\e}\,, 
\label{IntDefs}
\end{equation}
where
\begin{eqnarray}
\rg & = & e^{-\e \psi(1)} \,
{\Gamma(1+\e) \Gamma^2(1-\e) \over \Gamma(1-2\e)} 
\nonumber \\
& = & 1 - {1\over2} \zeta_2 \, \e^2 - {7\over3} \zeta_3 \, \e^3 -
{47\over16} \zeta_4 \, \e^4 + \Ord(\e^5) \,,
\end{eqnarray}
with
\begin{equation}
\zeta_s \equiv \sum_{n=1}^\infty n^{-s} \,, \qquad \quad
\zeta_2 = {\pi^2\over6} \,, \qquad \zeta_3 = 1.202057\ldots, \qquad
\zeta_4 = {\pi^4\over90} \,.
\label{ZetaValues}
\end{equation}
In the $s$-channel ($s>0$), the $\e$-expansion of~\eqn{IntDefs} 
is given by using the analytic continuation $\ln(-s) \to \ln s - i\pi$.
The $D=6-2\e$ scalar box integral is completely finite as
$\e\to0$.  Its expansions to $\Ord(\e^2)$ in the various kinematic
channels are given, for example, in 
refs.~\cite{BhabhaTwoLoop,BDDgggg}.
%


\subsection{Finite remainders}\label{OneloopRemainderSubsection}

Next we tabulate the finite remainders of the one-loop $\qqtogg$
and $\qgtogq$ amplitudes at $\Ord(\e^0)$, defined by 
$\cm_\qqtogg^{(1){\rm fin}}$ and $\cm_\qgtogq^{(1){\rm fin}}$
in~\eqn{OneloopCatani} and color decomposed into 
$M^{(1),[c]{\rm fin}}_{h}$ in \eqn{RemoveColorPhase}. 
We write,
\begin{eqnarray}
 M^{(1),[c]{\rm fin}}_{h} &=& 
 \Bigl[ - b_0 (\ln(s/\mu^2) - i\pi) 
        + {C_F\over2} (1-\delta_R) \Bigr] M^{(0),[c]}_{h} \nonumber\\
&& \hskip1cm
 + N \, a^{[c]}_{h} + {1\over N} \, b^{[c]}_{h} + \Nf \, d^{[c]}_{h} \,, 
  \quad c = 1,2,
\label{OneloopRemainderDefabd} \\
 M^{(1),[c]{\rm fin}}_{h} &=& 
  h^{[c]}_{h} + {\Nf \over N} \, j^{[c]}_{h} \,, 
  \quad c = 3.
\label{OneloopRemainderDefhj}
\end{eqnarray}
For helicity configuration $h=1$, Bose symmetry
under exchange of legs 3 and 4 ($t \lr u$) implies 
(see \eqn{oneloopM12}) that
\begin{eqnarray}
a_{1}^{[2]}(s,t,u) &=& - {t\over u} \, a_{1}^{[1]}(s,u,t) \,,
\\
b_{1}^{[2]}(s,t,u) &=& - {t\over u} \, b_{1}^{[1]}(s,u,t) \,,
 \\
d_{1}^{[2]}(s,t,u) &=& - {t\over u} \, d_{1}^{[1]}(s,u,t) \,.
\label{abdSym}
\end{eqnarray}

For the $h=1$ amplitude, the independent remainder functions 
$a$, $b$, $d$, $h$ and $j$ are
\begin{eqnarray}
a_{1}^{[1]} &=& 
 - {x \over 6 } - { 1\over 4 }
\,, \label{a11}\\[1pt plus 4pt]
%
%
b_{1}^{[1]} &=& 
 - { 1 \over 4 }
\,, \label{b11}\\[1pt plus 4pt]
%
%
d_{1}^{[1]} &=& 
 { x \over 6 }
\,, \label{d11}\\[1pt plus 4pt]
%
%
h_{1}^{[3]} &=& 
  0
\,, \label{h13}\\[1pt plus 4pt]
%
%
j_{1}^{[3]} &=& 
  0
\,, \label{j13}
\end{eqnarray}
where
\begin{equation}
x = {t\over s} \, , \qquad y = {u\over s} \, , \qquad 
X = \ln\biggl(-{t\over s}\biggr) \, , \qquad 
Y = \ln\biggl(-{u\over s}\biggr) \, .
\label{VariableNames}
\end{equation}

For $h=2$ the functions are
\begin{eqnarray}
a_{2}^{[1]} &=& 
- { (x-y) (1-x y) \over 4 y^3 } \, X^2 
- { 6 x^2 - 3 x y + 11 y^2 \over 12 y^2 } \, X 
+ { 1 \over 4 y } - { 3 \over 2 } 
   \PlusBreak{ } 
i \pi{} \biggl[  - { (x-y) (1-x y) \over 2 y^3 } \, X
             - { 6 x^2 - 3 x y + 11 y^2 \over 12 y^2 } \biggr]
\,, \label{a21}\\[1pt plus 4pt]
%
%
b_{2}^{[1]} &=&  
 { X^2 \over 4 y^3 } - { 2-y \over 4 y^2 } \, X + {1 \over 4 y } + 2 
  +    i \pi {} \biggl[ { X \over 2 y^3 } - { 2-y \over 4 y^2 } \biggr]
\,, \label{b21}\\[1pt plus 4pt]
%
%
d_{2}^{[1]} &=&  
 { X \over 6} + {i \pi \over 6}
\,, \label{d21}\\[1pt plus 4pt]
%
%
a_{2}^{[2]} &=&  
 { x-y \over 4 y } \, Y^2 - {5 \over 3 } {x \over y } \, Y 
  - {3 \over 2 } { x \over y } 
+ i \pi {}\biggl[ { x-y \over 2 y } \, Y - { 5 \over 3 } { x \over y } \biggr]
\,, \label{a22}\\[1pt plus 4pt]
%
%
b_{2}^{[2]} &=&  
  { Y^2 \over 4 y } + 2 { x \over y }
   + i \pi \, { Y \over 2 y }
\,, \label{b22}\\[1pt plus 4pt]
%
%
d_{2}^{[2]} &=&  
  { x \over 6 y } \, Y + i \pi \, { x \over 6 y }
\,, \label{d22}\\[1pt plus 4pt]
%
%
h_{2}^{[3]} &=&  
 - { x {} (1 - 2 x) \over 4 y^2 } \, X^2 - { X Y \over y } - { Y^2 \over 2 } 
     + { 5 \over 3 } { x \over y } \, X + { 5 \over 3 } \, Y 
     + { \pi^2 \over 2 y }
   \PlusBreak{ }
  i \pi {} \biggl[ { 2 - x y + x^2 \over 2 y^2 } \, X + { x \over y } \, Y 
                - { 5 \over 3 y } \biggr]
\,, \label{h23}\\[1pt plus 4pt]
%
%
j_{2}^{[3]} &=&  
 - { x \over 6 y } \, X - {Y \over 6} + { i \pi \over 6 y } 
\,. \label{j23}
\end{eqnarray}

For $h=3$ the functions are
\begin{eqnarray}
a_{3}^{[1]} &=& 
  - { 1 \over 6 x } - { 1 \over 4 }
\,, \label{a31}\\[1pt plus 4pt]
%
%
b_{3}^{[1]} &=&  
   - { 1 \over 4 }
\,, \label{b31}\\[1pt plus 4pt]
%
%
d_{3}^{[1]} &=&  
  { 1 \over 6 x }
\,, \label{d31}\\[1pt plus 4pt]
%
%
a_{3}^{[2]} &=&  
  { 1 \over 6 x } + { 1 \over 4 y }  
\,, \label{a32}\\[1pt plus 4pt]
%
%
b_{3}^{[2]} &=&  
  { 1 \over 4 y }
\,, \label{b32}\\[1pt plus 4pt]
%
%
d_{3}^{[2]} &=&  
 - { 1 \over 6 x }
\,, \label{d32}\\[1pt plus 4pt]
%
%
h_{3}^{[3]} &=&  
    0
\,, \label{h33}\\[1pt plus 4pt]
%
%
j_{3}^{[3]} &=&  
    0
\,. \label{j33}
\end{eqnarray}

For $h=4$ the functions are
\begin{eqnarray}
a_{4}^{[1]} &=& 
  - { (1-y) (1-x y) \over 4 y^3 } \, X^2 
   + { 6 x^2 + 15 x y - 2 y^2 \over 12 y^2 } \, X
   + { x \over 4 y } - { 3 \over 2 } 
   \PlusBreak{  }
  i \pi {} \biggl[ - { (1-y) (1-x y) \over 2 y^3 } \, X
             +  { 6 x^2 + 15 x y - 2 y^2 \over 12 y^2 } \biggr] 
\,, \label{a41}\\[1pt plus 4pt]
%
%
b_{4}^{[1]} &=&  
  { x^3 \over 4 y^3 } \, X^2 +  { x {} (2 x-y) \over 4 y^2 } \, X 
  + { x \over 4 y } + 2
+  i \pi {}\biggl[ { x^3 \over 2 y^3 } \, X + { x {} (2 x-y) \over 4 y^2 }
        \biggr]
\,, \label{b41}\\[1pt plus 4pt]
%
%
d_{4}^{[1]} &=&  
  { X \over 6 } + { i \pi \over 6 }
\,, \label{d41}\\[1pt plus 4pt]
%
%
a_{4}^{[2]} &=&  
  { 1-y \over 4 y }  \Bigl( (X-Y)^2 + \pi^2 \Bigr) 
   - { X + 10 Y \over 6 y } - { 3 \over 2 y }
   - i \pi {}\,  { 11 \over 6 y }
\,, \label{a42}\\[1pt plus 4pt]
%
%
b_{4}^{[2]} &=&  
  { x \over 4 y }  \Bigl( (X-Y)^2 + \pi^2 \Bigr) + { 2 \over y }
\,, \label{b42}\\[1pt plus 4pt]
%
%
d_{4}^{[2]} &=&  
  { X+Y \over 6 y } + { i \pi \over 3 y }
\,, \label{d42}\\[1pt plus 4pt]
%
%
h_{4}^{[3]} &=&  
  - { x {} (1-2 x) \over 4 y^2 } \, X^2 - { Y^2 \over 2 }
   - { X Y \over y }  +  { \pi^2 \over 2 y } 
   + { 5 \over 3 } { x \over y } \, X + { 5 \over 3 } \, Y
  \PlusBreak{ }
  i \pi {} \biggl[  
  \biggl( - { x {} (1-2 x) \over 2 y^2 } - { 1 \over y } \biggr) \, X 
   + { x \over y } \, Y - { 5 \over 3 y } \biggr]
\,, \label{h43}\\[1pt plus 4pt]
%
%
j_{4}^{[3]} &=&  
 - { x \over 6 y } \, X - { 1 \over 6 } \, Y + { i \pi \over 6 y }
\,. \label{j43}
\end{eqnarray}

For $h=5$ the functions are
\begin{eqnarray}
a_{5}^{[1]} &=& 
  - { 1-y \over 4 y } \, X^2 - { X \over 6 } - { 3 \over 2 }
+ i \pi {}\biggl[  - { 1-y \over 2 y } \, X - { 1 \over 6 } \biggr]
\,, \label{a51}\\[1pt plus 4pt]
%
%
b_{5}^{[1]} &=&  
 { x \over 4 y } \, X^2  +  2   +  i \pi { x \over 2 y } \, X
\,, \label{b51}\\[1pt plus 4pt]
%
%
d_{5}^{[1]} &=&  
  { X \over 6 } + { i \pi \over 6 }
\,, \label{d51}\\[1pt plus 4pt]
%
%
a_{5}^{[2]} &=&  
 { (1-y) (1-x y) \over 4 y } \Bigl( (X-Y)^2 + \pi^2 \Bigr)
  - { 11 x^2 + 19 x y + 2 y^2 \over 12 y } (X-Y) 
 -  { 11 \over 6 y } \, Y 
   \MinusBreak{ }
   { 7 \over 4 y } - { 1 \over 4 }
 - i \pi {} { 11 \over 6 y }
\,, \label{a52}\\[1pt plus 4pt]
%
%
b_{5}^{[2]} &=&  
 { x^3 \over 4 y } \Bigl( (X-Y)^2 + \pi^2 \Bigr) 
  - { x {} (1-2 x) \over 4 y } (X-Y) + { 7 \over 4 y } - { 1 \over 4 } 
\,, \label{b52}\\[1pt plus 4pt]
%
%
d_{5}^{[2]} &=&  
  { X+Y \over 6 y } + { i \pi \over 3 y }
\,, \label{d52}\\[1pt plus 4pt]
%
%
h_{5}^{[3]} &=&  
 { 2-3 x y \over 4 y } \Bigl( (X-Y)^2 + \pi^2 \Bigr) 
  + { 1 \over 2 } \, X^2  +  { x \over 2 y } \, Y^2 
  + { 5 \over 3 } { x \over y } \, X + { 5 \over 3 } \, Y 
   \PlusBreak{ }
  i \pi {} \biggl[  X + { x \over y } \, Y - { 5 \over 3 y } \biggr]
\,, \label{h53}\\[1pt plus 4pt]
%
%
j_{5}^{[3]} &=&  
 - { x \over 6 y } \, X - { 1 \over 6 } \, Y + { i \pi \over 6 y }
\,. \label{j53}
\end{eqnarray}
For the HV scheme ($\delta_R=1$), the 
results~(\ref{OneloopRemainderDefabd})--(\ref{j53}) for the
finite remainders of the one-loop helicity amplitudes are in complete 
agreement with those of ref.~\cite{GTYqqgg}.


\subsection{Comparison with CDR results}\label{Compare11CDRSection}

Results in the CDR scheme are usually phrased in terms of
amplitude interferences, summed over all polarizations and colors.
In the NNLO cross section, the one-loop amplitude enters 
interfered with itself.
The contribution of the one-loop finite remainders to the
NNLO $\qqtogg$ or $\qgtogq$ cross section, summed over all 
helicities and colors, is given by
\begin{equation}
 {\bar I}^{(1,1){\rm fin}} \equiv
 \sum_{\lambda_i = \pm 1} 
\langle \cm_{\lambda_1\lambda_2\lambda_3\lambda_4}^{(1){\rm fin}} 
      | \cm_{\lambda_1\lambda_2\lambda_3\lambda_4}^{(1){\rm fin}} \rangle
\,.
\label{I11finite}
\end{equation}
Using the color sum matrix ${\cal C\!C}_{ij}$ in~\eqn{ColorSumMatrix},
and \tab{heltable} to relate 
$\lambda_1\lambda_2\lambda_3\lambda_4$ to $h$ helicity configurations,
the color and helicity sum in ${\bar I}^{(1,1){\rm fin}}$ may be
evaluated in terms of the above explicit
expressions~(\ref{OneloopRemainderDefabd})--(\ref{j53}) for
$M^{(1),[c]{\rm fin}}_{h}$, in a general $\delta_R$ scheme. 

In ref.~\cite{GOTYqqgg}, the one-loop-squared contributions to the
$\qqtogg$ and $\qgtoqg$ cross sections were given in the CDR scheme, 
but not in a convenient form for comparing to our results, because 
the infrared poles were not organized in the same way.   (Also, the terms
related to ultraviolet renormalization in eq.~(4.1) of the original
version of ref.~\cite{GOTYqqgg} require correction.)
However, the authors of ref.~\cite{GOTYqqgg} have kindly supplied us
with their versions of the values of \eqn{I11finite} in both the 
HV and CDR schemes.   Both these expressions agree precisely with our 
result for \eqn{I11finite} in the HV scheme ($\delta_R=1$).


\section{Two-loop QCD amplitudes and finite remainders}
\label{TwoLoopFiniteQCDSection}

We generated the Feynman graphs for $\qqtogg$ using {\tt
QGRAF}~\cite{QGRAF}, from which a {\tt MAPLE} program was constructed
to evaluate each graph.  As a cross-check, some of the diagrams were
evaluated using {\tt FORM}~\cite{FORM}.  We employed the general
integral reduction algorithms developed for the all-massless
four-point
topologies~\cite{PBReduction,IntegralsAGO,NPBReduction,Lorentz}, in
order to reduce the loop integrals to a minimal basis of master
integrals.  To put the integrands into a form suitable for applying
the general reduction algorithms, spinor strings were converted to
traces over $\gamma$ matrices, by multiplying and dividing by
appropriate spinor inner products constructed from the external
momenta.  Evaluating the traces then gave dot products of momenta; any
terms containing an odd number of Levi-Civita tensors
$\varepsilon^{\mu\nu\sigma\rho}$ vanished upon integration. The gluon
polarization vectors of definite helicity can be incorporated using
some minor extensions of the integral reduction
techniques~\cite{BDDgggg}.  Here we incorporated the gluon
polarization vectors in a slightly differently fashion than described
there.  When forming the traces over $\gamma$ matrices we included
also the polarization vectors expressed in terms of
spinors~\cite{SpinorHelicity}, following the methods used in, for
example, refs.~\cite{Neq4,BernMorgan}.  Because the polarization vectors
are four-dimensional objects, they distinguish between 4-dimensional and
$(-2\e)$-dimensional components of the loop momentum.
We evaluated the resulting integrals, containing $(-2\e)$-dimensional 
components of the loop momentum, using the methods of ref.~\cite{BDDgggg}.  
(See refs.~\cite{BGMvdB,TwoLoopee3JetsHel,GTYqqgg}
for alternative multi-loop helicity techniques.)

After all the tensor loop integrals in the amplitudes have been
reduced to a linear combination of master integrals, the next step is 
to expand the master integrals in a Laurent series in $\e$, beginning 
at order $1/\e^4$, using results from 
refs.~\cite{PBScalar,NPBScalar,PBReduction,NPBReduction,IntegralsAGO}.
It is straightforward~\cite{NielsenIds} to 
express the results solely in terms of polylogarithms~\cite{Lewin},
\begin{eqnarray}
\Li_n(x) &=& \sum_{i=1}^\infty { x^i \over i^n }
= \int_0^x {dt \over t} \Li_{n-1}(t)\,,  
\\
\Li_2(x) &=& -\int_0^x {dt \over t} \ln(1-t) \,, 
\label{PolyLogDef}
\end{eqnarray}
with $n=2,3,4$.  The analytic properties of the non-planar double 
box integrals appearing in the amplitudes are somewhat
intricate~\cite{AllPlusTwo,NPBScalar}; there is no Euclidean
region in any of the three kinematic channels, $s$, $t$ or $u$.
So we do not attempt to give a crossing-symmetric representation, but
instead quote all our results in the physical $s$-channel 
$(s > 0; \; t, \, u < 0)$ for both the $\qqtogg$ and $\qgtogq$
kinematics, \eqns{qqgglabel}{qggqlabel}.


\subsection{Finite remainders}\label{TwoLoopFinRemSubsection}

The two-loop finite remainders are defined in~\eqn{TwoloopCatani} and
are color decomposed into 
$M^{(2),[c]{\rm fin}}_h$ in \eqn{RemoveColorPhase}.  
Their dependence on the renormalization scale $\mu$, color factors 
$N$ and $\Nf$, and scheme label $\delta_R$ may be extracted as
\begin{eqnarray}
 M^{(2),[c]{\rm fin}}_h 
&=& 
 \Bigl[- b_0^2 \, (\ln(s/\mu^2) - i\pi)^2 
        - b_1 \, (\ln(s/\mu^2) - i\pi)
        + \Bigl( - {1\over6} C_A + {1\over8}  C_F \Bigr) 
             C_F (1-\delta_R)^2
  \PlusBreak{ ~~~~~ }
           \Bigl( 2 R_q + 2 R_g
            + b_0 Q_1^{(qg)} (\ln(s/\mu^2) - i\pi)
            + {1\over2} C_F Q_0 i \pi \Bigr)
                (1-\delta_R) \Bigr]
    M^{(0),[c]}_h
  \PlusBreak{ }
 \Bigl[ - 2 b_0 \, (\ln(s/\mu^2) - i\pi) \, 
      + Q_1^{(qg)} (1-\delta_R) \Bigr]
    M^{(1),[c]{\rm fin}}_h
  \PlusBreak{ }
   Q_0 \Bigl[ M^{(1),[c]\,\e,\delta_R}_h
            - C_A V^{(1),[c]}_h \Bigr]
       (1-\delta_R)
  \PlusBreak{ }
   N^2 \, A^{[c]}_h
 +        B^{[c]}_h
 + {1\over N^2} \, C^{[c]}_h
 + N \Nf \, D^{[c]}_h 
 + {\Nf \over N} \, E^{[c]}_h 
  \PlusBreak{ }
  \Nfsq \, F^{[c]}_h 
 + {\Nfsq \over N^2} \, G^{[c]}_h 
\,, 
  \hskip3cm   c = 1,2,
\label{Twoloopc12} \\
 M^{(2),[c]{\rm fin}}_h 
&=& 
 \Bigl[ - 2 b_0 \, (\ln(s/\mu^2) - i\pi) \, 
      + Q_1^{(qg)} (1-\delta_R) \Bigr]
    M^{(1),[c]{\rm fin}}_h
  \PlusBreak{ }
   Q_0 \Bigl[ M^{(1),[c]\,\e,\delta_R}_h
            - C_A V^{(1),[c]}_h \Bigr]
       (1-\delta_R)
  \PlusBreak{ }
   N \, H^{[c]}_h 
 + {1 \over N} I^{[c]}_h
 + \Nf \, J^{[c]}_h 
  + {\Nf \over N^2} \, K^{[c]}_h 
 + {\Nfsq\over N} \, L^{[c]}_h 
\,, 
  \hskip0.6cm c = 3.
\label{Twoloopc3}
\end{eqnarray}
The $\mu$-dependence is a consequence of renormalization group invariance.

The tree and one-loop functions,
$M^{(0),[c]}_h$
and $M^{(1),[c]{\rm fin}}_h$,
are given in~\eqn{TreeAmps} and 
eqs.~(\ref{OneloopRemainderDefabd})--(\ref{j53}),
respectively, while $b_0$ and $b_1$ are given
in~\eqn{QCDBetaCoeffs}.
The following combinations of color constants also appear in 
\eqns{Twoloopc12}{Twoloopc3},
\begin{eqnarray}
  Q_0 &=& {5\over6} C_A - C_F + {1\over3} T_R \Nf \,, 
\label{Q0} \\
  Q_1^{(qg)} &=&  - {1\over6} C_A + {1\over2} C_F \,,
\label{Q1qg} \\
  R_q &=&  - {7\over48} C_A^2 
  + \Bigl( {\pi^2\over 192} + {617\over864} \Bigr) C_A C_F
  - \Bigl( {\pi^2\over24} + {1\over4} \Bigr) C_F^2 
  - {1\over16} C_F T_R \Nf \,,
\label{Rq} \\
 R_g &=& \Bigl( - {5\over576} \pi^2 + {5\over48} \Bigr) C_A^2
      + \Bigl( {2\over27} C_A - {1\over8} C_F \Bigr) T_R \Nf \,,
\label{Rg}
\end{eqnarray}
where the combination $4 R_g$ appears as well in the $\delta_R$ dependence
of the $\ggtogg$ amplitude, eq. (5.10) of ref.~\cite{BDDgggg}.
The quantities $M^{(1),[c]\,\e,\delta_R}_h$ are the $\delta_R$-dependent
parts of the $\Ord(\e^1)$ coefficients of the one-loop amplitude
remainders, after subtracting the poles according to \eqn{OneloopCatani}.
The explicit values of $M^{(1),[c]\,\e,\delta_R}_h$ are tabulated
in appendix~\ref{OrdepsdeltaRemainderAppendix}.
Finally, the quantities $V^{(1),[c]}_h$ are only non-vanishing in
the case of the simpler helicity configurations, $h=1$ and $h=3$.
They seem to be related to the $\delta_R$ dependence of one-loop
splitting amplitudes~\cite{OneloopSplit}.
Their explicit values are given by
\begin{eqnarray}
V^{(1),[1]}_1 &=& {x \over 6} \,,  \qquad
V^{(1),[2]}_1 = - {x \over 6} \,,  \qquad
V^{(1),[3]}_1 =  0 \,,  
\label{V1c} \\
V^{(1),[1]}_3 &=& {1 \over 6x} \,,  \qquad
V^{(1),[2]}_3 = - {1 \over 6x} \,,  \qquad
V^{(1),[3]}_3 =  0 \,.  
\label{V2c}
\end{eqnarray}
The coefficient functions $A,B,C,D,E,F,G,H,I,J,K,L$ depend only on
the Mandelstam variables.  
In appendix~\ref{QCDRemainderAppendix},
we give the explicit forms for the independent finite remainder 
functions appearing in~\eqns{Twoloopc12}{Twoloopc3}.

We have compared our results for the independent two-loop finite
remainder functions $M^{(2),[c]{\rm fin}}_h$ 
with corresponding results obtained contemporaneously in 
ref.~\cite{GTYqqgg}.  The results agree completely, once a 
correction is made for a slightly different definition of 
${\bom H}^{(2)}(\e)$ in \eqn{OurH}.  (We always dress the $1/\e$ pole
with $(\mu^2/(-s))^{2\e}$; ref.~\cite{GTYqqgg} sometimes dresses it
with $(\mu^2/(-t))^{2\e}$ or $(\mu^2/(-u))^{2\e}$.)


\subsection{Comparison with CDR results}\label{Compare2treeCDRSection}

Finally we discuss conversion from the HV scheme results
reported in section~\ref{TwoLoopFiniteQCDSection} to the CDR scheme used
in ref.~\cite{GOTYqqgg}.  In the CDR scheme, one usually computes 
the interference of amplitudes, summed over all external colors and 
($2-2\e$) polarizations.  The generic one-loop/tree interference encountered
at NLO is
\begin{equation}
2 \, \Re \, \bar{I}^{(1,0)}_\RS 
\equiv 2 \, \Re \, \sum_{\rm color, hel.} 
\Bigl[ \langle \cm_n^{(1)} | \cm_n^{(0)} \rangle \Bigr]_\RS
 \,.
\label{OneLoopInterference}
\end{equation}
Inserting the infrared decomposition~(\ref{OneloopCatani}) for 
$\cm_n^{(1)}$ into~\eqn{OneLoopInterference} gives
\begin{equation}
\bar{I}^{(1,0)}_\RS = 2 \, \Re \, \sum_{\rm color, hel.} 
\Bigl[ \langle \cm_n^{(0)} | {\bom I}^{(1)} | \cm_n^{(0)} \rangle\Bigr]_\RS
+ \bar{I}^{(1,0){\rm fin}}_\RS 
 \,,
\label{OneloopInterfExp}
\end{equation}
where
\begin{equation}
\bar{I}^{(1,0){\rm fin}}_\RS = 
2 \, \Re \, \sum_{\rm color, hel.} 
\Bigl[ \langle \cm_n^{(1){\rm fin}} | \cm_n^{(0)} \rangle 
\Bigr]_\RS \,.
\label{OneloopInterfFinite}
\end{equation}
It is well-established from explicit calculations and general 
arguments~\cite{BKgggg,KSTfourparton,CST} that the finite 
remainder~(\ref{OneloopInterfFinite}) has the same value in the HV and 
CDR schemes, in the limit $\e \to 0$.  Essentially, the treatment of
unobserved partons is the same in both schemes, so the infrared 
divergences should take the same form, when expressed in terms of the 
lower-order-in-$\alpha_s$ amplitudes.  

It is natural to expect the same pattern to hold at two loops.
(Indeed it does for the $\ggtogg$ amplitude~\cite{GOTYgggg,BDDgggg}.)
The two-loop/tree interference is
\begin{eqnarray}
2 \, \Re \, \bar{I}^{(2,0)}_\RS 
&\equiv& 2 \, \Re \, \sum_{\rm color, hel.} 
\Bigl[ \langle \cm_n^{(2)} | \cm_n^{(0)} \rangle \Bigr]_\RS
\label{Interference} \\
&=&  2 \, \Re \sum_{\rm color, hel.}
\Bigl[ \langle \cm_n^{(0)} | {\bom I}^{(2)} 
                                 | \cm_n^{(0)} \rangle 
     + \langle \cm_n^{(1)} | {\bom I}^{(1) \dagger} 
                                 | \cm_n^{(0)} \rangle 
\Bigr]_\RS
+ \bar{I}^{(2,0){\rm fin}}_\RS ,
\label{InterfExp}
\end{eqnarray}
where
\begin{equation}
\bar{I}^{(2,0){\rm fin}}_\RS = 
2 \, \Re \, \sum_{\rm color, hel.} 
\Bigl[ \langle \cm_n^{(2){\rm fin}} | \cm_n^{(0)} \rangle 
\Bigr]_\RS \,.
\label{InterfFinite}
\end{equation}
Note that ${\bom I}^{(1)}$ and ${\bom I}^{(2)}$ are the 
same operators in the HV scheme as in the CDR scheme.

We have interfered the color-decomposed finite remainders of the
two-loop $\qqtogg$ and $\qgtogq$ helicity amplitudes in the HV scheme,
as given in section~\ref{TwoLoopFinRemSubsection}, with the tree
amplitudes given in~\eqn{TreeAmps}, summing over all external
helicities and colors with the help of~\eqn{ColorSumMatrix}.  This sum
gives precisely the same result as the corresponding
quantity~(\ref{InterfFinite}) in the CDR scheme, as evaluated in
ref.~\cite{GOTYqqgg}, after accounting for the slightly different
definition of ${\bom H}^{(2)}$ that we used in~\eqn{OurH}.  This
result provides additional evidence that~\eqn{InterfFinite} should be
the same in the HV or CDR schemes for general two-loop QCD scattering
amplitudes.  A similar conclusion has been reached 
independently~\cite{GTYqqgg}.


\section{Amplitudes in pure $N=1$ super-Yang-Mills theory}
\label{N=1AmplitudesSection}

QCD amplitudes may be converted easily to amplitudes in $N=1$ pure
super-Yang-Mills theory by modifying the fermions to be in the adjoint
representation and by altering their multiplicity to be a single
Majorana fermion.  This theory is then supersymmetric with the fermion
the gluino superpartner of the gluon.  Besides the inherent interest
in supersymmetric theories, a practical consequence is that
supersymmetry imposes a set of powerful identities that provide
non-trivial checks on the amplitudes, including their finite parts.
The supersymmetry identities have been applied previously to the same
one-loop amplitudes discussed here~\cite{KSTfourparton}, but only
through $\Ord(\e^0)$, as needed in an NLO calculation.  Here we
extend the one-loop discussion to include all orders in the
dimensional regularization parameter $\e$, which are relevant at
two loops via Catani's formula (\ref{TwoloopCatani}). Then we 
verify the identities at two loops.


\subsection{Supersymmetry Ward identities}
\label{SWISubsection}

One set of identities implies that ``maximal helicity violating''
amplitudes vanish for any supersymmetric theory and any number of
loops,
\begin{eqnarray}
\cm^{\rm SUSY}(g_1^\pm,g_2^-,g_3^+,\ldots, g_n^+) &=& 0,
\label{SUSYVanish} \\
\cm^{\rm SUSY}(\tg_1^+,\tg_2^-,g_3^+,\ldots, g_n^+) &=& 0,
\label{SUSYVanishGluinos}
\end{eqnarray}
where $g$ and $\tg$ denote a gluon and gluino, and the superscripts
denote helicities.  In this paper we use the convention that 
legs 1 and 2 are incoming and the remaining ones are outgoing.  
Other identities relate the non-vanishing
supersymmetric helicity amplitudes for external gluons alone, to
amplitudes where some of the gluons are replaced by gluinos.  In
particular, the two-gluino two-gluon amplitude can be expressed in
terms of the four-gluon amplitude:
\begin{eqnarray}
\cm^{\rm SUSY}(\tg_1^+,\tg_2^-,g_3^-,g_4^+) &=& 
{ \spa2.3 \over \spa1.3 } \, \cm^{\rm SUSY}(g_1^+,g_2^-,g_3^-,g_4^+).
\label{SUSYNonVanish}
\end{eqnarray}
This identity is somewhat more stringent than the ones in
\eqns{SUSYVanish}{SUSYVanishGluinos} because it 
relates distinct amplitudes containing the most intricate infrared 
divergences, up to order $1/\e^4$ poles at two loops.
These relations are crossing symmetric, when a crossing symmetric
definition~\cite{GunionKunszt} of the spinor products is used. 

For the four-point case, the pure-gluon identities~(\ref{SUSYVanish}) 
have already been checked at one loop to all orders in $\e$, and at two
loops through $\Ord(\e^0)$ in the FDH regularization
scheme~\cite{TwoLoopSUSY,BDDgggg}.


\subsection{Color and infrared structure}
\label{IRSusySubsection}

Since the gluinos are in the adjoint representation we use the same
color basis as used for the four-gluon helicity amplitudes~\cite{BDDgggg}
\begin{equation}
\tilde {\cal M}^{(L)}_h = 
  S_h \times
  \sum_{c=1}^9 \trc^{[c]} \times 
       \tilde M^{(L),[c]}_h \,,
\label{RemoveColorPhaseAdjoint}
\end{equation}
where
\begin{eqnarray}
&& \trc^{[1]} = \Tr(T^{a_1} T^{a_2} T^{a_3} T^{a_4})\,, \nonumber \\
&& \trc^{[2]} = \Tr(T^{a_1} T^{a_2} T^{a_4} T^{a_3})\,, \nonumber \\
&& \trc^{[3]} = \Tr(T^{a_1} T^{a_4} T^{a_2} T^{a_3})\,, \nonumber \\
&& \trc^{[4]} = \Tr(T^{a_1} T^{a_3} T^{a_2} T^{a_4})\,, \nonumber \\ 
&& \trc^{[5]} = \Tr(T^{a_1} T^{a_3} T^{a_4} T^{a_2})\,, \nonumber\\
&& \trc^{[6]} = \Tr(T^{a_1} T^{a_4} T^{a_3} T^{a_2})\,, \nonumber\\
&& \trc^{[7]} = \Tr(T^{a_1} T^{a_2}) \Tr(T^{a_3} T^{a_4})\,, \nonumber\\
&& \trc^{[8]} = \Tr(T^{a_1} T^{a_3}) \Tr(T^{a_2} T^{a_4})\,, \nonumber\\
&& \trc^{[9]} = \Tr(T^{a_1} T^{a_4}) \Tr(T^{a_2} T^{a_3})\,.
\label{TraceBasisAdj}
\end{eqnarray}
A reflection identity implies that the $c=4,5,6$ coefficients are
equal to the $c=3,2,1$ coefficients (respectively), so there are
really only six different coefficients for each $h$, namely
$\tilde M^{(L),[c]}_h$, $c=1,2,3,7,8,9$.

Using $C$, $P$, and $T$, for the case of two external gluons and two external
gluinos, there are five independent processes to consider, 
including processes related by crossing:   
\begin{eqnarray}
h=1: \quad   \tg(p_1,+) + \tg(p_2,-) &\to& g(p_3,+) + g(p_4,+)\,, 
\label{susyhel1} \\ 
h=2: \quad   \tg(p_1,+) + \tg(p_2,-) &\to& g(p_3,-) + g(p_4,+)\,, 
\label{susyhel2} \\ 
h=3: \quad   \tg(p_1,+) + g(p_2,-) &\to& g(p_3,+) + \tg(p_4,+)\,, 
\label{susyhel3} \\ 
h=4: \quad   \tg(p_1,+) + g(p_2,-) &\to& g(p_3,-) + \tg(p_4,+)\,,
\label{susyhel4} \\
h=5: \quad   \tg(p_1,+) + g(p_2,+) &\to& g(p_3,+) + \tg(p_4,+)\,.
\label{susyhel5} 
\end{eqnarray}
The latter three processes are obtained from the first two by crossing.
However, just as for the quark case we keep them distinct, because 
the crossing properties at two loops are in principle nontrivial.

The infrared divergence structure is similar to that of gluon-gluon
scattering amplitudes~\cite{BDDgggg}.  For
the case of $N=1$ pure super-Yang-Mills theory, in the basis
(\ref{TraceBasisAdj}) the matrix ${\bom I}^{(1)}$
is~\cite{GOTYgggg,BDDgggg}
{\small 
\begin{eqnarray}
&&{\bom {\tilde I}}^{(1)}(\e) = - {e^{-\e\psi(1)} \over \Gamma(1-\e)}
 \biggl( {1\over\e^2} + {\tilde b_0\over N\e} \biggr) \times
\label{explicitI1Susy} \\ 
&\times&\left( {
\begin{array}{ccccccccc}
N(\tS+\tT) & 0 & 0 & 0 & 0 & 0 & (\tT-\tU) & 0 & (\tS-\tU) \\
0 & N(\tS+\tU) & 0 & 0 & 0 & 0 & (\tU-\tT) & (\tS-\tT) & 0 \\
0 & 0 & N(\tT+\tU) & 0 & 0 & 0 & 0 & (\tT-\tS) & (\tU-\tS) \\
0 & 0 & 0 & N(\tT+\tU) & 0 & 0 & 0 & (\tT-\tS) & (\tU-\tS) \\
0 & 0 & 0 & 0 & N(\tS+\tU) & 0 & (\tU-\tT) & (\tS-\tT) & 0 \\
0 & 0 & 0 & 0 & 0 & N(\tS+\tT) & (\tT-\tU) & 0 & (\tS-\tU) \\
(\tS-\tU) & (\tS-\tT) & 0 & 0 & (\tS-\tT) & (\tS-\tU) & 2N\tS & 0 & 0 \\
0 & (\tU-\tT) & (\tU-\tS) & (\tU-\tS) & (\tU-\tT) & 0 & 0 & 2N\tU & 0 \\
(\tT-\tU) & 0 & (\tT-\tS) & (\tT-\tS) & 0 & (\tT-\tU) & 0 & 0 & 2N\tT
\end{array}
}
\right) \nonumber
\end{eqnarray}
}

\noindent
where $\tS$, $\tT$ and $\tU$ are defined in \eqn{STUedef}.
For $N=1$ super-Yang-Mills theory the first two coefficients of the
$\beta$-function are
\begin{equation}
\tilde b_0 = {3 \over 2 } C_A \,, \hskip 2 cm 
\tilde b_1 = {3 \over 2} C_A^2  \,.
\label{N=1SYBetaCoeffs}
\end{equation}
The ${\bom I}^{(2)}$ operator for super-Yang-Mills theory is
\begin{eqnarray}
{ \bom {\tilde I}}^{(2)}_{\FDH}(\e,\mu;\{p\}) 
& =& -\frac{1}{2} {\bom {\tilde I}}^{(1)}(\e,\mu;\{p\})
\left({\bom {\tilde I}}^{(1)}(\e,\mu;\{p\}) + {2 \tilde b_0 \over \e} \right)
  \PlusBreak{}
 {e^{+\e \psi(1)} \Gamma(1-2\e) \over \Gamma(1-\e)}
\left( {\tilde b_0 \over \e} + K^{\Susy}_{\FDH} \right) 
   {\bom {\tilde I}}^{(1)}(2\e,\mu;\{p\})
  \PlusBreak{}
  {\bom {\tilde H}}^{(2)}_{\FDH}(\e,\mu;\{p\}) \,,
\label{CataniI2Susy}
\end{eqnarray}
where 
\begin{equation}
K_\FDH^\Susy = \left( 3 - {\pi^2 \over 6} - {4 \over 9} \e \right) C_A ,   
\label{SUSYK}
\end{equation}
\begin{equation}
{\bom {\tilde H}}^{(2)}_\FDH(\e,\mu;\{p\}) = 
{e^{-\e\psi(1)} \over 4\e \, \Gamma(1-\e) }
 \biggl( { \mu^2 \over -s } \biggr)^{2\e} 
   \Bigl(4 (H_g^{(2)})^\Susy_\FDH  \, {\bom 1} 
       + \hat{\bom H}^{(2)} \Bigr) \,,
\label{OurSusyH}
\end{equation}
and
\begin{equation}
(H_g^{(2)})_\FDH^\Susy = 
(H_{\tilde g}^{(2)})_\FDH^\Susy = 
 \biggl( {\zeta_3 \over 2} + {\pi^2 \over 16} - {2 \over 9} \biggr) C_A^2 \,.
\label{SUSYH}
\end{equation}
The equality of $(H_g^{(2)})_\FDH^\Susy$ and $(H_{\tilde
g}^{(2)})_\FDH^\Susy$ is a consequence of supersymmetry.
Equations~(\ref{SUSYK})--(\ref{SUSYH})
are obtained from the QCD formulas, 
eqs.~(\ref{CataniK})--(\ref{Hgluon}), by the replacements
$\delta_R \rightarrow 0$, $C_F \rightarrow C_A$ and $T_R N_F
\rightarrow C_A/2$ for converting to a single adjoint fermion in
the FDH scheme.   The operator
$\hat{\bom H}^{(2)}$ defined in \eqn{HExtra} does not explicitly
depend on the fermion representation.

The tree amplitudes in this color basis are given by $\tilde
M^{(0),[c]}_h$, where
\begin{eqnarray}
\tilde M^{(0),[c]}_{1} &=& \tilde M^{(0),[c]}_{3} = 0, 
                       \hskip 1.34 cm c=1,2,\ldots,9,  \nonumber \\
 \tilde M^{(0),[c]}_{h} &=& 0\,, \hskip 3.0 cm c=7,8,9, \hskip .4 cm 
                                            h=1,2,3,4,5, \nonumber \\
\tilde M^{(0),[c]}_{h} &=& 1\,,          \hskip 3.0 cm  c = 1,6, 
                                           \hskip .4 cm h = 2,4,5, \nonumber \\
\tilde M^{(0),[c]}_{h} &=& {t\over u} \,,  \hskip 2.9 cm  c = 2,5,
                                           \hskip .4 cm h = 2,4,5, \nonumber \\
\tilde M^{(0),[c]}_{h} &=& {s\over u} \,,  \hskip 2.9 cm  c = 3,4,
                                           \hskip .4 cm h = 2,4,5.
\label{TreeAmpsAdj}
\end{eqnarray}
%


\subsection{One-loop amplitudes in pure $N=1$ super-Yang-Mills theory}
\label{OneLoopN=1SubSection}

We now present the results for one-loop two-gluino two-gluon
scattering in a format valid to all orders in $\e$.  By comparing these
results to the corresponding $N=1$ pure Yang-Mills 
four-gluon amplitudes~\cite{BDDgggg}, we then verify that the
identities (\ref{SUSYVanishGluinos}) and (\ref{SUSYNonVanish}) do
indeed hold at one loop to all orders in $\e$ when the FDH scheme is
used.

To check the supersymmetry Ward identities at one loop, we use a 
crossing symmetric representation of the amplitudes directly in terms 
of one-loop scalar integral functions.  Thus it is sufficient 
to explicitly present only the $h=1,2$ helicity cases.
Using formulae from ref.~\cite{Oneloopqqggg} for obtaining the 
$N=1$ supersymmetric amplitudes from the primitive amplitudes, 
the coefficients of the first color structure $\trc^{[1]}$ 
in \eqn{TraceBasisAdj} for $h=1,2$ are given by
\begin{eqnarray}
\tilde M^{(1),[1]}_1(s,t,u)
& = & N \Bigl( A^L(1_q^{+}, 2_\qb^{-}, 3_g^{+}, 4_g^{+})
       +  A^R(1_q^{+}, 2_\qb^{-}, 3_g^{+}, 4_g^{+})
 \PlusBreak{~~~~} 
      A^{L,[1/2]}(1_q^{+}, 2_\qb^{-}, 3_g^{+}, 4_g^{+}) \Bigr)
\,,  \label{onelooptildeM11} \\[1pt plus 4pt]
\tilde M^{(1),[1]}_2(s,t,u)
& = & N \Bigl(   A^L(1_q^{+}, 2_\qb^{-}, 3_g^{-}, 4_g^{+})
       +  A^R(1_q^{+}, 2_\qb^{-}, 3_g^{-}, 4_g^{+})
 \PlusBreak{~~~~} 
      A^{L,[1/2]}(1_q^{+}, 2_\qb^{-}, 3_g^{-}, 4_g^{+})  \Bigr)
    - {\tilde b_0 \over \e} \tilde M^{(0),[1]}_2 
\,.  \label{onelooptildeM21}
\end{eqnarray}
For the second color structure in \eqn{TraceBasisAdj}, we have
\begin{eqnarray}
\tilde M^{(1),[2]}_1(s,t,u)
& = & -N  \, {t \over u}  \Bigl(A^L(1_q^{+}, 2_\qb^{-}, 4_g^{+}, 3_g^{+})
                          +  A^R(1_q^{+}, 2_\qb^{-}, 4_g^{+}, 3_g^{+})
 \PlusBreak{~~~~~~~} 
      A^{L,[1/2]}(1_q^{+}, 2_\qb^{-}, 4_g^{+}, 3_g^{+})  \Bigr)
\,,  \label{onelooptildeM12} \\[1pt plus 4pt]
\tilde M^{(1),[2]}_2(s,t,u)
& = & N  \, {t \over u}  \Bigl(A^L(1_q^{+}, 2_\qb^{-}, 4_g^{+}, 3_g^{-})
                          +  A^R(1_q^{+}, 2_\qb^{-}, 4_g^{+}, 3_g^{-})
 \PlusBreak{~~~~~~~} 
      A^{L,[1/2]}(1_q^{+}, 2_\qb^{-}, 4_g^{+}, 3_g^{-})  \Bigr)
    - {\tilde b_0 \over \e} \tilde M^{(0),[2]}_2 
\,.  \label{onelooptildeM22} 
\end{eqnarray}
The third color configuration $\trc^{[3]}$ has coefficients given by,
\begin{eqnarray}
\tilde M^{(1),[3]}_1(s,t,u)
   & = & 0
\,,  \label{onelooptildeM13}  \\[1pt plus 4pt]
\tilde M^{(1),[3]}_2(s,t,u)
& = & 2 N \, {s\over u} \, A^{R}(1_q^+, 3_g^-, 2_\qb^-, 4_g^+)  
  - {\tilde b_0 \over \e} \tilde M^{(0),[3]}_2 
  \label{onelooptildeM23} \,.
\end{eqnarray}
In the above, $\tilde b_0$ is the first $\beta$-function coefficient for $N=1$
super-Yang-Mills theory.
The remaining color coefficients are given in terms of 
these~\cite{OneloopColor,Neq4},
\begin{eqnarray}
\tilde M^{(1),[4]}_h(s,t,u) &=& \tilde M^{(1),[3]}_h(s,t,u) \nonumber \,, \\
\tilde M^{(1),[5]}_h(s,t,u) &=& \tilde M^{(1),[2]}_h(s,t,u) \nonumber \,, \\
\tilde M^{(1),[6]}_h(s,t,u) &=& \tilde M^{(1),[1]}_h(s,t,u) \nonumber \,, \\
\tilde M^{(1),[7]}_h(s,t,u) &=& {2 \over N}  \Bigl(
         \tilde M^{(1),[3]}_h(s,t,u) + 
         \tilde M^{(1),[2]}_h(s,t,u) +
         \tilde M^{(1),[1]}_h(s,t,u)  \Bigr) \,, \hskip 2 cm \nonumber \\
\tilde M^{(1),[8]}_h(s,t,u) &=& \tilde M^{(1),[9]}_h(s,t,u) = 
        \tilde M^{(1),[7]}_h(s,t,u) \,, 
\label{OtherSusyHelicities}
\end{eqnarray}
for all helicity configurations $h$.

Inserting the explicit values of the primitive amplitudes and taking
the FDH scheme ($\delta_R = 0$), yields for the three independent
color factors of the $h=1$ helicity configuration in \eqn{susyhel1},
\begin{eqnarray}
\tilde M^{(1),[1]}_1(s,t,u) & = & 0
\,,  \label{ExplicitHel11Susy} \\
\tilde M^{(1),[2]}_1(s,t,u) & = & 0
\,,  \label{ExplicitHel12Susy} \\
\tilde M^{(1),[3]}_1(s,t,u) & = & 0
\,. \label{ExplicitHel13Susy}
\end{eqnarray}
For the $h=2$ helicity configuration (\ref{susyhel2}), 
\begin{eqnarray}
\tilde M^{(1),[1]}_2(s,t,u) & = &
N \biggl[ - {s\over 2 u}  { 2 u + \e {} (4 s + t) \over 1-2 \e} \,  \Trifour(s)
 - {t \over 2 u}  { 2 u + \e{} (4 t + s)  \over 1-2 \e} \, \Trifour(t)
  \PlusBreak{N \bigglB } 
 {1\over 2 u} \Bigl[2 s^2  + t s + 2 t^2  - \e {}( 4 s^2 
          + 5 t s + 4 t^2 )\Bigr] \Boxsix(s,t)  \biggr] 
 \MinusBreak{ } 
 {\tilde b_0 \over \e} \tilde M^{(0),[1]}_2
\,,  \label{ExplicitHel21Susy} \\[1pt plus 4pt]
\tilde M^{(1),[2]}_2(s,t,u) & = &  
 N \biggl[ - { 2 - \e \over 1 - 2 \e} {s t \over 2 u} \, \Trifour(s)
  -t \, \Trifour(u) 
  \PlusBreak{ N \bigglB} 
  { t {} (2 t + \e {} (4 u + s) ) \over 2 u} \,\Boxsix(s,u)  \biggr]
  - {\tilde b_0 \over \e} \tilde M^{(0),[2]}_2
 \,,  \label{ExplicitHel22Susy} \\[1pt plus 4pt]
\tilde M^{(1),[3]}_2(s,t,u) & = & 
 N \biggl[
 - {2- \e \over 1 - 2 \e}  {s t \over 2 u} \, \Trifour(t)
 -s \Trifour(u)
  \PlusBreak{ N \bigglB} 
  {s {} (2 s + \e {} (4 u + t)) \over 2 u}  \, \Boxsix(t,u) \biggr]
  - {\tilde b_0 \over \e} \tilde M^{(0),[3]}_2
\,.
\label{ExplicitHel23Susy}
\end{eqnarray}

For helicity $h=1$, the amplitudes clearly satisfy the supersymmetry
Ward identity~(\ref{SUSYVanishGluinos}) exactly. To check 
identity~(\ref{SUSYNonVanish}), we compared the amplitudes in
eqs.~(\ref{ExplicitHel21Susy}), (\ref{ExplicitHel22Susy}) 
and~(\ref{ExplicitHel23Susy}) to the corresponding four-gluon amplitudes.
The latter amplitudes may be obtained from section~3.1 of 
ref.~\cite{BDDgggg}, by letting $N_f \rightarrow N$ 
and $b_0 \rightarrow \tilde b_0$ in
eq.~(3.1) of that paper.  (Note that the ``$N=1$'' there refers to a
chiral multiplet consisting of a scalar and a fermion, while here we are
considering a pure super-Yang-Mills multiplet consisting of a gluon
and a gluino.)  The result of this comparison is that the
supersymmetry identity~(\ref{SUSYNonVanish}) holds to all orders in
$\e$ when the FDH scheme is used.

We note that the symmetry relation evident between
\eqns{ExplicitHel22Susy}{ExplicitHel23Susy},
\begin{equation}
\tilde M^{(1),[3]}_2(s,t,u) = \tilde M^{(1),[2]}_2(t,s,u) \,,
\end{equation}
which involves a swap of a gluon and gluino leg, is a consequence of
the Bose symmetry of the corresponding four-gluon amplitude, because
the gluon legs being exchanged in the latter amplitude have the 
same helicity (when both are considered as outgoing states).

After performing the subtraction of infrared divergences using 
\eqns{OneloopCatani}{explicitI1Susy}, we obtain
\begin{eqnarray}
\null \hskip -1 cm 
 M^{(1),\Susy,[c]{\rm fin}}_h &=& 
\Bigl[ - \tilde b_0 (\ln(s/\mu^2) - i\pi) 
     + {1\over 2} N \Bigr]  M^{(0),[c]}_h
 + N \, a^{\Susy,[c]}_h \,, 
\label{SUSYOneloopRemainderDef123} \\
&& \hskip 7.6 cm   \,  c = 1,2,3,
\nonumber \\
 M^{(1),\Susy,[c]{\rm fin}}_h &=& 
  g^{\Susy,[c]}_h \,,
  \hskip6.3cm c = 7,8,9,
\label{SUSYOneloopRemainderDef789}
\end{eqnarray}
where the one-loop supersymmetric remainder functions are given in
terms of the QCD ones,
\begin{eqnarray}
 a^{\Susy,[c]}_h &=&
  a^{[c]}_h - b^{[c]}_h +  d^{[c]}_h
\,, \hskip 3.5cm  c = 1,2,3,
\label{aSUSY} \\
  g^{\Susy,[c]}_h &=& 
    2 \Bigl( a^{\Susy,[1]}_h
           + a^{\Susy,[2]}_h
           + a^{\Susy,[3]}_h \Bigr)
\,,
 \qquad c = 7,8,9.
\label{gSUSY}
\end{eqnarray}
The $N M^{(0),[c]}_h/2$ term in \eqn{SUSYOneloopRemainderDef123} 
is a consequence of the finite shift between the FDH scheme 
used in super-Yang-Mills theory and the HV scheme used in QCD.


\subsection{Two-loop amplitudes in pure $N=1$ super-Yang-Mills theory}
\label{TwoLoopFiniteN=1Section}

An immediate consequence of the one-loop supersymmetry identities
holding to all orders in $\e$ is that all infrared divergent terms at
two loops also satisfy the same identities.  Since neither ${\bom
I}^{(1)}$ in \eqn{explicitI1Susy} nor ${\bom I}^{(2)}$ in
\eqn{CataniI2Susy} depend on whether the external lines are gluons or
gluinos, the Catani formula (\ref{TwoloopCatani}) dictates that
divergent terms at two loops must satisfy the same supersymmetry Ward
identities as the tree and one-loop amplitudes do.

This then leaves the question of whether the two-loop finite terms satisfy 
supersymmetry Ward identities.  For the case of four-gluon 
scattering, the finite remainder terms, after subtracting the
Catani terms (including their finite parts), have been presented 
previously~\cite{BDDgggg}.  We have carried out the
analogous subtraction using eqs.~(\ref{TwoloopCatani}), (\ref{explicitI1Susy})
and (\ref{CataniI2Susy}).   The result of this is, 
\begin{eqnarray}
 M^{(2),\Susy,[c]{\rm fin}}_h 
&=& 
 - \Bigl[ \bigl( \tilde b_0 \bigr)^2 \, (\ln(s/\mu^2) - i\pi)^2 
        + \tilde b_1 \, (\ln(s/\mu^2) - i\pi) \Bigr]
    M^{(0),[c]}_h
  \MinusBreak{ }
  2 \tilde b_0 \, (\ln(s/\mu^2) - i\pi) \, 
    M^{(1),\Susy,[c]{\rm fin}}_h
 + N^2 \, A^{\Susy,[c]}_h
 +        B^{\Susy,[c]}_h \,, 
\nonumber \\
&&  \hskip8cm   c = 1,2,3,
\label{SUSYTwoloopSingleTrace} \\
 M^{(2),\Susy,[c]{\rm fin}}_h 
&=& 
 - 2 \tilde b_0 \, (\ln(s/\mu^2) - i\pi) \, 
    M^{(1),\Susy,[c]{\rm fin}}_h
 + N \, G^{\Susy,[c]}_h \,, 
\nonumber \\
&&  \hskip8cm c = 7,8,9.
\label{SUSYTwoloopDoubleTrace} 
\end{eqnarray}
where we have verified that the finite remainder functions match those of
the pure gluon case, {\it i.e.}
\begin{eqnarray}
X_1^{\Susy,[c]} &=& X_3^{\Susy,[c]} = 0,
\label{h13SYM} \\
X_2^{\Susy,[c]} &=& X_4^{\Susy,[c]}
                 = - {s t \over u^2} X_{-+-+}^{\Susy,[c]},
\label{h24SYM} \\
X_5^{\Susy,[c]} &=& - {t\over s} X_{--++}^{\Susy,[c]},
\label{h5SYM}
\end{eqnarray}
where $X \in \{ A, B, G \}$.  The functions
$X_{-+-+}^{\Susy,[c]}$  and  $X_{--++}^{\Susy,[c]}$ for $\ggtogg$ 
scattering in pure $N=1$
super-Yang-Mills theory are given in ref.~\cite{BDDgggg}. (Note that
in that reference an all outgoing definition of helicity is used.)
The relations~(\ref{h13SYM})--(\ref{h5SYM}) are precisely equivalent to the
content of the supersymmetry Ward identities~\cite{SWI}, after removing
overall factors and the divergent terms, and extracting the $N$ and
$\mu$ dependence.  Thus they provide a direct, nontrivial check on the 
finite remainders.

Because the adjoint color indices of the gluino fields are identical 
to those of gluons, and only structure constants $f^{abc}$ appear
in the two-loop Feynman diagrams, the color coefficients for
two-gluino two-gluon scattering obey the same group theory relations 
identified for $\ggtogg$ in ref.~\cite{BDDgggg}, 
\begin{eqnarray}
G^{\Susy,[7]}_h & = &
 2 \Bigl( A^{\Susy,[1]}_h
   + A^{\Susy,[2]}_h
   + A^{\Susy,[3]}_h \Bigr) 
   - B^{\Susy,[3]}_h
\,, \label{GSUSY7elim}\\[1pt plus 4pt]
%
%
G^{\Susy,[8]}_h & = &
 2 \Bigl( A^{\Susy,[1]}_h
   + A^{\Susy,[2]}_h
   + A^{\Susy,[3]}_h \Bigr) 
   - B^{\Susy,[1]}_h
\,, \label{GSUSY8elim}\\[1pt plus 4pt]
%
%
G^{\Susy,[9]}_h & = &
 2 \Bigl( A^{\Susy,[1]}_h
   + A^{\Susy,[2]}_h
   + A^{\Susy,[3]}_h \Bigr) 
   - B^{\Susy,[2]}_h 
\,, \label{GSUSY9elim}
\end{eqnarray}
and
\begin{equation}
B^{\Susy,[3]}_h = 
    - B^{\Susy,[1]}_h
    - B^{\Susy,[2]}_h \,.
\label{BSUSY3elim}
\end{equation}
 

\section{Conclusions}
\label{ConclusionsSection}

In this paper we have presented the two-loop amplitudes for
quark-gluon scattering in QCD and gluino-gluon scattering in $N=1$
super-Yang-Mills theory, including the full dependence on external
colors and helicities.  We confirmed that, as in the case of 
gluon-gluon scattering, there is an additional
$1/\e$ pole term, $\hat{\bom H}^{(2)}$
in~\eqn{HExtra} having nontrivial color dependence.
This additional term vanishes after interfering it with the tree
amplitude and summing over colors.  We investigated the dependence of
the amplitudes on the variety of dimensional regularization employed.
The FDH scheme respects the supersymmetry Ward identities, in what
represents the most detailed test to date of these identities, 
and of the FDH scheme. 
The scheme dependence of the quark-gluon amplitudes is much more 
intricate than was the case for the four-gluon helicity 
amplitudes~\cite{BDDgggg}.

The two-loop QCD results, when interfered with the tree amplitude,
summed over all external colors and helicities, and converted to the CDR
scheme, are in complete agreement with the previous results of
Anastasiou, Glover, Oleari, and Tejeda-Yeomans~\cite{GOTYqqgg}.  We
also expressed the one-loop-squared contribution to the NNLO $\qqtogg$
and $\qgtogq$ cross sections in terms of one-loop finite remainders.
These also agree with suitably subtracted expressions in the CDR 
scheme~\cite{AGTYPrivate}.

So far the new two-loop amplitudes have been implemented in only a
handful of phenomenological studies~\cite{PhotonPaper,Higgs}.  We may
anticipate that once general algorithms for dealing with infrared
divergent phase space integrations at next-to-next-leading-order 
are developed (see {\it e.g.} ref.~\cite{NNLOPS}), many
more phenomenological applications will follow.  These applications
would include the implementation of the two-loop amplitudes of this 
paper, or those of ref.~\cite{GOTYqqgg}, as ingredients in 
a numerical program for computing dijet production cross sections
at hadron colliders at NNLO in QCD.
When this task is accomplished, the intrinsic precision on the
QCD predictions should reach the few percent level, providing a
stringent test of the Standard Model at short distances.

\acknowledgments We thank Babis Anastasiou, David Kosower and Henry
Wong for helpful comments. We also thank Babis Anastasiou, Nigel
Glover and Maria Elena~Tejeda-Yeomans for discussions regarding the
one-loop-squared interference, and Nigel Glover and Maria
Elena~Tejeda-Yeomans for communicating their results to us prior to
publication.  We thank Vittorio Del Duca, Giulio Falcioni, Lorenzo
Magnea and Leonardo Vernazza for pointing out errors in
eq.~(\ref{oneloopM22}) in an earlier version of this paper.
Z.B. thanks SLAC, and L.D. thanks UCLA, for hospitality while this
paper was being completed.


\appendix

\section{Finite remainder functions for QCD}
\label{QCDRemainderAppendix}

In this appendix, we present the explicit forms for the independent
finite remainder functions for the processes $\qqtogg$ and $\qgtogq$
in QCD, which appear in~\eqns{Twoloopc12}{Twoloopc3}.  

For the helicity $h=1$ configuration in \eqn{hel1} and color
factor $\trc^{[1]}$ in \eqn{TraceBasisqqgg}, the finite remainder 
functions are:
%
\begin{eqnarray}
A^{[1]}_1 & = &
 - {1 \over 48 y^2}( 30 x^3 + 51 x^2 y + 22 x y^2 - 3 y^3 ) X^2 
   \MinusBreak{}
     {1 \over 144 y} (180 x^2 + 239 x y + 99 y^2 ) X 
      + {25 \over 864} x + {475 \over 288} y 
   \MinusBreak{}
     i \pi {} \biggl[ 
        {1 \over 24 y^2}( 30 x^3 + 51 x^2 y + 22 x y^2 - 3 y^3 ) X 
   \PlusBreak{\null - i \pi \BigglBl}
         {1 \over 144 y}(180 x^2 + 239 x y + 99 y^2) \Biggl]
\,, \label{AA11}\\[1pt plus 4pt]
%
B^{[1]}_1 & = &
        {x{}(y-4) \over 16 y^2} X^2 + {x-y \over 16 x} Y^2
        + {x \over 16} (1-6 y) \Bigl( (X-Y)^2 + \pi^2 \Bigr)
 \MinusBreak{}
          {1 \over 8 y} (4 x^2 + 5 x y + 7 y^2) X - {3 \over 8} (x-y) Y
        + {115 \over 72} x + {139 \over 72} y
 \PlusBreak{}
  i\pi {} \biggl( -{x{}(4-y) \over 8 y^2} X +{x-y \over 8 x} Y 
       -{1 \over 2 y}
\biggr)
\,, \label{BB11}\\[1pt plus 4pt]
%
C^{[1]}_1 & = &
         {y^2 - 2 x^2 \over 16 y^2} X^2 + {x-y \over 16x} Y^2
        + {x{}(1-6y) \over 16} \Bigl( (X-Y)^2 + \pi^2 \Bigr)
 \PlusBreak{}
    {1 \over 16 y}(4 x^2 + 13 x y - 3 y^2) X 
         - {3 \over 8} (x-y) Y - {17 \over 32}
 \PlusBreak{}
    i\pi {} \biggl( { y^2 - 2x^2 \over 8 y^2} X + 
               {x-y \over 8 x} Y + {3-x \over 16 y} \biggr)
\,, \label{CC11}\\[1pt plus 4pt]
%
D^{[1]}_1 & = &
     {x \over 24 y^2} (5 - x y + x^2) X^2 + {1 \over 72 y} 
      (9 + 5 x y + 27 x^2) X
      + {143 \over 108} x 
- {37 \over 72} y 
 \PlusBreak{}
 i\pi {} \biggl( {x \over 12 y^2} (5 - x y + x^2) X 
           + {1 \over 72 y}(9 + 5 x y + 27 x^2) \biggr) 
\,, \label{DD11}\\[1pt plus 4pt]
%
E^{[1]}_1 & = &
          {x{}(2 x - 1) \over 8 y^2} X^2 + {1 - 2 y \over 8 x} Y^2
         + {x{}(x-y) \over 8} \Bigl( (X-Y)^2 + \pi^2 \Bigr) 
 \PlusBreak{}
           {1 \over 8 y}(y^2 + 7 x y - 6 x^2) X - {3 \over 4}(x-y) Y 
         + {7 \over 36} x - {19 \over 72} y
 \PlusBreak{}
 i\pi {} \biggl( -{x{}(1-2x) \over 4 y^2} X + {1-2y \over 4 x} Y + 
               {6 x-7 y \over 8 y} \biggr)  
\,, \label{EE11}\\[1pt plus 4pt]
%
F^{[1]}_1 & = &  
{x \over 36} X - {5 \over 54} x + {i \pi \over 36} x 
\,, \label{FF11}\\[1pt plus 4pt]
%
G^{[1]}_1 & = & 
0
%
\,. \label{GG11}
\end{eqnarray}

\noindent
Bose symmetry under exchanges of legs 3 and 4 $(t \leftrightarrow u )$
implies that for $h=1$ the finite remainders of the second color
configuration in \eqn{TraceBasisqqgg} can be expressed in terms
of the first ones, {\it i.e.} 
\begin{equation}
Z_1^{[2]}(s,t,u) =
  - {t \over u}\,  Z_1^{[1]}(s,u,t)
\,, \label{ZZ12}
\end{equation}
where $Z \in \{ A, B, C,D,E,F,G \}$.

\noindent
For $h=1$ in \eqn{hel1}
and color factor $\trc^{[3]}$ in \eqn{TraceBasisqqgg}:
\begin{eqnarray}
H^{[3]}_1 & = &
 {1 \over 48 y^2} (-6 - 25 x + 7 x^2 -4 x^3) X^2
  -{x \over 18 y} (11 + 32 x) X
 \MinusBreak{}
   {1 \over 48 x y} (- 6 - 25 y + 7 y^2 - 4 y^3 ) Y^2
  +{1 \over 18} (11 + 32 y) Y
 \MinusBreak{}
   {x - y \over 8 y} (1 - x y) \Bigl( (X-Y)^2 + \pi^2 \Bigr) 
 \PlusBreak{}
    i\pi {} \biggl[
                   {1 \over 24 y^2} (- 6 -25 x + 7 x^2 -4 x^3) X
 \MinusBreak{\null +  i\pi \bigglB }
                   {1 \over 24 x y} (- 6 - 25 y + 7 y^2 -4 y^3) Y
                 + {7 (x - y) \over 6 y} \biggr]
\,, \label{HH13}\\[1pt plus 4pt]
I^{[3]}_1 & = & 
 - {1 - x \over 16 y} Y^2
 + { 1 - y \over 16 y} X^2
 - {x - y \over 8 y} (1 + 2 x y) \Bigl((X - Y)^2 + \pi^2 \Bigr)
 \PlusBreak{}
   {1 \over 12} (1 + 12 x) Y
 - {x \over 12 y} (1 + 12 y) X
 \PlusBreak{}
  i\pi {} \biggl(
                 {1 - y \over 8 y} X
               - {1-x \over 8 y} Y
               - {x - y \over 12 y} \biggr)
\,, \label{II13}\\[1pt plus 4pt]
J^{[3]}_1 & = & 
 {x{} (2 x^2 + x + 5) \over 24 y^2} X^2
 -{2 y^2 + y + 5 \over 24 x} Y^2
+{x \over 72 y} (39 x - 55 y) X
 \MinusBreak{}
  {1 \over 72} (39 y - 55 x) Y
 -{1 \over 4} x{}(x - y) \Bigl( (X - Y)^2 + \pi^2 \Bigr)
 \PlusBreak{}
  i\pi {} \biggl(
                {x \over 12 y^2} (2 x^2 + x + 5) X
              - {2 y^2 + y + 5 \over 12 x} Y
              - {13 (x - y) \over 24 y} \biggr)
\,, \label{JJ13}\\[1pt plus 4pt]
K^{[3]}_1 & = & 
  {x{} (1 - 2 x) \over 8 y^2} X^2
 -{1 - 2 y \over 8 x} Y^2
 -{1 \over 8} x{}(x-y) \Bigl( (X - Y)^2 + \pi^2 \Bigr)
 \PlusBreak{}
  {1 \over 24} (17 x - 19 y) Y
 -{x \over 24 y} (17 y - 19 x) X
 \PlusBreak{}
  i\pi {} \biggl(
                {x{} (1 - 2 x) \over 4 y^2} X
               -{1 - 2 y \over 4 x} Y
               -{19 (x - y) \over 24 y} \biggr)
\,, \label{KK13}\\[1pt plus 4pt]
L^{[3]}_1 & = & 
{1 \over 36} x{} (X - Y)
\,. \label{LL13}
\end{eqnarray}


\noindent
For $h=2$ in \eqn{hel2}
and color factor $\trc^{[1]}$ in \eqn{TraceBasisqqgg}:
\begin{eqnarray}
A_2^{[1]} & = & 
 {1 \over 2 y^3} \biggl[
     \left( x^3 \li3(-x) - \zeta_3 (x^3+2 y^3) \right) X
    -{1 \over 2} (3 x^3 + y^3) \li4(-x)
 \MinusBreak{ {1 \over 2 y^3} \bigglP }
     {1 \over 4} (x - y) (1 - x y) X^2 \li2(-x)
    -{1 \over 48} (7 x^3 - 3 y^3) X^4
 \PlusBreak{ {1 \over 2 y^3} \bigglP }
     {\pi^2 \over 3} (2 x^3 - y^3) X^2 
    +{\pi^4 \over 1440} (24 x^3 + 95 y^3)  
\biggr]
 \MinusBreak{}
  {1 \over 6} \pi^2 \li2(-x)
 +{1 \over 12} X Y{} \Bigl(X^2 - 2 \pi^2 \Bigr)
%
 \PlusBreak{} 
  {1 \over 144 y^3} \biggl[ 
    (- 24 x^2 y - 77 y^3 + 21 x y^2 + 86 x^3) X^3 
 \PlusBreak{\null + {1 \over 144 y^3} \bigglP }
   (576 y + 401 y^3 + 684 y^2 + 172) \pi^2 X \biggr]
 \PlusBreak{}
  {1 \over 48 y^2} \biggl[
   (35 y^2 + 6 x^2 - 3 x y) ( 2\li2(-x) X + X^2 Y  - 2 \li3(-x) )
 \PlusBreak{\null +  {1 \over 48 y^2} \bigglP }
    (12 x y + 241 y^2 - 24 x^2) \zeta_3  
\biggr]
 - {1 \over 72 y^2} (45 y^2 - 49 x y + 86 x^2) \pi^2
%
 \MinusBreak{}
 {1 \over 288 y^3} (99 x^2 y - 505 y^3 + 279 x y^2 + 475 x^3) X^2
%
 \MinusBreak{}
 {1 \over 864 y^2} (2850 x^2 + 609 x y - 2273 y^2) X
%
 -{475 x \over 288 y}
   -{36077 \over 3456}
%
 \PlusBreak{}
 i\pi {} \biggl[
      {x^3 \over 2 y^3} (\li3(-x) - \zeta_3)
    - {7 x^3 \over 24 y^3} X^3
 \PlusBreak{ \null + i \pi \bigglB }
   {1 \over 12 y^3} (y - x) (1 - x y) X{} (3 \li2(-x) - \pi^2) 
  +{1 \over 8} X^3
 \PlusBreak{\null +  i \pi \bigglB }
   {1 \over 4} Y X^2  
  - \zeta_3    
%
   - {\pi^2 \over 48 y^2} (5 x^2 + 20 x + 11)    
 \PlusBreak{\null + i \pi \bigglB }
  {1 \over 48 y^3} (-24 x^2 y - 77 y^3 + 21 x y^2 + 86 x^3) X^2
 \PlusBreak{\null + i \pi \bigglB  }
  {1 \over 24 y^2} (35 y^2 + 6 x^2 - 3 x y) (\li2(-x) + X Y)
%
 \MinusBreak{\null +  i \pi \bigglB }
  {1 \over 144 y^3} (475 x^3 + 99 x^2 y + 279 x y^2 - 505 y^3) X
%
 \PlusBreak{ i \pi \bigglB }
  {1 \over 864 y^2} (32 x^2 + 5155 x + 2273)  
\biggr]
\,, \label{AA21}\\[1pt plus 4pt]
%
B_2^{[1]} & = &
 {1 \over 2 y^3} \biggl[ 
      {x - y \over 2} (4 - 5 x - 5 x^2)\li4(-x)
     + (5 x^3+x^2-4 x+1) X \li3(-x)
 \MinusBreak{ {1 \over 2 y^3} \bigglP }
     (11 x y + x^2 + 2 y^2) \zeta_3 X 
     + {\pi^2 \over 6} (17 x y^2 + 18 x^2 y - y^3 + 2 x^3) \li2(-x)
 \PlusBreak{ {1 \over 2 y^3} \bigglP }
      {1 \over 48} (9 y + 11 y^2 + 2 y^3 + 6) X^4
     +{\pi^2 \over 12} (3 x^3 + 11 x y^2 + 4 y^3) X^2
 \MinusBreak{ {1 \over 2 y^3} \bigglP }
      {\pi^4 \over 1440} (112 x^3 + 1128 x^2 y + 916 x y^2 + 45 y^3)
\biggr]
 \PlusBreak{}
 {1 \over y^2} \biggl[  
         {1 \over 2} (1 + 13 x) \li4 \left( -{x \over y} \right)
        +{1 \over 2} (12 x^2 + 15 x y + 7 y^2) \li4(-y)
 \MinusBreak{\null +  {1 \over y^2} \bigglP }
         (1 - 5 x) \biggl( X \li3(-y) 
        -{1 \over 2} Y{} (\li3(-x) - \zeta_3) \biggr) 
 \PlusBreak{\null + {1 \over y^2} \bigglP }
          {5 \over 8} (3 x^2 + 2 y^2 + 3 x y) X^2 \li2(-x)
        -{1 \over 8} (31 x y + 9 y^2 + 24 x^2) X^2 Y^2
 \MinusBreak{\null +  {1 \over y^2} \bigglP }
          2 y^2 Y \li3(-y)
        -(2 - x) X Y \li2(-x)
        -{1 \over 48} (-3 y^2 + 12 x^2 + 11 x y) Y^4 
 \PlusBreak{\null + {1 \over y^2} \bigglP }
          x y \li2(-x) Y^2
        +{1 \over 12} (6 x^2 + 11 x y + 10 y^2) Y X^3
 \MinusBreak{\null +  {1 \over y^2} \bigglP }
         {1 \over 6} (2 - y) (3 x - y) X Y^3
        -{1\over 6} (2 x^2 - 2 x - 1) \pi^2 Y^2  
 \PlusBreak{\null +  {1 \over y^2} \bigglP }
         {1 \over 6} (15 x^2 + 13 x y + 3 y^2) \pi^2 Y X 
\biggr]  
+ {1 \over 2 y} \pi^2 X^2                        
%
 \PlusBreak{}
  {1 \over 4 y^3} \biggl[ 
        (4 x y + 3 x^2 + 9 y^2) \biggl( \li3(-x) - X \li2(-x) \biggr)
 \PlusBreak{\null + {1 \over 4 y^3} \bigglP }
        {1 \over 36} (720 x y^2 - 11 y^3 + 396 x^2 y + 108 x^3) \zeta_3
 \MinusBreak{\null + {1 \over 4 y^3} \bigglP }
        {1 \over 18} (120 y^2 + 60 y^3 + 43 + 78 y) X^3
 \MinusBreak{\null + {1 \over 4 y^3} \bigglP }
        {\pi^2 \over 18} (91 x^2 + 140 x y + 160 y^2 - 132) X 
\biggr]
 \PlusBreak{}
 {1 \over y^2}
\biggl( -{1 \over 12} (1 - x) \pi^2 Y
       +{1 \over 24} (20 x^2 - 23 x - 13) X^2 Y 
\biggr)
 \PlusBreak{}
 {1 \over y}
\biggl( 3 \li3(-y)
       + 3 \li2(-x) Y
       - {1 \over 12} (10 x + 3) X Y^2 
\biggr)
-{5 \over 6} Y^3
%
 \MinusBreak{}
{1 \over 144 y^3} ( 416 x^3 + 27 x^4 + 742 x^2 + 190 + 902 x) X^2
 \PlusBreak{}
 {1 \over 144} (191 + 27 x) Y^2
-{\pi^2 \over 288 y^2} (463 + 366 x - 35 x^2 - 54 x^3)
 \PlusBreak{}
 {1 \over 72 y^2} (54 x^3 + 17 x^2 y + 100 y^3 + 36 x y^2) X Y
%
-{3 (y-1) \over 8 y}Y
 \MinusBreak{}
 {1 \over 72 y^2} (291 y^2 + 305 x^2 + 383 x y) X
%
-{139 x \over 72 y} + {14135 \over 2592}
%
 \PlusBreak{}
i \pi {} \biggl[
{1 \over y^3} 
\biggl( -y{} (7 x y + 3 y^2 + 6 x^2) \li3(-y)
        -{x^2 \over 2} (6 y + x) \li3(-x)
 \MinusBreak{\null +  i \pi \bigglB {1 \over y^3} \bigglP }
         {1\over 2} (1 - 3 y - 3 y^2) \zeta_3    
        -{\pi^2 \over 12} (2 y^2 + 8 x y + x^2) X
 \MinusBreak{\null + i \pi \bigglB {1 \over y^3} \bigglP }
         {1 \over 24} (5 x y^2 + 4 y^3 + 4 x^3 + 3 x^2 y) X^3 
\biggr)
 \MinusBreak{\null + i \pi \bigglB }
 {1 \over y^2}
\biggr({1 \over 4} (3 + 5 y) (x - y) X  \li2(-x)
 \PlusBreak{\null +  i \pi \bigglB {1 \over y^2} \bigglP }
         (3 x y + 3 x^2 + 2 y^2) \li2(-x) Y
 \PlusBreak{\null +  i \pi \bigglB {1 \over y^3} \bigglP }
         {1 \over 2} (6 x^2 + 5 y^2 + 8 x y) X Y^2 
        -{\pi^2 \over 12} (4 x^2 + 7 x + 9) Y    
\biggl) 
 \PlusBreak{\null + i \pi \bigglB }
 {1 \over 4 y}
\biggl( (3 y - 4) Y X^2     
       + 2 X Y^2   
      -{1 \over 3} (x - y) Y^3 
        + 2 \pi^2 X   
\biggr)
%
 \PlusBreak{\null +  i \pi \bigglB }
 {1 \over 4 y^3}
\biggl( - \li2(-x) (3 x^2 - 3 y^2 + 4 x y)
 \MinusBreak{\null + i \pi \bigglB {1 \over 4 y^3} \bigglP }
        {1 \over 6} (34 + 42 y + 33 y^2 + 40 y^3) X^2
 \PlusBreak{\null + i \pi \bigglB {1 \over 4 y^3} \bigglP }
       {1 \over 24} (83 x + 47 x^2 + x^3 + 49) \pi^2  
\biggr)
 \PlusBreak{\null +  i \pi \bigglB }
         {1 \over 12 y^2} (29 y^2 + 30 x^2 + 19 x y) X Y
        +{1 \over 12 y} (11 x - 9 y) Y^2
       + {1\over 96} \pi^2
%
 \MinusBreak{\null + i \pi \bigglB }
  {1 \over 72 y^3} (-504 y - 199 y^2 - 144 y^3 - 359) X
 \PlusBreak{\null +  i \pi \bigglB }
 {1 \over 72 y^2} (-91 y + 54 x) Y
%
 -{5 \over 72 y^2} (48 y^2 + 40 y + 61)   
\biggr]
\,, \label{BB21}\\[1pt plus 4pt]
%
C_2^{[1]} & = & 
{1 \over 4 y^3} 
\bigg[
    (2 x^2 - 8 x + 2) 
             \biggl( \li4(-y)  - \li4 \left( -{x \over y} \right) \biggr)
 \PlusBreak{ {1 \over 4 y^3} \bigglP }
     (2 x^2 - 8 x - 5) \li4(-x)
    -{1 \over 6} (2 x^2 - 8 x - 3) \pi^2 X^2
 \PlusBreak{ {1 \over 4 y^3} \bigglP }
      (1 + 4 x - x^2) \biggl( 2 (\li3(-x) - \zeta_3 ) X
                             -{\pi^2 \over 3} \li2(-x) \biggr)
 \MinusBreak{ {1 \over 4 y^3} \bigglP }
     (1 - 4 x + x^2) Y{} \biggl( {\pi^2 \over 6} Y 
                          - 2 ( \li3(-x) - \zeta_3)  
                          - {1 \over 3} Y^2 X + {1 \over 12} Y^3  \biggr)
 \MinusBreak{ {1 \over 4 y^3} \bigglP }
 {1 \over 2} X^2 \li2(-x)
-{1 \over 24} (2 x^2 - 8 x + 3) X^4
 \PlusBreak{ {1 \over 4 y^3} \bigglP }
 {1 \over 180} (11 x^3 + 12 x^2 + 117 x + 24) \pi^4
\biggr]
%
 \PlusBreak{}
{1 \over 8 y^3} 
\biggl[
    3 (2 x^2 + y^2 - 2 x y) X \li2(-x)
   +12 (1 - x^2) \li3(-y)
 \MinusBreak{ \null +  {1 \over 8 y^3} \bigglP }
    3 (5 x^2 + 1 + 4 x) \li3(-x)
   +(29 x + 41 x^2 + 15 x^3 + 9) \zeta_3
 \PlusBreak{\null +  {1 \over 4 y^3} \bigglP }
   (4 y^2 + 5 x^2 + 4 x y) \pi^2 X 
\biggr]
 \PlusBreak{}
 {1 \over 2 y^2}
\biggl[
    3 (x - 1) Y \li2(-x)
   -{3 \over 2} (1 - x) X Y^2
   +{1 \over 8} (y - 2) X^2 Y
 \MinusBreak{\null +{1 \over 2 y^3} \bigglP }
    {1 \over 6} (7 x - 3) \pi^2 Y
   +{1 \over 24} (9 - 13 x) X^3 
\biggr]
%
 \PlusBreak{}
 {1 \over 32 y^3} (-16 x^3 - 6 x^4 + 5 - 18 x - 22 x^2) X^2
 \MinusBreak{}
 {1 \over 96 y^2} \biggl(
 12 x{} (3 x^2 + 5 x + 8) X Y
+ (15 - 118 x - 59 x^2 - 18 x^3) \pi^2 \biggr)
 \MinusBreak{}
 {1 \over 16 y} (2 x + 3 x^2 + 11) Y^2
%
+{1 \over 32 y^2} (12 x^2 + 19 x - 39) X
 \MinusBreak{}
 {3 \over 8 y} (y - 1) Y
+{187 \over 128} - {17 x \over 32 y}
 \PlusBreak{}
 i \pi {} \biggl[
   {1\over 8 y^3} \Bigl(
    8 \li3(-x)    
    - 2 X \li2(-x)
    -  X^3
    - 8 \zeta_3  
       \Bigr)
%
+{1 \over 8 y^2} (y - 2) X Y
 \MinusBreak{\null + i \pi \bigglB }
 {1 \over 16 y^3} (x^2 + 4 x + 9) X^2
 + {3 \over 8 y^3}(x^2 + 4 x + 5) \li2(-x) 
- {1 \over 8 y^3}  \pi^2    
%
 \PlusBreak{\null + i \pi \bigglB }
 {1 \over 16 y^3} (4 x^2 - 2 x + 5) X
 - {1 \over 8 y^2} (5 y - 6) Y
- {1\over 32 y^2} (17 x + 63)    
\biggr]
\,, \label{CC21}\\[1pt plus 4pt]
%
D_2^{[1]} & = & 
-{1 \over 3} \li2(-x) X
-{7 \over 72 y^3} (x - y) (1 - x y) X^3
+{1 \over 3} \li3(-x)
 \PlusBreak{}
 {\pi^2 \over 72 y^3} (11 + 33 x + 33 x^2 + 25 x^3) X
-{17 \over 24} \zeta_3
-{1 \over 6} X^2 Y
 \PlusBreak{}
 {1 \over 72 y^3} (18 x y^2 + 27 x^2 y + 37 x^3 - 43 y^3) X^2
 \PlusBreak{}
 {7 \pi^2 \over 216 y^2} (20 x^2 + 25 x + 11) 
%
+{1 \over 36 y^2} (10 + x + 28 x^2) X
%
+{37 x \over 72 y} + {1307 \over 864}
%
 \PlusBreak{}
  i \pi{} \biggl[
   {7 \over 24 y^3} (y - x) (x^2 + x + 1) X^2
  -{1 \over 3} \li2(-x)
  -{1 \over 3} X Y
  +{6 \over 144} \pi^2   
%
 \PlusBreak{\null + i \pi  \bigglB }
 {1 \over 36 y^3} ( 71 x^3 + 138 x^2 + 147 x  + 43) X
%
 + {1\over 36 y^2} (28 x^2 + x + 10)    
\biggr]
\,, \label{DD21}\\[1pt plus 4pt]
%
E_2^{[1]} & = & 
 {x{} (x-1) \over y^3} 
\biggl[
    2 \biggl( \li4(-y) - \li4 \left( -{x \over y} \right) -\li4(-x) 
    +( \li3(-x) - \zeta_3 ) Y \biggr)
 \MinusBreak{ {x (x-1) \over y^3} \bigglP }
     {1 \over 3} \li2(-x) \pi^2
    +{1 \over 3} Y^3 X
    -{1 \over 12} Y^4
    -{1 \over 6} \pi^2Y^2
    +{7 \over 180} \pi^4 
\biggr]
%
 \PlusBreak{}
 {1 \over y^3}  
\biggl[ {1 \over 3} x^3 (\li3(-x) - \zeta_3) 
         -{1 \over 3} \li2(-x) x^3 X
 \MinusBreak{\null + {1 \over y^3} \bigglP } 
          {1 \over 72} (6 x^3 + 3 x^2 - 7 - 3 x) X^3 
         -{\pi^2 \over 36} (4 - 6 x - 3 x^2 + 10 x^3) X
\biggr]
 \PlusBreak{}
  {1 \over 24 y^2} \biggl[
   - 8 (12 x^2 + 12 y x - y^2 ) (\li3(-y) - \zeta_3 
               + Y \li2(-x) )
 \PlusBreak{ \null + {1 \over 24 y^2} \bigglB }
     (3 x y + 4 y^2 - 3 x^2) Y X^2
   - 2 (26 x y - y^2 + 24 x^2) X Y^2 
\biggr]
 \MinusBreak{}
  {1 \over 12 y} (1 + 4 y) \pi^2 Y
 +{1 \over 12} Y^3
 +{1 \over 72} \zeta_3
 + {1 \over 72 y^3} (-33 x^2 y + x^3 - 4 x y^2 - 8 y^3) X^2
 \PlusBreak{}
 {1 \over 18 y^2} (6 x^2 - 8 x y + 5 y^2) Y X
+{1 \over 144 y^2} (84 + 6 y - 43 y^2) \pi^2
 \PlusBreak{}
 {1 \over 36 y x} (-10 x y - 27 y^2 + 36 x^2) Y^2
%
-{1 \over 36 y^2} (14 x y - 33 y^2 + 11 x^2) X 
 \MinusBreak{}
 {1 \over 6 y} (5 x + 9 y) Y
%
+{19 x \over 72 y} - {3401 \over 2592}
%
 \PlusBreak{}
 i \pi {} \biggl[
%
  -{1 \over 3 y^3} \li2(-x) (12 x y^2 - y^3 + x^3 + 12 x^2 y)
 \PlusBreak{\null + i \pi \bigglB }
   {2 x{} (x-1) \over y^3} (\li3(-x) - \zeta_3 )
  +{1 \over 8 y^3} (1 + 2 x y) X^2
 \PlusBreak{\null + i \pi \bigglB }
    {1\over 18 y^3} \pi^2 x^3         
   +{1 \over 12 y^2} (3 x - 2 y) X Y
 \MinusBreak{\null + i \pi \bigglB }
   {1 \over 6 y} (x - y) Y^2
%
  -{1 \over 18 y^2 x} (6 x^2 + 22 x y - 27 y^2) Y
 \MinusBreak{\null + i \pi \bigglB }
   {1 \over 36 y^3} (x^2 + 14 y^2 - 10 x y + 12 y) X
%
 + {1\over 36 y^2} (12 x^2 + 2 x - 21)  
\biggr]
\,, \label{EE21}\\[1pt plus 4pt]
%
F_2^{[1]} & = & 
{1 \over 24} X^2 - {5 \over 54} X - {\pi^2 \over 27} 
 + i \pi {} \biggl( {1 \over 12} X - {5 \over 54} \biggr)   
\,, \label{FF21}\\[1pt plus 4pt]
%
G_2^{[1]} & = & 
- {x \over 72 y} X^2 +  {x-y \over 72 y} X Y + {1 \over 72} Y^2 
+ {i \pi  \over 72 y} (X-Y)
\,. \label{GG21} 
\end{eqnarray}


\noindent
For $h=2$ in \eqn{hel2}
and color factor $\trc^{[2]}$ in \eqn{TraceBasisqqgg}:
\begin{eqnarray}
A_2^{[2]} & = & 
{1 \over y^3}  
\biggl[
  x{} (4 x - 1) \biggl(
            \li3(-x) Y - \li4(-x) - \li4 \biggl( -{x \over y} \biggr) 
              \biggr)
 \MinusBreak{ {1 \over y^3} \bigglB }
   {1 \over 24} (4 x^3 - 5 x^2 + 14 x + 3) X Y^3
  + {1 \over 96} (10 x^3 + 11 x^2 + 28 x + 7) Y^4
 \PlusBreak{ {1 \over y^3} \bigglB }
   {\pi^2 \over 6} (x^2 - 2 x + 2) x \li2(-x)
 +  {1 \over 4} (2 x^3 + 23 x^2 + 4 x + 3) \li4(-y)
 \MinusBreak{ {1 \over y^3} \bigglB }
   {\pi^2 \over 48} (22 x^3 + 75 x^2 + 48 x + 15) Y^2
 - {1 \over 2} (x^3 + 9 x^2 - 3 x - 1) \zeta_3 Y
 \MinusBreak{ {1 \over y^3} \bigglB }
   {\pi^4 \over 2880} (9 x^3 - 182 x^2 + 113 x + 24) 
\biggr]
 - {x - y \over 8 y} Y^2 \li2(-x)
+{1 \over 2} Y \li3(-y)
%
 \PlusBreak{}
 {1 \over y^2} \biggl[
   {1 \over 6} (11 x^2 + 50 x + 9) \biggl( \li3(-y) + \li2(-x) Y
                                   + {1 \over 2} X Y^2 \biggr)
 \MinusBreak{\null + {1 \over y^2} \bigglB } 
   {1 \over 48} (205 x^2 + 517 x + 72) \zeta_3
         \biggr]
 \MinusBreak{}
 {13 (x - y) \over 18 y} Y^3
+{\pi^2 \over 144 y} (445 x + 208) Y
%
 +  {1 \over 288 x y} (1160 x^2 - 713 x - 360) Y^2
 \PlusBreak{}
 {5 x \over 6 y^2} (3 x + 2) \pi^2
%
 + {2 x + 135 \over 54 y} Y
%
+{x {} (30377 x + 30593) \over 3456 y^2}
%
 \PlusBreak{}
 i \pi {}  
\biggl[
   {1 \over y^3} \biggl( (4 x - 1) x \li3(-x)
     -{1\over 2} (x^3 + 9 x^2 - 3 x - 1) \zeta_3  
           \biggr)
 \MinusBreak{\null - i \pi \bigglB }
    {x - y \over 4 y} Y {} \biggl( \li2(-x)  +  {\pi^2 \over 6} \biggr)
   +{10 x + 7 \over 24 y} Y^3
   +{1 \over 2} \li3(-y)
 \PlusBreak{\null - i \pi \bigglB }
    {1 \over 4} X Y^2
%
   +{1 \over 6 y^2} (11 x^2 + 50 x + 9) \li2(-x)
   -{13 (x - y) \over 6 y} Y^2
 \PlusBreak{\null - i \pi \bigglB }
    {29 x \over 144 y} \pi^2    
%
   -{1 \over 144 x y} (-1160 x^2 + 713 x + 360) Y
%
 + {133 \over 54 y} - {1\over 27}
\biggr]
\,, \label{AA22}\\[1pt plus 4pt]
%
B_2^{[2]} & = & 
{1 \over 2 y^3}   
\biggl[
    (4 x^3 + 12 x^2 - 3) \li4(-x)
   +(5 x^3 + 4 x^2 - x + 7) \li4(-y)
 \PlusBreak{ {1 \over 2 y^3} \bigglB } 
    (14 x^2 + 22 x + 1) \li4 \biggl( -{x \over y} \biggr)
   -(5 x^2 + 10 x - 2) Y \li3(-x)
 \MinusBreak{ {1 \over 2 y^3} \bigglB } 
    2 (2 x^2 + x - 4) x X \li3(-x)
   +{x \over 12} (x^2 - x + 1) X^4
 \PlusBreak{ {1 \over 2 y^3} \bigglB } 
    {\pi^2 \over 6} (2 x^3 - 6 x^2 - 12 x + 1) \li2(-x)
 \PlusBreak{ {1 \over 2 y^3} \bigglB } 
    {1 \over 24} (x^3 + 20 x^2 + 31 x + 5) Y^4
  - {1 \over 6} (10 x^3 + 39 x^2 + 42 x + 6) X Y^3
 \MinusBreak{ {1 \over 2 y^3} \bigglB }
    {\pi^2 \over 12} (5 x^3 + 10 x^2 + 11 x + 13) Y^2
 - {\pi^2 \over 4} (4 x^3 + 4 x^2 - 3 x + 1) X^2
 \MinusBreak{ {1 \over 2 y^3} \bigglB }
    (5 x^2 + 10 x - 1) \zeta_3 X
   +(5 x^2 + 10 x - 2) \zeta_3 Y
 \MinusBreak{ {1 \over 2 y^3} \bigglB }
    {\pi^4 \over 1440} (225 x^3 - 130 x^2 - 719 x + 112) 
\biggr]
 \PlusBreak{}
 {1 \over 2 y^2}   
\biggl[
   -(11 x - 1) X \li3(-y)
   -(2 x^2 + x + 2) X^2 \li2(-x)
 \MinusBreak{\null -  {1 \over 2 y^2} \bigglB }
    2 (x - 2) X Y \li2(-x)
   -{1 \over 6} (4 x^2 + 7 x + 9) X^3 Y
 \PlusBreak{\null - {1 \over 2 y^2} \bigglB }
     {1 \over 2} (x^2 - 7 x + 4) X^2 Y^2
    -{\pi^2 \over 3} (6 x^2 - 5 x + 4) X Y
\biggr]
 \PlusBreak{}
  {1 \over 2 y}  \biggl( 
  -(5 x + 4) Y \li3(-y)
  -{5 \over 2} x Y^2 \li2(-x) 
  + \pi^2 (Y^2 - X^2)    
              \biggr)
%
 \PlusBreak{}
 {1 \over 4 y^3}  
\biggl[ 
    (21 x^2 + 26 x + 12) ( X \li2(-x) - \li3(-x) )
 \PlusBreak{\null - {1 \over 4 y^3} \bigglB }
    {5 \pi^2 \over 6} (7 x^2 + 10 x + 4) X
   -{x \over 3} (10 x^2 + 5 x + 1) X^3
 \PlusBreak{\null - {1 \over 4 y^3} \bigglB }
    {1 \over 6} (20 x^3 + 15 x^2 - 12 x - 4) X^2 Y 
 \MinusBreak{\null -  {1 \over 4 y^3} \bigglB }
    {1 \over 36} (443 x^3 + 706 x^2 + 299 x - 216) \zeta_3
\biggr]
 \PlusBreak{}
 {1 \over 2 y^2} \biggl[
    (10 x + 3) (\li3(-y) + Y \li2(-x))
 \MinusBreak{\null - {1 \over 2 y^2} \bigglB }
    {1 \over 6} (10 x^2 + x + 12) X Y^2
   -{2 \pi^2 \over 9} (19 x + 10) Y
        \biggr]
 \MinusBreak{}
 {15 x + 28 \over 18 y} Y^3
%
-{x \over 144 y^3} (27 x^3 - 92 x^2 - 184 x - 164) X^2
 \MinusBreak{}
 {1 \over 288 y^2}   \biggl[
      4 (27 x^3 - 119 x^2 - 138 x - 100) X Y
   - \pi^2 x{} (54 x^2 + 217 x + 343) 
          \biggr]
 \MinusBreak{}
 {1 \over 144 x y} (27 x^3 - 290 x^2 - 199 x + 72) Y^2
%
-{3 x{} (y - 1) \over 8 y^2} X
 \MinusBreak{}
 {71 x - 24 \over 24 y} Y
%
-{x \over 2592 y^2} (19139 x + 18815)
%
 \PlusBreak{}
  i \pi {}
\biggl[
    - {1 \over 2 y^3}   \biggl(
      (4 x^3 + 7 x^2 + 2 x - 2) \li3(-x)
     + \zeta_3         
             \biggr)
 \PlusBreak{\null + i \pi \bigglB }
   {1 \over 2 y^2}  \biggl[
      (5 x^2-2 x + 5) \li3(-y)
      +(5 x^2 + 3 x + 4) Y \li2(-x)
 \MinusBreak{\null + i \pi \bigglB + {1 \over 2 y^2}  \bigglP }
       {1 \over 6} (2 x - 1) (x - 1) X^3
      +{1 \over 2} (7 x^2 + 7 x + 12) X Y^2   
 \PlusBreak{\null + i \pi \bigglB + {1 \over 2 y^2}  \bigglP }
       {\pi^2 \over 6} (8 x + 5) X           
      -{\pi^2 \over 6} (5 x^2 + 7 x + 8) Y   
                 \biggr]
 \PlusBreak{\null + i \pi \bigglB }
    {1 \over y}  \biggl(
      2 x X \li2( - x)
     +{1 \over 12} (x + 3) Y^3
     -{1 \over 2} (y - 2) X^2 Y   
          \biggr)
%
 \PlusBreak{\null + i \pi \bigglB }
   {1 \over 4 y^3}   \biggl[
     (x^2 + 6) \li2(-x)
    -{1 \over 3} (20 x^3 + 39 x^2 + 48 x + 20) X^2
 \PlusBreak{\null + i \pi \bigglB + {1 \over 4 y^3} \bigglB }
     {1 \over 3} (40 x^3 + 97 x^2 + 92 x + 38) X Y
 \MinusBreak{\null + i \pi \bigglB + {1 \over 4 y^3} \bigglB }
     {\pi^2 \over 24} (x^3 + 86 x^2 + 153 x + 96) 
        \biggr]
 \MinusBreak{\null + i \pi \bigglB }
     {5 (4 x + 7) \over 12 y} Y^2
     + {x \over 96 y} \pi^2
%
   -{1 \over 72 y^3} (73 x^2 + 74 x + 100) X
 \MinusBreak{\null + i \pi \bigglB }
    {1 \over 8 x y^2} (16 x^3 + 39 x^2 + 3 x - 8) Y
%
 + {1\over 24 y^2} (80 x^2 + 65 x - 24)   
\biggr]
\,, \hskip 1.3 cm \label{BB22}\\[1pt plus 4pt]
%
C_2^{[2]} & = &
{1 \over 2 y^3}   
\biggl[ 
 (2 x^2 + 4 x + 1) \biggl( \li4(-x) + \li4 \biggl( -{x \over y}\biggr) 
                          + {\pi^2 \over 6} \li2(-x) \biggl)
 \MinusBreak{ {1 \over 2 y^3} \bigglB }
   {1 \over 2} (7 x^2 + 14 x + 5) \li4(-y)
  +x{} (y - 1) Y \li3(-x)
 \PlusBreak{ {1 \over 2 y^3} \bigglB }
   {\pi^2 \over 24} (9 x^2 + 18 x + 7) Y^2
  -\zeta_3 Y
 - {1 \over 12} (x^2 + 2 x - 1) \biggl( X Y^3 - {1 \over 4} Y^4 \biggr)
 \MinusBreak{ {1 \over 2 y^3} \bigglB }
   {\pi^4 \over 360} (11 x^3 + 19 x^2 + 5 x - 10) 
\biggr]
 \PlusBreak{}
  {1 \over 2 y}   \biggl(
    X{} (\li3(-y) - \zeta_3)
  +Y \li3(-y)
  +{1 \over 4} Y^2 \li2(-x)
  +{1 \over 4} X^2 Y^2 
             \biggr)
%
 \PlusBreak{}
 {x{} (3 y + 1) \over 4 y^3}  \biggl(
   \li3(-x)
  -X \li2(-x)
  -{1 \over 2} X^2 Y
  +{\pi^2 \over 6} X
        \biggr)
 \MinusBreak{}
 {5 x + 4 \over 8 y^3} (3 x^2 + 4 x + 3) \zeta_3
+{x \over 96 y^2} (18 x^2 + 25 x - 5) \pi^2 
 \PlusBreak{}
 {2 y - 1 \over 4 y^2} (
    2 \li3(-y)
   +2 \li2(-x) Y + X Y^2
         )
 \MinusBreak{}
 {x \over 16 y^3} (3 x^3 + 12 x^2 + 12 x + 4) X^2
-{3 x{} (y - 1) \over 8 y^2} X
 \MinusBreak{}
 {1 \over 32 x y} (6 x^3 + 12 x^2 - 15 x - 8) Y^2
-{3 x{} (y - 1) \over 8 y} X Y
%
 \PlusBreak{}
 {3 x - 4 \over 8 y} Y
%
-{x{} (255 x + 247) \over 128 y^2}
%
 \PlusBreak{}
 i \pi {}
\biggl[
  {1 \over 2 y^3}   \biggl( 
    (y - 1) x \li3(-x)
    -(3 x^3 + 7 x^2 + 5 x + 2) \zeta_3
         \biggr)
 \PlusBreak{\null + i  \pi \bigglB }
   {1 \over 4 y}  \biggl(
     4 \li3( - y)
    +\li2(-x) Y
    -{\pi^2 \over 6} Y
    +X Y^2
    -{1 \over 2} Y^3
    + 6 x \zeta_3  
         \biggr)
%
 \PlusBreak{\null + i  \pi \bigglB }
  {1 \over 4 y^3}   \biggl( 
    -(3 y + 1) x X Y
    - {\pi^2 \over 6} x {} (3 x + 2)    
    + (7 x^2 + 12 x + 6) \li2(-x)
           \biggr)
%
 \MinusBreak{\null + i  \pi \bigglB }
 {x{} (3 y + 1) \over 8 y^3} X
 + {15 x + 8 \over 16 x y} Y
%
 + {1\over 8 y^2} (7 x + 4) 
\biggr]
\,, \label{CC22}\\[1pt plus 4pt]
%
D_2^{[2]} & = & 
{x{} (x - 1) \over y^3}  
\biggl[
  \li4(-x) - \li4(-y) + \li4 \biggl( -{x \over y} \biggr)
  +{\pi^2 \over 6} \li2(-x)
 \PlusBreak{ {x (x - 1) \over y^3} \bigglB }
   {1 \over 24} Y^4
  -{1 \over 6} X Y^3
  +{\pi^2 \over 12} Y^2
  -Y{} (\li3(-x) - \zeta_3)
  -{7 \pi^4 \over 360} 
\biggr]
%
 \PlusBreak{}
  {x \over 6 y^2}   \biggl[
  -{1 \over 4} (8 x + 56) \li3(-y)
  -2 (x + 7) \li2(-x) Y
 \PlusBreak{\null + {x \over 6 y^2} \bigglP } 
   {1 \over 4} (17 x + 65) \zeta_3
  -(x + 7) X Y^2
          \biggr]
+{7 (x - y) \over 72 y} Y^3
 \MinusBreak{}
 {\pi^2 \over 72 y} (29 x + 14) Y
%
-{1 \over 72 x y} (71 x^2 - 71 x - 36) Y^2
 \MinusBreak{}
 \pi^2 {x{} (35 x + 17) \over 54 y^2} 
%
+{7 x - 9 \over 9 y} Y
%
+{863 x \over 864 y}
%
 \PlusBreak{}
  i \pi {}
\biggl[
   {(1 - x) x \over y^3}   (\li3(-x) - \zeta_3)
%
  -{x{} (x + 7) \over 3 y^2} \li2(-x)
  + {7 (x - y) \over 24 y} Y^2
 \MinusBreak{\null + i \pi \bigglB }
   \pi^2 {x \over 72 y}         
%
-{1 \over 36 x y} (71 x^2 - 71 x - 36) Y
%
-  {16 \over 9 y} - {7\over 9}     
\biggr]
\,, \label{DD22}\\[1pt plus 4pt]
%
E_2^{[2]} & = &
{x{} (x - 1) \over y^3}   
\biggl[
   -2 \li4(-x)
   +2 \li4(-y)
   -2 \li4 \biggl( -{x \over y} \biggr)
   + 2 Y{} (\li3(-x) - \zeta_3)
 \MinusBreak{ {x (x - 1) \over y^3} \bigglB }
    {\pi^2 \over 3} \li2(-x)
   -{1 \over 12} Y^4
   + {1 \over 3} X Y^3
   -{\pi^2 \over 6} Y^2
   +{7 \pi^4 \over 180} 
\biggr]
%
 \PlusBreak{}
 {1 \over 3 y^3}   \biggl[
    x^3 ( \li3(-x) - X \li2(-x) )
-{\pi^2 \over 12} (10 x^3 + 3 x^2 + 6 x + 3) X
 \PlusBreak{\null + {1 \over 3 y^3} \bigglB }
 {1 \over 24} (x^3 + 362 x^2 + 361 x + 24) \zeta_3
        \biggr]
 \PlusBreak{}
 {1 \over 3 y^2}  \biggl[
    (x^2 + 14 x + 1) ( \li3(-y) + \li2(-x) Y )
   -{1 \over 8} x{} (2 x - 1) X^3
 \PlusBreak{\null + {1 \over 3 y^2}  \bigglB } 
    {1 \over 8} (2 x^2 - 5 x - 4) X^2 Y
    +{1 \over 4} (x + 7) (5 x + 1) X Y^2
             \biggr]
 \PlusBreak{}
 {1 \over 9 y}  \biggl(
   {1 \over 8} (6 x + 13) Y^3
  +(3 x + 2) \pi^2 Y
      \biggr)
%
-{x \over 36 y^3} (x^2 + 20 x + 28) X^2
 \MinusBreak{}
 {1 \over 18 y^2} (x^2 + 12 x + 5) X Y
-\pi^2 {x \over 144 y^2} (29 x + 85) 
+{x{} (4 x + 9) \over 6 y^2} X
 \MinusBreak{}
 {1 \over 72 x y} (2 x^2 + 147 x + 54) Y^2
%
+{2 x + 3 \over 2 y} Y
%
-{4085 x \over 2592 y}
%
 \PlusBreak{}
  i \pi {}
\biggl[ 
   2 {x{} (x - 1) \over y^3} (\li3(-x) - \zeta_3)
   +{1 \over 12 y^2} (8 x^2 + 11 x + 6) X Y
%
 \MinusBreak{\null - i \pi \bigglB }  
    {1 \over 3 y^3}  
\biggl(
        (2 x^3 + 15 x^2 + 15 x + 1) \li2(-x)
 \MinusBreak{X \null - i \pi \bigglB {1 \over 3 y^3} \bigglP }  
    {1 \over 4} (4 x^3 + 3 x^2 + 3 x + 2) X^2
   -{\pi^2 \over 6}  x^3      
\biggr)
 \PlusBreak{\null - i \pi \bigglB } 
    {1 \over 8 y} Y^2
%
 + {1 \over 36 x y^2} (125 x^2 + 191 x + 54) Y
 \MinusBreak{\null - i \pi \bigglB } 
   {1 \over 18 y^3} (7 x^2 + 11 x - 5) X
%
- {1\over 6 y^2} (2 x^2 + 6 x + 9)  
\biggr]
\,, \label{EE22}\\[1pt plus 4pt]
%
F_2^{[2]} & = &
 {x \over y} 
 \biggl[
     {1 \over 24} Y^2
     -{\pi^2 \over 27} 
     -{5 \over 54} Y
   + i \pi {} \biggl( 
       {1 \over 12} Y
       -{5 \over 54}   
         \biggr)
   \biggr]
\,, \label{FF22}\\[1pt plus 4pt]
%
G_2^{[2]} & = &
{1 \over 72}  
\biggl[
   {x \over y} X^2
  -{2 x + 1 \over y} X Y
  -Y^2
  +i {\pi \over y} (Y - X)
\biggr]
\,. \label{GG22}
\end{eqnarray}


\noindent
For $h=2$ in \eqn{hel2}
and color factor $\trc^{[3]}$ in \eqn{TraceBasisqqgg}:
\begin{eqnarray}
H_2^{[3]} & = &  
 {1 \over 2 y^3}  
\biggl[
      (5 x^3 + 56 x^2 + 31 x + 13) \li4(-y)
   -  {1 \over 2} (10 x^3 - 16 - 47 x + 23 x^2) \li4(-x)
 \PlusBreak{ {1 \over 2 y^3} \bigglB }
     (16 x^3 + 7 x^2 + 32 x + 8)    \li4 \biggl( -{x \over y} \biggr)
 \MinusBreak{ {1 \over 2 y^3} \bigglB }
      (12 x^3 + 2 x^2 + 28 x + 7) Y \li3(-x)
 \PlusBreak{ {1 \over 2 y^3} \bigglB }
      (12 x^3 + 21 x^2 + 18 x + 5) \biggl( X \li3(-y) 
                                   + {1 \over 2} X^2 Y^2 \biggr)
 \PlusBreak{ {1 \over 2 y^3} \bigglB }
     (13 x^3 + 10 x^2 - 2 + 2 x) X \li3(-x)
 \MinusBreak{ {1 \over 2 y^3} \bigglB }
     {1 \over 4} (10 x^3 + 3 x^2 - 9 x - 8) X^2 \li2(-x)
 \PlusBreak{ {1 \over 2 y^3} \bigglB }
    2 (x - y) (2 x^2 + 3 x + 2) X Y \li2(-x)
 \MinusBreak{ {1 \over 2 y^3} \bigglB }
    {\pi^2 \over 6} (44 x^3 + 109 x^2 + 68 x + 22) \li2(-x)
 \MinusBreak{ {1 \over 2 y^3} \bigglB }
    {1 \over 6} (32 x^3 + 63 x^2 + 96 x + 32) X Y^3
   + {1 \over 6} (x^3 + 10 x^2 + 11 x + 4) X^3 Y
 \PlusBreak{ {1 \over 2 y^3} \bigglB }
    {1 \over 24} (19 x^3 + 16 x^2 + 41 x + 11) Y^4
   + {\pi^2 \over 4} (10 x^3 + 11 x^2 + 13 x + 8) X^2
 \MinusBreak{ {1 \over 2 y^3} \bigglB }
    {\pi^2 \over 3} (17 x^3 + 30 x^2 + 24 x + 6) X Y
   - {\pi^2 \over 12} (9 x^3 + 40 x^2 - 13 x - 11) Y^2
 \MinusBreak{ {1 \over 2 y^3} \bigglB }
    (14 x + 13 x^2 + 11 x^3 + 5) \zeta_3 X
   + (11 x^3 - x^2 + 25 x + 6) \zeta_3 Y
 \PlusBreak{ {1 \over 2 y^3} \bigglB }
    {\pi^4 \over 360} (276 x^3 + 147 x^2 - 378 x - 202) 
\biggl]
 + {x{} (2 x - 1) \over 32 y^2} X^4
 \MinusBreak{}
  {1 \over 2 y} \biggl(
    (13 x + 15 ) Y \li3(-y)
  + {1\over 2} (5 x + 9)  Y^2 \li2(-x)
  + (2 x + 1) \pi^2 (X^2 - Y^2)           
          \biggr)
%
 \PlusBreak{}
 {1 \over 12 y^3}   
\biggl[
     (62 x^3  +  67 x^2  +  62 x  +  26) (\li3(-x) - X \li2(-x))
 \PlusBreak{\null + {1 \over 12 y^3} \bigglB }
       (242 x^2  +  238 x  +  27) \zeta_3
     + {x \over 3} (22 x^2 + 14 x - 11) X^3
 \MinusBreak{\null + {1 \over 12 y^3} \bigglB }
       (21 x^3 + 48 x^2 + 39 x + 16) X^2 Y
 \MinusBreak{\null + {1 \over 12 y^3} \bigglB }
       {\pi^2 \over 12} (321 x^3 - 182 x^2 - 139 x - 18) X
\biggr]
 \PlusBreak{}
  {1 \over 12 y^2}
\biggl[
     (62 x^2 + 263 x + 27) (\li3(-y) + Y \li2(-x))
 \PlusBreak{\null + {\pi^2 \over 12 y^2} \bigglB }
       (41 x^2 + 132 x + 4) X Y^2
    - {\pi^2 \over 12} (445 x^2 + 418 x - 75) Y
\biggr]
 \PlusBreak{}
   {11 \over 18} Y^3
%
 - {x \over 72 y^3} (172 x^2 + 404 x + 97) X^2
 + {x - 2 \over 12 y^2} (6 x - 11) X Y
 \PlusBreak{}
   {1 \over 72 x y} (208 x^2 - 161 x - 108) Y^2
 - {\pi^2 \over 72 y^2} (18 x^2 - 108 x - 49) 
%
 \PlusBreak{}
   {x \over 54 y^2} (38 x + 83) X
 + {1 \over 108 y} (76 x + 211) Y
%
%
 \PlusBreak{}
  i \pi {}
\biggl[
  {1 \over 2 y^3}  \biggl(
     (x^3 + 8 x^2 - 26 x - 9) \li3(-x)
    - (x^3 + 20 x^2 + 25 x + 10) \li3(-y)
 \PlusBreak{\null + i \pi \bigglB {1 \over 2 y^3}  \bigglP }
     (3 x^3 - 3 x^2 - 9 x - 5) Y \li2(-x)
 \PlusBreak{\null + i \pi \bigglB {1 \over 2 y^3}  \bigglP }
     {1 \over 2} (6 x^3 + 29 x^2 + 37 x + 16) X \li2(-x)
 \PlusBreak{\null + i \pi \bigglB {1 \over 2 y^3}  \bigglP }
     {1 \over 2} (5 x^3 - 6 x^2 - 15 x - 8) X Y^2
 \MinusBreak{ \null +i \pi \bigglB {1 \over 2 y^3}  \bigglP }
     {1 \over 12} (10 x^3 - 3 x^2 - 3 x + 4) X^3
   - (14 x^2 - 11 x - 1) \zeta_3
 \MinusBreak{\null + i \pi \bigglB {1 \over 2 y^3}  \bigglP }
     {\pi^2 \over 6} (3 x^3 - 15 x^2 - 33 x - 17) Y
   +  {\pi^2 \over 3} (5 x^2 + 10 x + 6) X
          \biggr)
 \PlusBreak{\null + i \pi \bigglB }
     {1 \over 12 y}  \biggl(
     (5 x + 3) Y^3
    - 15 y X^2 Y                                     
    - 6 (2 x + 1) \Bigl(X Y^2  + \pi^2 (X-Y) \Bigr)  
         \biggr)
%
 \PlusBreak{\null + i \pi \bigglB }
    {1 \over 12 y^3} \biggl(
    - (124 x^3 + 392 x^2 + 352 x + 53) \li2( - x)
 \PlusBreak{\null + i \pi \bigglB  {1 \over 12 y^3} \bigglP }
      {1 \over 2} (64 x^3 - x^2 - 38 x - 6) X^2
 \MinusBreak{\null + i \pi \bigglB  {1 \over 12 y^3} \bigglP }
      (62 x^3 + 117 x^2 + 60 x + 13) X Y
 \PlusBreak{\null + i \pi \bigglB  {1 \over 12 y^3} \bigglP }
      {\pi^2 \over 12} (124 x^3 + 29 x^2 - 110 x - 77) 
          \biggr)
 \MinusBreak{\null + i \pi \bigglB }
      {64 x + 25 \over 24 y} Y^2
%
 - {1 \over 36 y^3} (190 x^3 + 353 x^2 + 94 x + 66) X
 \MinusBreak{\null + i \pi \bigglB }
   {1 \over 36 x y^2} (190 x^3 + 116 x^2 - 335 x - 108) Y
%
 - {121 x + 211 \over 108 y^2}
       \biggr]
\,, \label{HH23}\\[1pt plus 4pt]
%
I_2^{[3]} & = &  
{1 \over 2 y^3}   
\biggl[
      (x^3 + 4 x^2 + 5 x + 5) \li4(-y)
    - (3 x^2 + 9 x - 1) \zeta_3 X
 \MinusBreak{\null - {1 \over 2 y^3} \bigglB }
      {1 \over 2} (2 x^3 + 3 x^2 + 9 x - 8) \li4(-x)
    - 3 (y - 1) x   \li4 \biggl( -{x \over y} \biggr)
 \MinusBreak{\null - {1 \over 2 y^3} \bigglB }
      (2 x^2 + 4 x + 1) Y{} (\li3(-x) - \zeta_3)
    - (x + 3) y^2 Y \li3(-y)
 \PlusBreak{\null - {1 \over 2 y^3} \bigglB }
      (3 x^2 + 6 x - 1) X \li3(-y)
    + {1 \over 2} y^3 Y^2 \li2(-x)
 \PlusBreak{\null - {1 \over 2 y^3} \bigglB }
     (x^3 + 4 x^2 + 8 x - 2) X \li3(-x)
    - 2 x{} (y - 1) \li2(-x) X Y
 \PlusBreak{ \null - {1 \over 2 y^3} \bigglB }
      {\pi^2 \over 6} (4 x^3 - x^2 - 14 x + 2) \li2(-x)
    - {1 \over 24} (x^3 - 3 x + 1) Y^4
 \MinusBreak{\null -  {1 \over 2 y^3} \bigglB }
      {x \over 4} (2 x^2 + 7 x + 11) X^2 \li2(-x)
    + {1 \over 4} (5 x^2 + 10 x - 3) X^2 Y^2
 \MinusBreak{ \null - {1 \over 2 y^3} \bigglB }
      {1 \over 6} (x^3 + 3 x^2 + 3 x - 1) X^3 Y
    - {\pi^2 \over 3} (y - 1) (x^2 - 3 x + 1) X Y
 \MinusBreak{\null -  {1 \over 2 y^3} \bigglB }
      {\pi^2 \over 12} (2 x^3 + x^2 + 2 x + 15) X^2
    + {\pi^2 \over 4} (x^3 - x^2 - 5 x - 4) Y^2
 \MinusBreak{\null -  {1 \over 2 y^3} \bigglB }
      {1 \over 6} (2 x^3 + 8 x^2 + 10 x + 1) X Y^3
    - {x y \over 48} (2 x - 1) X^4
 \MinusBreak{\null -  {1 \over 2 y^3} \bigglB }
      {\pi^4 \over 360} (20 x^3 - 85 x^2 - 248 x + 20) 
\biggl]
%
 \PlusBreak{}
   {1 \over 4 y^3}  
\biggl[
   - (x^2 + 4 x - 6) (\li3(-x)  - X \li2(-x))
   + {1 \over 3} x^2 X^3
 \PlusBreak{\null + {1 \over 4 y^3} \bigglB }
     {1 \over 3} (9 x + 5) X^2 Y
   + 9 (x - y) \zeta_3
   - {\pi^2 \over 6} (14 x + 47) X
\biggr]
 \PlusBreak{}
   {1 \over y^2}   
\biggl[
       {1 \over 4} (7 x + 9) ( \li3( - y) + \li2(-x) Y )
     +  {1 \over 12} (10 x + 13) X Y^2
     + {\pi^2 \over 3} Y
\biggr]
%
 \PlusBreak{}
   {x \over 8 y^3} (2 x^3 + 6 x + 9) X^2
 + {1 \over 12 y^2} (6 x^3 + 18 x^2 + 25 x + 6) X Y
 \PlusBreak{}
   {1 \over 8 x y} (2 x^3 + 12 x^2 + 9 x - 2) Y^2
 - {\pi^2 \over 24 y^2} (6 x^3 + 18 x^2 + 14 x - 9)
%
 \MinusBreak{}
   {x \over 6 y^2} (23 x + 26) X
 - {1 \over 12 y} (46 x + 47) Y
%
%
 \PlusBreak{}
  i \pi {}
\biggl[
  {1 \over 2 y^3}   \biggl(
   - (x^3 + 2 x^2 + x + 4) \li3(-y)
   - {\pi^2 \over 6} (x + 3) (x - y) X
 \PlusBreak{\null + i \pi \bigglB {1 \over 2 y^3} \bigglP}
     (x^3 + 2 x^2 + 4 x - 3) \li3(-x)
   - (x^2 + 5 x - 2) \zeta_3
 \PlusBreak{\null + i \pi \bigglB {1 \over 2 y^3} \bigglP }
     {1 \over 12} (2 x^3 + x^2 - x + 6) X^3
   + {1 \over 2} (y - 1) (x - 1)^2 X Y^2
 \MinusBreak{\null + i \pi \bigglB {1 \over 2 y^3} \bigglP}
     {x \over 2} (2 x^2 + 3 x + 3) X \li2(-x)
 \MinusBreak{\null + i \pi \bigglB {1 \over 2 y^3} \bigglP }
     (x^3 + x^2 - x + 1) Y \li2(-x)
 \PlusBreak{ \null + i \pi \bigglB {1 \over 2 y^3} \bigglP }
     {\pi^2 \over 6} (x^3 - 2 x^2 - 7 x - 2) Y
               \biggr)
   - {x - 2 \over 12 y} Y^3
   + {1 \over 4} X^2 Y
%
 \PlusBreak{\null +  i \pi \bigglB }
     {1 \over 4 y^3}  \biggl(
   - 3 (2 x^2 + 4 x + 5) \li2(-x)
   + {1 \over 6} (3 x^2 + 6 x + 28) X^2
 \PlusBreak{\null +  i \pi \bigglB {1 \over 4 y^3}  \bigglP }
     {1 \over 3} (x^2 + 20 x + 11) X Y
   - {\pi^2 \over 2} (2 x - 1) 
         \biggr)
 \PlusBreak{\null + i \pi \bigglB }
     {1 \over 24 y} Y^2
%
 - {1 \over 12 y^3} (24 x^3 + 25 x^2 + 4 x + 6) X
 \MinusBreak{\null +  i \pi \bigglB }
   {1 \over 12 x y^2} (24 x^3 + 38 x^2 + 15 x - 6) Y
%
 + {1 \over 12 y^2} (41 x + 47)
\biggr]
\,, \label{II23}\\[1pt plus 4pt]
%
J_2^{[3]} & = &  
{(x - 1) x \over y^3}  
\biggl[
   \li4(-x) + \li4 \biggl( -{x \over y} \biggr) - \li4(-y)
 - Y \li3(-x)
 \PlusBreak{ {(x - 1) x \over y^3} \bigglB }  
   {\pi^2 \over 6} \li2(-x)
 - {1 \over 6} X Y^3
 + {1 \over 24} Y^4
 + {\pi^2 \over 12} Y^2  
 + \zeta_3 Y
 - {7 \pi^4 \over 360}
\biggr]
%
 \PlusBreak{}
   {1 \over 24 y^3} \biggl[
     (6 x^3 + 15 x^2 + 15 x + 5) X^2 Y
    + {\pi^2 \over 3} (x^3 + 8 x^2 + 19 x + 18) X
      \biggr]
 \PlusBreak{}
   {1 \over 3 y^2} \biggl[
     - (x^2 + 6 x - 1) (\li3(-y)  +  Y \li2(-x))
     + {1 \over 6} (2 x - 1) x X^3
 \MinusBreak{\null - {1 \over 3 y^2} \bigglB }
       {1 \over 8} (x + 9) (2 x - 1) X Y^2
     + (5 x - 1) \zeta_3
        \biggr]
 \PlusBreak{}
   {1 \over 3 y}  \biggl(
    (y - 1) (\li3(-x)-X \li2( - x))
     - {\pi^2 \over 24} (5 x - 21) Y
          \biggr)
 \MinusBreak{}
   {1 \over 9} Y^3
%
 + {x \over 24 y^3} (6 x^3 + 41 x^2 + 48 x + 19) X^2
 \PlusBreak{}
   {1 \over 72 y^2} (36 x^3 + 108 x^2 + 37 x - 29) X Y
 + {(5 x - 4) x \over 18 y^2} X
 \PlusBreak{}
   {1 \over 24 x y} (6 x^3 - 5 x^2 + 13 x + 12) Y^2
 + {(20 x - 19) \over 72 y} Y
 \MinusBreak{}
   {\pi^2 \over 36 y^2} (9 x^3 + 27 x^2 + 14 x + 11) 
%
%
%
 \PlusBreak{}
  i \pi {}
\biggl[
   - {(x - 1) x \over y^3} (\li3(-x) - \zeta_3)
%
    - {1 \over 3 y^2} (2 x^2 + 9 x + 1) \li2(-x)
 \PlusBreak{\null + i \pi \bigglB }
     {1 \over 24 y^3} \Bigl(
     2 (4 x^2 + 6 x + 3) x X Y
      - (6 x^3 + 5 x^2 + x + 3) X^2
               \Bigr)
%
 \PlusBreak{\null + i \pi \bigglB }
      {1 \over 72 y^3} (102 x^3 + 143 x^2 + 106 x + 29) X
 \PlusBreak{\null + i \pi \bigglB }
      {1 \over 72 x y^2} (102 x^3 - 11 x^2 - 179 x - 72) Y
 \PlusBreak{\null + i \pi \bigglB }
      {x - y \over 8 y} Y^2
    - {\pi^2 \over 72 y} (4 x - 7)
%
    - {17 x - 19 \over 72 y^2} 
\biggr]
\,, \label{JJ23}\\[1pt plus 4pt]
%
K_2^{[3]} & = &  
{1 \over y^3}  
\biggl[
2\biggl( \li4(-x) +\li4\biggl( -{x \over y}\biggr) -\li4(-y) \biggr) (x -
1){}x
 \PlusBreak{ {1 \over y^3} \bigglB }
      2 ( - \li3(-x) + \zeta_3) (x - 1) x Y
    + {\pi^2 \over 3} (x - 1) x \li2(-x)
 \MinusBreak{ {1 \over y^3} \bigglB }
      {x \over 3} (x - 1) X Y^3
    + {x \over 12} (x - 1) Y^4
    + {\pi^2 \over 6} (x - 1) x Y^2
    - {7 \pi^4 \over 180} (x - 1) x
\biggr]
%
 \PlusBreak{}
   {1 \over 3 y^3} 
\biggl[
    - x^3 \li3(-x)
    + x^3 X \li2(-x)
    - {1 \over 8} X^2 Y
 \MinusBreak{\null - {1 \over 3 y^3} \bigglB }
      (15 x^2 + 15 x + 1) \zeta_3
    + {\pi^2 \over 12} (10 x^3 + 3) X
\biggr]
 \MinusBreak{}
   {1 \over 3 y^2} (x^2 + 14 x + 1) (\li3(-y) + Y \li2(-x))
 \MinusBreak{}
   {1 \over 24 y^2} (8 x^2 + 63 x + 7) X Y^2
 + {\pi^2 \over 12 y} (4 y + 1) Y
%
 \MinusBreak{}
  {1 \over 8 y^2} 
    \biggl( (4 x^2 + x - 1) X Y
        -{\pi^2 \over 9} (18 x^2 + 43 x - 9)
           \biggr)
 \MinusBreak{}
   {1 \over 4 y^3} (x^2 - 2) x X^2
 - {1 \over 4 x y} (x^2 - 6 x - 3) Y^2
%
 - {8 x + 29 \over 24 y} Y
 - {x \over 6 y^2} (2 x + 7) X
%
 \PlusBreak{}
  i \pi {}
\biggl[
    2 { (x - 1) x \over y^3} ( -\li3(-x) + \zeta_3)
%
 \PlusBreak{\null +  i \pi \bigglB }
      {1 \over 3 y^3}  \biggl(
     (2 x^3 + 15 x^2 + 15 x + 1) \li2(-x)
      - {1 \over 8} (4 x^3 + 1) X^2
 \PlusBreak{\null + i \pi \bigglB {1 \over 3 y^3}  \bigglP }
        {1 \over 4} (4 x^3 + 11 x^2 + 10 x + 2) X Y
      - {\pi^2 \over 6} x^3
           \biggr)
   - {4 y + 1 \over 24 y} Y^2
%
 \PlusBreak{\null +  i \pi \bigglB }
     {1 \over 8 y^3} (5 x^2 + 8 x - 1) X
    -  {1 \over 8 y^2 x} (21 x^2 + 35 x + 12) Y
%
 + {9 x + 29 \over 24 y^2}   
\biggr]
\,, \hskip 1.4 cm \label{KK23}\\[1pt plus 4pt]
%
L_2^{[3]} & = &  
   {1 \over 72 y} X Y
 - {x \over 36 y} X^2
 - {1 \over 36} Y^2
 + {5 x \over 54 y} X
 - {\pi^2 \over 24 y}
 + {5 \over 54} Y
 \PlusBreak{}
    {i \pi \over 72 y}  \biggl( (5 + 4 x) Y
                + ( 1 - 4 x) X
                - {20 \over 3} \biggr)
\,. \label{LL23}
\end{eqnarray}

\noindent
For $h=3$ in \eqn{hel3}
and color factor $\trc^{[1]}$ in \eqn{TraceBasisqggq}:
\begin{eqnarray}
A_3^{[1]} & = & 
 {1 \over 48 x y^2} (-3 x^3 - 31 x^2 - 2 x - 4) X^2
 \PlusBreak{}
 {1 \over 48 x y} (55 x y - 16 y^2 + 60 x^2) X
+{175 y \over 108 x} 
 -{25 \over 864} 
 \PlusBreak{}
 i \pi {} \biggl(
    {1 \over 24 x y^2} (-3 x^3 - 31 x^2 - 2 x - 4) X
   + {1 \over 48 x y} (55 x y - 16 y^2 + 60 x^2) 
\biggr)
\,, \hskip 1.9 cm \label{AA31}\\[1pt plus 4pt]
%
B_3^{[1]} & = & 
    -{1 \over 16 y^2} (4 x^3 + y^3 + 7 x^2 y + 3 x y^2) X^2
 \MinusBreak{}
     {y \over 16 x^2} (6 x^2 + 6 y^2 + 11 x y) Y^2
    -{1\over 8} (1 - y) Y X
 \MinusBreak{}
     {2 \over 24 y} (6 - 5 y) X
    -{3 (1-y) \over 8 x} Y 
    +{\pi^2 \over 16} (1 - y)
    -{24 \over 72 x} 
    -{139 \over 72} 
 \PlusBreak{}
  i {\pi \over 8} \biggl(
   -{5 x + 1 \over y^2} X
   -{7 y + 1 \over x^2} Y
   +{1 \over 3 x y} (18 y^2 + 31 x y + 12 x^2) 
\biggr)
\,, \label{BB31}\\[1pt plus 4pt]
%
C_3^{[1]} & = &
-{y \over 16 x^2} (6 x^2 + 6 y^2 + 11 x y) Y^2
+{y - 1 \over 8} Y X
-{3 x - 1 \over 16 y} X
 \PlusBreak{}
 {1 \over 16 y^2} (6 x^2 y + 4 x y^2 - y^3 + 2 x^3) X^2
 +(y - 1) \biggl( {3 \over 8 x} Y - {\pi^2 \over 16} \biggr) 
- {17 \over 32}
 \PlusBreak{}
  i {\pi \over 8} \biggl(
   - {1 \over y^2} (y^2 + 4 x y + 2 x^2) X
    +{6 + 7 x \over x^2}Y
    - {1 \over 2 x y} (-12 y^2 + 4 x^2 - 5 x y) 
\biggr)
\,, \label{CC31}\\[1pt plus 4pt]
%
D_3^{[1]} & = &
{5 x^2 + x + 2 \over 24 x y^2} X^2
+{x^2 + 23 x + 10 \over 24 x y} X
-{397 y\over 216 x} - {143 \over 108}
 \PlusBreak{}
   i \pi {}\biggl(
   {5 x^2 + x + 2 \over 12 x y^2} X
  +{x^2 + 23 x + 10 \over 24 y x}
\biggr)
\,, \label{DD31}\\[1pt plus 4pt]
%
E_3^{[1]} & = &
 -{\pi^2 \over 8} (1 + 3 y)
 +{1 - y \over 8 y^2} (3 x^2 + 2 x y + 2 y^2) X^2
 \PlusBreak{}
  {18 x + 35 y \over 24 y} X
 -{y \over 8 x^2} (2 y^2 + 6 x^2 + 5 x y) Y^2
 \PlusBreak{}
  {1 \over 4} (1 + 3 y) Y X
 -{3 (1 - y) \over 4 x} Y
 +{19 x + 33 \over 72 x}
 \MinusBreak{}
  i \pi {} \biggl(
       {x - 2 \over 4 y^2} X
     - {1 - y \over 4 x^2} Y
     - {1 \over 24 y x} (36 y^2 + 18 x^2 + 53 x y)
\biggr)
\,, \label{EE31}\\[1pt plus 4pt]
%
F_3^{[1]} & = &
{1 \over 12 x} X
-{5 \over 54 x}
+i \, {\pi  \over 12 x}
\,, \label{FF31}\\[1pt plus 4pt]
%
G_3^{[1]} & = & 
0
\,. \label{GG31}
\end{eqnarray}


\noindent
For $h=3$ in \eqn{hel3} 
and color factor $\trc^{[2]}$ in \eqn{TraceBasisqggq}:
\begin{eqnarray}
A_3^{[2]} & = &
{1 \over x y}   
\biggl[
-{1 \over 48} (30 x^3 + 39 x^2 + 10 x + 4) \Bigl((X - Y)^2 + \pi^2 \Bigr)
 \MinusBreak{ {1 \over x y} \bigglB }
 {1 \over 48} (60 x^2 + 55 x - 16) X
+{1 \over 144} (180 x^2 + 121 x + 40) Y
 \PlusBreak{ {1 \over x y} \bigglB }
{25 \over 864} (x - 56)
\biggr]
-i \pi {} {11 (x - 2) \over 36 x y}
\,, \label{AA32}\\[1pt plus 4pt]
%
B_3^{[2]} & = & 
{6 - x \over 16 x^2} Y^2
+{1 - y \over 16 y^2} X^2
+{1 \over 16} (4 x - 1) \Bigl((X - Y)^2 + \pi^2\Bigr)
 \MinusBreak{}
 {6 x + 11 \over 12 y} X
+{1 \over 8 x y} (4 x^2 + 3 x + 6) Y
+{115 x - 24 \over 72 x y}
 \MinusBreak{}
  i {\pi \over 8}   \biggl(
    {x - 6 \over x^2} Y
   +{y - 1 \over y^2} X
   +{13 x - 18 \over 3 x y}
          \biggr)
\,, \label{BB32}\\[1pt plus 4pt]
%
C_3^{[2]} & = &
 {1 \over 16}  
\biggl[
  -{x - 6 \over x^2} Y^2
  -{y - 1 \over y^2} X^2
  +{1 \over y} (2 x^2 + 4 x + 1)  \Bigl((X - Y)^2 + \pi^2 \Bigr)
 \PlusBreak{ {1 \over 16} \bigglB }
   {4 x + 1 \over y} X
  -{1 \over x y} (4 x^2 - 5 x - 12) Y
  +{17 \over 2 y}
\biggr]
 \MinusBreak{}
  i {\pi \over 8}  \biggl(
   {x - 6 \over x^2} Y
  +{y - 1 \over y^2} X
  +3 {y - 1 \over x y} \biggr)
\,, \label{CC32}\\[1pt plus 4pt]
%
D_3^{[2]} & = & 
{1 \over 24 x y}   
\biggl[
   (12 x^2 + 3 x - 10) X
  -{1 \over 3} (36 x^2 + 49 x + 22) Y
  + {1 \over 9} (286 x + 397)
\biggr]
 \MinusBreak{}
 {1 \over 24 x} (6 x^2 + 3 x + 2) \Bigl((X - Y)^2 + \pi^2 \Bigr)
 -i \pi {} \, {10 x + 13 \over 18 x y}
\,,\hskip 4.5 cm \label{DD32}\\[1pt plus 4pt]
%
E_3^{[2]} & = & 
{1 \over 8}  
\biggl[
-{x - 2 \over y^2} X^2
-{y  - 1 \over x^2} Y^2
+(3 x + 2) \Bigl((X - Y)^2 +\pi^2\Bigr)
 \MinusBreak{ {1 \over 8} \bigglB }
 {18 x + 35 \over 3 y} X
+{1 \over x y} (12 + 19 x + 6 x^2) Y
+{33 + 14 x \over 9 x y}
\biggr]
 \MinusBreak{}
 i \,{\pi \over 4}  \biggl(
    {x - 2 \over y^2} X
   +{y - 1 \over x^2} Y
   -{11 x + 18 \over 3 x y} 
              \biggr)
\,, \label{EE32}\\[1pt plus 4pt]
F_3^{[2]} & = & 
-{1 \over 12 x}   \biggl( X + {1 \over 3} Y - {10 \over 9} \biggr)
           -i \, {\pi \over 9 x}
\,, \label{FF32}\\[1pt plus 4pt]
G_3^{[2]} & = & 
0
\,. \label{GG32}
\end{eqnarray}

\noindent
For $h=3$ in \eqn{hel3}
and color factor $\trc^{[3]}$ in \eqn{TraceBasisqggq}:
\begin{eqnarray}
H_3^{[3]} & = & 
{y - 1 \over 8 x^2 y} (1 - x y) Y^2
-{1 \over 48 y^2 x} (6 x^3 + 25 x^2 - 7 x + 4) X^2
-{21 x + 32 \over 18 x} Y
 \MinusBreak{}
 {1 \over 48 x y} (30 x^3+51 x^2+19 x+4) \Bigl((X-Y)^2 + \pi^2 \Bigr)
+{7 (y-1) \over 6 y} X
 \PlusBreak{}
 i \pi {} \biggl(
    {y - 1 \over 4 x^2 y} (1-x y) Y 
   -{1 \over 24 y^2 x} (6 x^3 + 25 x^2 - 7 x + 4) X
 +   {11 x + 32 \over 18 x y}
            \biggr)
\,,  \hskip 1.5 cm \label{HH33}\\[1pt plus 4pt]
%
I_3^{[3]} & = & 
 {y - 1 \over 8 x^2 y} (x^2 - 2 x - 2) Y^2
-{x - 1 \over 16 y} \Bigl( (X - Y)^2 + \pi^2 \Bigr)
 \PlusBreak{}
{x - y \over 16 y} X^2
-{y - 1 \over 12 y} X
+{x + 12 \over 12 x} Y
 \PlusBreak{}
  i \pi {} \biggl(
 {x - y \over 8 y} X
+{y - 1 \over 4 x^2 y} (x^2 - 2 x - 2) Y
-{11 x + 12 \over 12 x y}
\biggr)
\,, \label{II33}\\[1pt plus 4pt]
%
J_3^{[3]} & = & 
{1 \over 24 x y^2} (5 x^2 + x + 2) X^2
+{y - 1 \over 4 x^2} Y^2
+{39 x + 94 \over 72 x} Y
 \MinusBreak{}
 {13 (y - 1) \over 24 y} X
-{1 \over 24 x} (6 x^2 + 3 x +2) \Bigl( (X - Y)^2 + \pi^2 \Bigr)
 \PlusBreak{}
 i \pi {} \biggl(
    {y - 1 \over 2 x^2} Y
   +{1 \over 12 x y^2} (5 x^2 + x + 2) X
   -{55 x + 94 \over 72 x y}
          \biggr)
\,, \label{JJ33}\\[1pt plus 4pt]
%
K_3^{[3]} & = & 
{x - 2 \over 8 y^2} X^2
+ {y - 1 \over 8 x^2} Y^2
-{1 \over 8} (3 x + 2) \Bigl( (X - Y)^2 + \pi^2 \Bigr)
 \PlusBreak{}
  {19 x + 36 \over 24 x} Y
- {19 (y-1) \over 24 y} X
 \PlusBreak{}
  i \pi {} \biggl(
  {y - 1 \over 4 x^2} Y
+ {x - 2 \over 4 y^2} X
- {17 x + 36 \over 24 x y} 
\biggr)
\,, \label{KK33}\\[1pt plus 4pt]
L_3^{[3]} & = & 
-{1 \over 36 x} (Y + i \pi) 
\,. \label{LL33}
\end{eqnarray}


\noindent
For $h=4$ in \eqn{hel4}
and color factor $\trc^{[1]}$ in \eqn{TraceBasisqggq}:
\begin{eqnarray}
A_4^{[1]} & = &
{1 \over 2 y^3} \biggl[
     {1 \over 2} (x^3 + 3 x^2 + 3 x - 2) 
         \Bigl(\pi^2 X^2 - \li4(-x) \Bigr)
      - X \li3(-x)
 \PlusBreak{ {1 \over 2 y^3} \bigglP }
     (1 - x  y) X^2 \biggl( {1 \over 4} (2 + x) \li2(-x) +
                            {1 \over 8} (y - 1) X^2 \biggr)
 \PlusBreak{ {1 \over 2 y^3} \bigglP }
    {1 \over 2} (15 x + 5 x^3 + 15 x^2 + 7) \zeta_3 X
   +{13 \over 6} \pi^2 X^2
 \PlusBreak{ {1 \over 2 y^3} \bigglP }
    {\pi^4 \over 480} (3 x + 1) (19 + 24 x + 9 x^2) 
\biggr]
 \PlusBreak{}
 {\pi^2 \over 6} \li2(-x)
+{\pi^2 \over 6} Y X
-{1 \over 12} Y X^3
%
 \PlusBreak{}
 {1 \over 288 y^3} 
\biggl[
    4 (50 + 108 x + 126 x^2 + 45 x^3) X^3
 \MinusBreak{\null + {1 \over 288 y^3} \bigglP }  
    (579 x + 795 x^2 + 297 x^3 + 265) \pi^2 X  
\biggr]
 \PlusBreak{}
{1 \over 48 y^2}  
\biggl[
     (44 + 73 x + 35 x^2) ( 2 X \li2(-x) + Y X^2 - 2 \li3(-x) )
 \PlusBreak{\null + {1 \over 48 y^2} \bigglP }
    (241 x^2 + 470 x + 205) \zeta_3
\biggr]
%
 \MinusBreak{}
 {1 \over 288 y^3} (1530 x^2 + 505 x^3 + 368 + 918 x) X^2
 \PlusBreak{}
 {1 \over 288 y^2} (18 x^2 - 431 x + 501) X
%
 \PlusBreak{}
{\pi^2 \over 72 y^2} (18 - 73 x - 45 x^2)
%
-{475 \over 288 y} - {36077 \over 3456}
%
 \PlusBreak{}
 i\, {\pi \over 4} {}
\biggl[
    {1 \over y^3} \biggl(
    (2 + x) (1 - x y) X \li2(-x)
   -(1 - x y) ( 1 - y) X^3 
 \MinusBreak{\null + i {\pi \over 4} \bigglB {1 \over y^3} \bigglP }
    2 \li3(-x)
   +{2 \pi^2 \over 3} X
    +(5 x^3 + 15 x^2 + 15 x + 7) \zeta_3  
         \biggr)
 \MinusBreak{ i {\pi \over 4} \bigglB }
  Y X^2
%
+{1 \over 6 y^3} (50 + 108 x + 126 x^2 + 45 x^3) X^2
 \PlusBreak{ i {\pi \over 4} \bigglB }
 {1 \over 6 y^2} (44 + 73 x + 35 x^2) (\li2(-x) + X Y)
%
 \MinusBreak{ i {\pi \over 4} \bigglB }
 {1 \over 36 y^3} (505 x^3 +1530 x^2 + 368 + 918 x) X
 \MinusBreak{ i {\pi \over 4} \bigglB }
  {\pi^2 \over 24 y^2} ( 21 x^2 + 50 x + 45)  
%
+ {475 \over 36 y^2} + {467 \over 72 y} + {1 \over 4} 
\biggr]
\,, \hskip 5 cm \label{AA41}\\[1pt plus 4pt]
%
B_4^{[1]} & = &  
{1 \over 4 y^3} 
\biggl[
    (y-1) (4 x^2 - 5 x- 5) \li4(-x)
   +{x \over 12} (x^2 - 3 x - 3) X^4
 \MinusBreak{ {1 \over 4 y^3} \bigglP }
    2 (4 x - 4 x^2 - 6 x^3 + 1) X \li3(-x)
 \MinusBreak{ {1 \over 4 y^3} \bigglP }
    {\pi^2 \over 3} \li2(-x) (25 x^3 + 44 x^2  + 31 x + 14)
 \MinusBreak{ {1 \over 4 y^3} \bigglP }
    (11 x^3 + 17 x^2 + 13 x + 9) \zeta_3 X
  -  {\pi^2 \over 6} (4 x^3 + 3 x^2 - 6 x - 12) X^2  
 \PlusBreak{ {1 \over 4 y^3} \bigglP }
    {\pi^4 \over 720} (653 x^3 - 133 x^2 - 1241 x  - 343) 
\biggr]
 \PlusBreak{}
 {1 \over 2 y^2} 
\biggl[
   -x{} (x + 13) \li4(-y)
   -(7 x^2 - x + 4) \li4 \left( -{x \over y} \right)
 \PlusBreak{\null + {1 \over 2 y^2}  \bigglP }
    (3 x + 4 + 5 x^2) (\li3(-x) - \zeta_3) Y
   - 2 (2 - x + 3 x^2) X \li3(-y)
 \PlusBreak{\null + {1 \over 2 y^2}  \bigglP }
    {1 \over 4} (6 x^2 - 5 x - 2) X^2 \li2(-x)
   - 2 \li2(-x) (2 x^2 + x + 2) X Y
 \MinusBreak{\null + {1 \over 2 y^2}  \bigglP }
   2 y Y^2 \li2(-x)
   + 4 y^2 Y \li3(-y)
   -{1 \over 4} (-3 x + 13 x^2 + 8) X^2 Y^2
 \PlusBreak{\null + {1 \over 2 y^2}  \bigglP }
    {1 \over 3} (5 x^2 + 4 x + 5) Y^3 X
   -{1 \over 6} (1 + 7 x) Y X^3
 \MinusBreak{\null + {1 \over 2 y^2}  \bigglP }
    {1 \over 24} (5 x^2 - 5 x + 2) Y^4
   -{\pi^2 \over 6} (8 x^2 + 9 x + 7) Y^2   
 \PlusBreak{\null + {1 \over 2 y^2}  \bigglP }
    {\pi^2 \over 3} (6 + 4 x + 13 x^2) Y X
\biggr]
%
 \PlusBreak{}
  {1 \over y^3} 
\biggl[
    {x \over 4} (4 + 10 x + 3 x^2) ( X \li2(-x) - \li3(-x) )
 \MinusBreak{\null - {1 \over y^3} \bigglP }
    {x \over 72} (6 x + 32 x^2 + 15) X^3
 \PlusBreak{\null + {1 \over y^3} \bigglP }
    {1 \over 144} (1509 x + 443 x^3 + 443 + 1617 x^2) \zeta_3
 \PlusBreak{\null + {1 \over y^3} \bigglP }
    {\pi^2 \over 288} \Bigl(477 + 891 x + 999 x^2 + 733 x^3 \Bigr) X  
\biggr]
 \PlusBreak{}
 {1 \over 24 y^2} \left(
    x{} (11 x - 19) Y X^2
   -2 (x^2 + 9 x + 10) \pi^2 Y
           \right)
 \MinusBreak{}
 {1 \over y} \biggl(
     3 x {} ( \li3(-y) + Y \li2(-x) )
   +{1 \over 12} (27 x + 20) X Y^2
        \biggr)
 \MinusBreak{}
 {5 \over 6} Y^3
%
-{1 \over 144 y^3} (246 x^2 - 215 x^3 + 246 x + 144) X^2
 \MinusBreak{}
 {1 \over 288 y^2} \left(
    4 x{} (145 + 91 x) X Y
   -(311 x^2 - 340 x - 455) \pi^2
          \right)
 \PlusBreak{}
 {1 \over 144 x} (191 x + 27) Y^2
%
-{1 \over 216 y^2} (269 x^2 + 1966 x + 782) X
 \PlusBreak{}
 {3 (x - y) \over 8 y} Y
%
-{139 \over 72 y} + {14135 \over 2592}
%
%
 \PlusBreak{}
 i \pi {} \biggl[
  {1\over y^3}   \biggl(                            
    {1\over 2} (x^3 - 4 x^2 - 11 x - 5) \li3(-x)     
    -{\pi^2\over 12} (-7 x^2 + 2 x^3 - 6 - 14 x)  X
 \PlusBreak{\null + i \pi \bigglB {1\over y^3} \bigglP }
     {1\over 4} (y^3 + 2 x^2 + 4 x) \zeta_3          
    - {1 \over 12} x {} (x y - 2) X^3
            \biggr)
 \PlusBreak{\null + i \pi \bigglB }
   {1\over y^2}   \biggl(
     - (x-5) x \li3(-y)                    
     + {1\over 4} (2 x + 5) (y - 1) X  \li2(-x)   
 \MinusBreak{\null + X i \pi \bigglB {1\over y^2} \bigglP }
      x {} (2 x-1) Y \li2(-x)
   -{\pi^2\over 12} Y {} (17 x + 3 x^2 + 8)        
 \MinusBreak{\null + X i \pi \bigglB {1\over y^2} \bigglP }
    {1\over 2} (1 - 2 x + 3 x^2) X Y^2
           \biggr)
 \PlusBreak{\null + i \pi \bigglB }
  {1\over y} \biggl(
    -{1\over 12} x Y^3
    - {1\over 2} x X Y^2          
    + {1\over 4} (3 x + 1) X^2 Y
    - {\pi^2 \over 2} x X         
          \biggr)
%
 \MinusBreak{\null + i \pi \bigglB }
   {x \over 4 y^3} (9 x^2 + 14 x + 8) \li2(-x)
   + {\pi^2 \over 96 y^3} (23 x^3 + 93 x^2 + 161 x + 79)    
 \MinusBreak{\null + i \pi \bigglB }
   {x \over 6 y^3} (13 x^2 + 7 x + 2) X^2
   +{1 \over 12 y^2} (29 x^2 + 39 x + 40) X Y 
 \PlusBreak{\null + i \pi \bigglB }
    {21 x + 10 \over 12 y} Y^2
%
   + {1 \over 72 y^3} (306 x^3 - 10 x^2 - 101 x - 144) X
 \PlusBreak{\null + i \pi \bigglB }
    {1 \over 72 x y^2} (100 x^3 + 264 x^2 + 27 + 245 x) Y
%
 \MinusBreak{\null + i \pi \bigglB }
     {1\over 216 y^2} (431 x^2 + 2209 x + 863)   
\biggr]
\,, \label{BB41}\\[1pt plus 4pt]
%
C_4^{[1]} & = &
{1 \over 2 y^3} \biggl[
                    x{} (x^2  - 4 x + 1) \biggl( 
                              \li4(-y) - \li4 \left( -{x \over y} \right) 
                             + ( \li3(-x) - \zeta_3 ) Y 
 \MinusBreak{\null + {1 \over 2 y^3} \bigglB x (1 - 4 x + x^2) \bigglP }
                              {1 \over 24} Y^4
                             -{\pi^2 \over 12} Y^2
                             +{1 \over 6} Y^3 X
                                      \biggr)
 \PlusBreak{\null + {1 \over 2 y^3} \bigglB }
      {1 \over 2} x{} (5 x^2 + 8 x  - 2) \li4(-x)
     -2 x^3 X \li3(-x)
 \PlusBreak{\null + {1 \over 2 y^3} \bigglB }
      {x^3 \over 4} \li2(-x) X^2
     +{\pi^2 \over 6} x{} (x^2 + 4 x - 1) \li2(-x)
     -{x^3 \over 12} X^4
 \MinusBreak{\null + {1 \over 2 y^3} \bigglB }
      (x^3  + 9 x^2 + 9 x + 3) \zeta_3 X
     +{\pi^2 \over 12} x^3 X^2
 \MinusBreak{\null + {1 \over 2 y^3} \bigglB }
      {\pi^4 \over 360} (16 x^3  - 5 x^2 - 40 x - 11)
\biggr]
%
 \PlusBreak{}
{1 \over 24 y^3} 
\biggl[
    9 x{} (5 x^2 + 1 + 4 x) ( \li2(-x) X - \li3(-x) )
 \PlusBreak{\null + {1 \over 24 y^3} \bigglP }
    3 (15 + 29 x + 21 x^3 + 29 x^2) \zeta_3
   + x{} (1 + 4 x + 9 x^2) X^3
 \PlusBreak{\null + {1 \over 24 y^3} \bigglP  }
    (3 + 10 x + 13 x^2 + 9 x^3) \pi^2 X 
\biggr]
 \PlusBreak{}
 {1 \over 4 y^2} 
\biggl[ 
     3 x{}(x-1) (2 \li3(-y) + 2 \li2(-x) Y + X Y^2)
 \MinusBreak{\null + {1 \over 4 y^2} \bigglP }
    {1 \over 12} x{} (3 x + 1) Y{} (3 X^2 + 4 \pi^2)
\biggr]
%
 \MinusBreak{}
 {x \over 32 y^3} (17 x^2 + 6 + 34 x) X^2
-{1 \over 16 y x} (2 x + 11 x^2 + 3) Y^2
 \MinusBreak{}
 {1 \over 96 y^2} \Bigl(
   12 x{}(5 + 11 x) X Y
  -(51 x^2 + 100 x + 29) \pi^2
             \Bigr)
%
 \PlusBreak{}
 {1 \over 32 y^2} (66 x^2 + 23 x + 3) X
+{3 (x - y) \over 8 y} Y
%
-{17 \over 32 y} + {187 \over 128}
%
%
%
 \PlusBreak{}
 i \pi {}
\biggl[
    {1 \over 12 y^3} \biggl(
    6 x{} ( 1 - 4 x - x^2) (\li3(-x) - \zeta_3)
 \PlusBreak{\null + i \pi \bigglB {1 \over 12 y^3} \bigglB }
    3 x^3 \li2(-x) X
   +x^3 \pi^2 X
   -x^3 X^3    
  \biggr)
   + {3\over 2} \zeta_3    
%
 \PlusBreak{\null + i \pi \bigglB }
   {1 \over 24 y^3} \biggl(
    9 x{} (x^2 + 4 x + 5) \li2(-x)
  + 9 x^3 X^2
   -  (3 x^3 - 5 x^2 - 8 x - 3) \pi^2    
             \biggr)
 \MinusBreak{\null + i \pi \bigglB }
   {x \over 8 y^2} (3 x + 1) X Y
%
  +{x \over 16 y^3} (5 x^2 + 4 - 2 x) X
 \PlusBreak{\null + i \pi \bigglB }
   {1 \over 8 x y^2} (8 x^2 + 5 x + 3) Y
%
   +{1 \over 32 y^2} (3 x + 1) (14 x - 9)   
\biggr]
\,, \label{CC41}\\[1pt plus 4pt]
%
D_4^{[1]} & = &
- {1 \over 144 y^3} \biggl[
    2 (9 x^3 + 27 x + 27 x^2 + 14) X^3
  - (47 + 81 x + 27 x^3 + 81 x^2) \pi^2 X
        \biggr] 
 \PlusBreak{}
{1 \over 3} (\li3(-x) - \li2(-x) X)
-{17 \over 24} \zeta_3
-{1 \over 6} Y X^2
%
 \PlusBreak{}
 {1 \over 72 y^3} (43 x^3 + 90 x + 135 x^2 + 35) X^2
+{\pi^2 \over 216 y^2} (77 x^2 + 139 x + 32)
%
 \PlusBreak{}
 {1 \over 216 y^2} (53 + 400 x + 125 x^2) X
%
+{37 \over 72 y} + {1307 \over 864}
 \PlusBreak{}
 i \pi {} \biggl[
  -{1 \over 3} \li2(-x)
  -{1 \over 3} X Y
  -{1 \over 24 y^3} (9 x^3 + 27 x + 27 x^2 + 14) X^2
 \PlusBreak{\null + i \pi \bigglB }
   {1 \over 16} \pi^2    
%
 + {1 \over 36 y^3} (43 x^3 + 90 x + 135 x^2 + 35) X
%
 \PlusBreak{\null + i \pi \bigglB }
   {1\over 216 y^2} (125 x^2 + 400 x + 53)  
\biggr]
\,, \label{DD41}\\[1pt plus 4pt]
%
E_4^{[1]} & = & 
 {x{} (x - 1) \over y^3} 
\biggl[
    -2 \biggl( \li4(-x) + \li4(-y) - \li4 \left( -{x \over y} \right) \biggr)
 \PlusBreak{ {x (x - 1) \over y^3} \bigglP }
     2 (X - Y) ( \li3(-x) - \zeta_3 )
    -{\pi^2 \over 3} \li2(-x)
 \MinusBreak{ {x (x - 1) \over y^3} \bigglP }
     {1 \over 3} X Y^3
    +{1 \over 12} Y^4
    +{1 \over 12} X^4
    +{\pi^2 \over 6} Y^2
    + {\pi^2 \over 3} X^2
    +{7 \over 60} \pi^4
\biggr]
%
 \PlusBreak{}
  {1 \over 144 y^3}
\biggl[
     48 (15 x + 15 x^2 + x^3 + 2) ( \li3(-x) - \li2(-x) X )
 \PlusBreak{\null + {1 \over 144 y^3} \bigglP }
     2 x{} (13 x^2 - 24 - 24 x) X^3
   - 2 (x^3 + 3 x^2 + 25 + 3 x) \zeta_3
 \MinusBreak{\null +  {1 \over 144 y^3} \bigglP }
     3 \pi^2 (13 x^3 + 9 + 71 x + 71 x^2) X 
\biggr]
 \PlusBreak{}
  {1 \over 6 y^2} 
\biggl[
      -2 (x^2 + 14 x + 1) ( \li3(-y) + Y \li2(-x) ) 
     + {1 \over 4} (6 x^2 + 17 x + 8) Y X^2
 \MinusBreak{\null +  {1 \over 6 y^2} \bigglP }
     (2 x^2 + 17 x + 3) X Y^2
     - {\pi^2 \over 6} (2 x^2 - 53 x - 7) Y
\biggr]
 \PlusBreak{}
 {1 \over 12} Y^3
%
-{x \over 72 y^3} (123 + 312 x + 244 x^2 + 54 x^3) X^2
 \PlusBreak{}
{\pi^2 \over 144 y^2} (108 x^3 + 287 x^2 + 128 x + 33)
-{1 \over 36 y} (44 x + 27 x^2 - 19) Y^2
 \MinusBreak{}
 {x \over 18 y^2} (43 + 76 x + 27 x^2) X Y
%
+{1 \over 216 y^2} (353 x^2 + 478 x + 191) X
 \PlusBreak{}
 {9 x + 4 \over 6 y} Y
%
+ {19 \over 72 y} - {3401 \over 2592}
%
%
 \PlusBreak{}
  i \pi {}
\biggl[
    {1 \over 3 y^3} \biggl(
    {\pi^2 \over 48} (21 x^3 + 51 x^2 + 39 x + 17)  
    -\li2(-x)
 \PlusBreak{\null + i \pi \bigglB {1 \over 3 y^3} \bigglP }
     {x \over 8} (13 x + 11 x^2 + 11) X^2
                    \biggr)
 \PlusBreak{\null + i \pi \bigglB }
     {x{} (2 x + 5) \over 12 y^2} X Y
    -{(x - 1) \over 12 y} Y^2
%
    - {x \over 36 y^3} (38 x^2 + 37 + 74 x) X
 \MinusBreak{\null + i \pi \bigglB }
     {1 \over 18 y^2} (18 x + 5 x^2 + 19) Y
%
    -{8 x - 3 \over 36 y^2} + 
     {29\over 216}  
\biggr]
\,, \label{EE41}\\[1pt plus 4pt]
%
F_4^{[1]} & = &
{1 \over 24} X^2 - {\pi^2 \over 27} - {5 \over 54} X
+ i \pi {} \biggl( {1 \over 12} X - {5 \over 54} \biggr)     
\,, \label{FF41}\\[1pt plus 4pt]
%
G_4^{[1]} & = &
{1 \over 72 y} (
     x Y X + Y^2 y + \pi^2
        )
+ i \, {\pi \over 72 y} (
 x X + (y-1) Y 
       )
\,. \hskip 4.7 cm \label{GG41}
\end{eqnarray}


\noindent
For $h=4$ in \eqn{hel4}
and color factor $\trc^{[2]}$ in \eqn{TraceBasisqggq}:
\begin{eqnarray}
A_4^{[2]} & = & 
{1 \over 4 y^3}   
\biggl[
  -4 (x - 4) x {} (\li4(-x) + \li4(-y))
 \MinusBreak{ {1 \over 4 y^3} \bigglB }
   (3 x^3 + 4 x^2 + 23 x + 2) \biggl(
     \li4 \biggl( -{x \over y} \biggr) - {\pi^2 \over 6} \li2(-x) \biggr)
 \MinusBreak{ {1 \over 4 y^3} \bigglB }
   2 (x^3 + x^2 + 11 x + 1) (X - Y) \li3(-x)
 \PlusBreak{ {1 \over 4 y^3} \bigglB }
   {1 \over 12} (3 x^3 + 14 x^2 + 7 x + 6) X^4
  +{1 \over 6} (x^3 + 5 x^2 + x + 2) Y^4
 \MinusBreak{ {1 \over 4 y^3} \bigglB }
    {1 \over 6} (x^3 + 10 x^2 - 7 x + 4) \Bigl( X Y^3 - \pi^2 X^2 \Bigr)
 \PlusBreak{ {1 \over 4 y^3} \bigglB }
   {\pi^2 \over 12} (x - 1) (x^2 + 9 x - 2) Y^2
  -4 (6 x + 1) \zeta_3 Y
 \MinusBreak{ {1 \over 4 y^3} \bigglB }
   (9 x^2 - 6 x + 5) \zeta_3 X
  -{\pi^4 \over 240} (12 x^3 - 53 x^2 + 194 x - 21) 
\biggr]
 \MinusBreak{}
 {y - 1 \over 4 y}  \biggl(
  {1 \over 2} (X - Y)^2 \li2(-x)
+ {3 \over 4} X^2 Y^2
     \biggr)
 \MinusBreak{}
 {1 \over y} \biggl(
 {1 \over 12} (3 x + 5) X^3 Y
+{\pi^2 \over 6} X Y
       \biggr)
+{1 \over 2} (X - Y) \li3(-y)
%
 \MinusBreak{}
 {1 \over 6 y^2}  
\biggl[
   (9 x^2 + 50 x + 11) \biggl( \li3(-x) + \li3(-y) - X \li2(-x) 
                      +  Y \li2(-x) \biggr)
 \MinusBreak{\null - {1 \over 6 y^2} \bigglB }
   {1 \over 6} (16 x^2 + 71 x + 25) X^3
  +{1 \over 2} (24 x^2 + 95 x + 41) X Y^2
 \MinusBreak{ \null - {1 \over 6 y^2} \bigglB }
   {\pi^2 \over 24} (176 x^2 + 683 x + 267) Y
 \MinusBreak{\null -  {1 \over 6 y^2} \bigglB }
   {5 \over 48} \pi^2 (40 x^2 + 203 x + 67) X
  -{117 \over 8} y \zeta_3
\biggr]
 \PlusBreak{}
  {y - 1 \over 3 y}  \biggl(
       X^2 Y + {13 \over 6} Y^3
           \biggr)
%
+{\pi^2 \over 288 y^2} (360 x^3 + 1073 x^2 + 1001 x + 528) 
 \MinusBreak{}
 {1 \over 288 y}   \biggl(
    x {} (360 x + 713)   (X - Y)^2 
   -368 X^2  +  560 X Y  - 1160 Y^2 
         \biggr)
%
 \PlusBreak{}
 {1 \over y}   \biggl(
   {1 \over 54} (135 x + 2) Y
  -{1 \over 96} (240 x - 167) X
    \biggr)
%
+{30593 x + 30377 \over 3456 y^2}
%
%
 \PlusBreak{}
  i \pi {} \biggl[
    {11 (y - 1) \over 12 y} \Bigl((X - Y)^2  +  \pi^2\Bigr)
  -{9 \over 4 y} \zeta_3     
  +{11 \pi^2 \over 288 y}    
%
  \PlusBreak{\null + i \pi \bigglB}
   {11 \over 18 y} (X + 10 Y)
%
  + {1535 \over 864 y}  
\biggr]
\,, \label{AA42}\\[1pt plus 4pt]
%
B_4^{[2]} & = & 
{1 \over 2 y^3}   
\bigg[
    (3 x^3 - 12 x - 4) \li4(-x)
  -(x^2 + 22 x + 14) x \li4(-y)
 \MinusBreak{ {1 \over 2 y^3} \bigglB }
   (7 x^3 - x^2 + 4 x + 5) \li4 \biggl( -{x \over y} \biggr)
  - (7 x^3 + x^2 - 4 x + 1) X \li3(-x)
 \PlusBreak{ {1 \over 2 y^3} \bigglB }
   (6 x^3 + 3 x^2 + 9 x + 5) Y \li3(-x)
  + {\pi^2 \over 6} (23 x^3 + 39 x^2 + 24 x + 13) \li2(-x)
 \PlusBreak{ {1 \over 2 y^3} \bigglB }
   {x \over 24} (x^2 + 4 x + 2) X^4
  - {1 \over 24} (3 x+2) (x^2 - 4 x + 2) Y^4
 \PlusBreak{ {1 \over 2 y^3} \bigglB }
    {1 \over 6} (8 x^3 + 10 x^2 + 23 x + 14) X Y^3
   - {\pi^2 \over 12} (x^3 - 17 x^2 - 4 x + 12) X^2  
 \MinusBreak{ {1 \over 2 y^3} \bigglB }
   {\pi^2 \over 12} (15 x^3 + 27 x^2 + 36 x + 17) Y^2  
  -(6 x^3 + 3 x^2 + 9 x + 5) \zeta_3 Y
 \PlusBreak{ {1 \over 2 y^3} \bigglB }
   {1 \over 2} (12 x^3 + 7 x^2 + 8 x + 11) \zeta_3 X
  - {\pi^4 \over 1440} (592 x^3 - 551 x^2 - 1570 x - 343) 
\biggr]
 \PlusBreak{}
 {1 \over 2 y^2}   
\biggl[
   (5 x^2 - 2 x + 5) X \li3(-y)
  -{1 \over 2} (4 x^2 - x + 1) X^2 \li2(-x)
 \PlusBreak{\null +  {1 \over 2 y^2} \bigglB }
   (4 x^2 + 3 x + 5) X Y \li2(-x)
  +{1 \over 6} (5 x - 1) X^3 Y
 \PlusBreak{ \null + {1 \over 2 y^2} \bigglB }
   {1 \over 4} (10 x^2 - 3 x + 11) X^2 Y^2
  -{\pi^2 \over 2} (7 x^2 + x + 4) X Y
\biggr]
 \PlusBreak{}
  {1 \over 2 y}  \biggl(
   (4 x + 5) Y \li3(-y)
   +{5 \over 2} Y^2 \li2(-x)
\biggr)
%
 \PlusBreak{}
 {1 \over 4 y^3}   
\biggl[
   -x {} (6 x^2 + 1) (\li3(-x) - X \li2(-x))
 \MinusBreak{\null + {1 \over 4 y^3} \bigglB }
    {x \over 18} (14 x^2 + 40 x + 29) X^3
   +{x \over 2} (4 x^2 - 3) X^2 Y
 \MinusBreak{\null +  {1 \over 4 y^3} \bigglB }
    {\pi^2 \over 72} (176 x^3 + 853 x^2 + 782 x + 477) X
 \MinusBreak{\null + {1 \over 4 y^3} \bigglB }
    {1 \over 36} (1235 x^2 + 1426 x + 443) \zeta_3
\biggr]
 \PlusBreak{}
  {1 \over 2 y^2}   
\biggl[
  -(3 x + 10) x {} (\li3(-y) + \li2(-x) Y)
 \MinusBreak{ {1 \over 2 y^2} \bigglB }
   {1 \over 6} (33 x^2 + 74 x + 20) X Y^2
  +{\pi^2 \over 9} (10 x^2 + 28 x + 15) Y
\biggr]
%
 \MinusBreak{}
 {1 \over 144 y^3} (72 x^4 + 145 x^3 - 304 x^2 - 422 x - 144) X^2
 \MinusBreak{}
  {28 x + 15 \over 18 y} Y^3
 -{1 \over 8 y^2} (8 x^3 - 3 x^2 - 39 x - 16) X Y
 \PlusBreak{}
 {\pi^2 \over 288 y^2} (144 x^3 + 146 x^2 - 83 x + 167) 
 + {1 \over 216 y^2} (216 x^2 + 917 x + 782) X
 \MinusBreak{}
 {1 \over 144 x y} (72 x^3 - 199 x^2 - 290 x + 27) Y^2
%
 +{24 x - 71 \over 24 y} Y
%
-{18815 x + 19139 \over 2592 y^2}
%
 \PlusBreak{}
  i \pi {}
\biggl[
   {1 \over y^3}   \biggl( 
     -{1 \over 2} (x^3 - 2 x^2 - 13 x - 4) \li3(-x)
     -3 \zeta_3 x
          \biggr)
 \PlusBreak{\null +  i \pi \bigglB }
   {1 \over 2 y^2}   \biggl(    
     (x - 11) x \li3(-y)
   +2 (2 x - 1) x Y \li2(-x)
   +x X^3 
 \PlusBreak{\null + i \pi \bigglB {1 \over 2 y^2} \bigglP }
    (3 x^2 - 2 x + 1) X Y^2
   +{\pi^2 \over 2} (3 x + 2) X
   - {\pi^2 \over 6} (3 x^2 - 11 x - 8) Y
             \biggr)
 \PlusBreak{\null + i \pi \bigglB }
   {1 \over y} \biggl(
   -2 X \li2(-x)
   -{x\over 4} X^2 Y          
   +{1\over 2} x X Y^2        
   +{x \over 12} Y^3
   +{\pi^2\over 2} x {} (X - Y)  
   +{1\over 4} \zeta_3        
         \biggr)
%
 \PlusBreak{\null + i \pi \bigglB }
    {1 \over 4 y^3}   \biggl(
    (12 x^2 + 26 x + 21) x \li2(-x)
   -{x \over 6} (20 x^2 + 40 x + 41) X^2
 \PlusBreak{\null + i \pi \bigglB  {1 \over 4 y^3} \bigglP }
    {1 \over 3} (60 x^3 + 136 x^2 + 119 x + 40) X Y
 \MinusBreak{\null + i \pi \bigglB  {1 \over 4 y^3} \bigglP }
    {\pi^2 \over 24} (112 x^3 + 79 + 258 x + 263 x^2)        
           \biggr)
 \MinusBreak{\null + i \pi \bigglB }
    {16 x + 5 \over 6 y} Y^2
%
   -{x \over 72 y^3} (100 x^2 + 74 x + 73) X
 \MinusBreak{\null + i \pi \bigglB }
    {1 \over 72 x y^2} (100 x^3 + 138 x^2 + 119 x - 27) Y
%
   + {1\over 216 y^2} (1340 x + 1421) 
\biggr]
\,, \hskip 1.8 cm \label{BB42}\\[1pt plus 4pt]
%
C_4^{[2]} & = &
{1 \over 2 y^3}   
\biggl[
    {x \over 2} (5 x^2 + 14 x + 7) \biggl( 
          \li4 \biggl( -{x \over y} \biggr)
         -{\pi^2 \over 6} \li2(-x) - {1 \over 6} X Y^3 \biggr)
 \MinusBreak{ {1 \over 2 y^3} \bigglB }
   (x^2 + 4 x + 2) x {} ( \li4(-y) + \li4(-x) + Y{} \li3(-x) )
 \PlusBreak{ {1 \over 2 y^3} \bigglB }
   (2 x^2 + 6 x + 3) x X \li3(-x)
  -{x \over 24} (2 x^2 + 2 x + 1) X^4
 \PlusBreak{ {1 \over 2 y^3} \bigglB }
   {x \over 24} (x^2 + 4 x + 2) Y^4
   +{\pi^2 \over 24} x {} (3 x^2 + 10 x + 5) Y^2
 \PlusBreak{ {1 \over 2 y^3} \bigglB }
   {\pi^2 \over 12} x {} (x^2 + 6 x + 3) X^2
  +(x^2 + 5 x + 3) \zeta_3 X
 \PlusBreak{ {1 \over 2 y^3} \bigglB }
   (x - y) x \zeta_3 Y
  +{\pi^4 \over 360} (41 x^3 + 113 x^2 + 40 x - 11) 
\biggr]
 \PlusBreak{}
  {x \over 2 y} 
\biggl[
  (2 X - Y) \li3(-y)
  -{1 \over 4} (X - Y)^2 \li2(-x)
 \PlusBreak{\null + {x \over 2 y} \bigglB }
   {1 \over 3} X^3 Y
  +{3 \over 8} X^2 Y^2
  -{\pi^2 \over 2} X Y
\biggr]
%
 \PlusBreak{}
  {1 \over y^3}   
\biggl[
  -{x \over 4} (6 x^2 + 12 x + 7) (\li3( - x) - X \li2(-x))
 \PlusBreak{\null + {1 \over y^3} \bigglB } 
   {x \over 24} (6 x^2 + 8 x + 1) X^3
  -{1 \over 8} (11 x^2 + 24 x + 15) \zeta_3
 \PlusBreak{\null + {1 \over y^3} \bigglB } 
   {x \over 8} (2 x + 3) X^2 Y
  +{\pi^2 \over 24} (12 x^3 + 19 x^2 + 5 x - 3) X
\biggr]
 \PlusBreak{}
  {x {} (3 x + 2) \over 2 y^2}   
\biggl(
    \li3( - y)
  +Y \li2(-x)
  -{\pi^2 \over 3} Y
  +{1 \over 2} X Y^2
\biggr)
%
 \PlusBreak{}
 {x \over 32 y^3} (8 x^3 + 31 x^2 + 42 x + 21) X^2
-{x {} (8 x + 15) \over 16 y} X Y
 \MinusBreak{}
 {\pi^2 \over 96 y^2} (24 x^3 + 69 x^2 + 86 x + 29) 
 + {1 \over 32 x y} (8 x^3 + 15 x^2 - 12 x - 6) Y^2
%
 \MinusBreak{}
 {1 \over 32 y^2} (16 x^2 + 31 x + 3) X
-{4 x - 3 \over 8 y} Y
%
-{247 x + 255 \over 128 y^2}
%
 \PlusBreak{}
  i \pi {} \biggl[
   {x \over 2 y}  \biggl(
      \li3(-x) + \li3(-y)
   -{1 \over 6} X^3
   +{1 \over 2} X^2 Y
   -{\pi^2 \over 6} X
        \biggr)
  +{3\over 2 y} \, \zeta_3 
%
 \PlusBreak{\null + i \pi  \bigglB }
 {x {} (2 x + 3) \over 24 y^3} (6 \li2(-x)  +  6 X Y -3 X^2 -\pi^2)
  - {\pi^2\over 8 y}   
%
 \PlusBreak{\null + i \pi  \bigglB }
 {x {} (2 x + 3) \over 8 y^3} X
-{3 (x - y) \over 8 x y} Y
%
- {3 \over 32 y^2} (9 x + 5)   
\biggr]
\,, \label{CC42}\\[1pt plus 4pt]
%
D_4^{[2]} & = &
{x {} (x - 1) \over y^3}   
\biggl[
   \li4(-x)
  +\li4(-y)
  -\li4 \biggl( -{x \over y} \biggr)
  + X {} (-\li3(-x) + \zeta_3)
 \PlusBreak{ {x (x - 1) \over y^3} \bigglB }
   Y {} (\li3(-x) - \zeta_3)
  +{\pi^2 \over 6} \li2(-x)
  -{1 \over 24} X^4
  +{1 \over 6} X Y^3
 \MinusBreak{ {x (x - 1) \over y^3} \bigglB }
   {1 \over 24} Y^4
  -{\pi^2 \over 6} X^2
  -{\pi^2 \over 12} Y^2
  -{7 \over 120} \pi^4
\biggr]
%
 \PlusBreak{}
 {1 \over 3 y^2}   
\biggl[
   (7 x + 1) ( \li3( - x) + \li3(-y) + (Y-X) \li2(-x) )
 \MinusBreak{\null + {1 \over 3 y^2} \bigglB } 
   {1 \over 24} (5 x^2 + 43 x + 14) X^3
  +{1 \over 8} (3 x^2 + 37 x + 10) X Y^2
 \MinusBreak{\null + {1 \over 3 y^2} \bigglB } 
   {\pi^2 \over 24} (7 x^2 + 76 x + 21) Y
  -{\pi^2 \over 48} (10 x^2 + 143 x + 37) X
\biggr]
 \PlusBreak{}
  {1 \over 8 y}  \biggl( 
   {1 \over 3} (y - 1) X^2 Y
  -{7 \over 9} (y - 1) Y^3
  -3 \zeta_3 \biggr)
%
 \MinusBreak{}
  {\pi^2 \over 216 y^2} (108 x^3 + 321 x^2 + 332 x + 191) 
 \PlusBreak{}
  {1 \over 72 y}  \biggl(
    x {} (36 x + 71) (X - Y)^2
    -35 X^2  + 18 X Y - 71 Y^2
          \biggr)
%
 \PlusBreak{}
 {216 x + 53 \over 216 y}  X
-{9 x - 7 \over 9 y} Y
%
+{863 \over 864 y}
%
%
 \PlusBreak{}
 i \pi {}
\biggl[
   {1 - y \over 6 y} \Bigl((Y - X)^2 + \pi^2\Bigr)
   - {\pi^2 \over 144 y}   
%
  -{13 \over 18 y} X
  -{31 \over 18 y} Y
%
    + {221 \over 216 y}  
\biggr]
\,, \label{DD42}\\[1pt plus 4pt]
%
E_4^{[2]} & = & 
{x {} (x - 1) \over y^3}   
\biggl[
  -2 \biggl( \li4(-x)
  - \li4 \biggl( -{x \over y} \biggr)
  + \li4(-y) \biggr)
  -{\pi^2 \over 3} \li2(-x) 
 \PlusBreak{ {x (x - 1) \over y^3} \bigglB }
   2 (X - Y) (\li3(-x) - \zeta_3)
  +{1 \over 12} (X^4  +  Y^4)
 \MinusBreak{ {x (x - 1) \over y^3} \bigglB }
   {1 \over 3} X Y^3
  +{\pi^2 \over 6} (Y^2 + 2 X^2)
  +{7 \over 60} \pi^4
\biggr]
%
 \PlusBreak{}
  {1 \over 3 y^3}   
\biggl[
   (x^3 + 15 x^2 + 15 x + 2) (\li3(-x) - X \li2(-x))
 \PlusBreak{ {1 \over 3 y^3} \bigglB }
   {\pi^2 \over 48} (42 x^3 - 109 x^2 - 116 x + 27) X
 \PlusBreak{ {1 \over 3 y^3} \bigglB }
   {x \over 24} (13 x^2 - 14 x - 19) X^3
  +{1 \over 24} (x^2 + 2 x - 23) \zeta_3
\biggr]
 \PlusBreak{}
 {1 \over 3 y^2}  
\biggl[ 
  -(x^2 + 14 x + 1) (\li3( - y) + Y \li2(-x))
 \PlusBreak{\null + {1 \over 3 y^2} \bigglB }
   {1 \over 8} (11 x^2 + 16 x + 8) X^2 Y
  -{1 \over 8} (5 x^2 + 57 x + 4) X Y^2
 \PlusBreak{\null + {1 \over 3 y^2} \bigglB }
   {\pi^2 \over 24} (13 x^2 + 111 x + 2) Y
\biggr]
 \PlusBreak{}
 {13 x + 6 \over 72 y} Y^3
%
-{x \over 72 y^3} (54 x^3 + 235 x^2 + 338 x + 139) X^2
 \MinusBreak{}
 {x \over 36 y^2} (54 x^2 + 191 x + 125) X Y 
-{1 \over 72 y} (54 x^2 + 147 x + 2) Y^2
 \PlusBreak{}
 {\pi^2 \over 144 y^2} (108 x^3 + 362 x^2 + 117 x - 33)
%
+{3 x + 2 \over 2 y} Y
 \PlusBreak{}
{1 \over 216 y^2} (324 x^2 - 47 x - 191) X
%
-{4085 \over 2592 y}
%
%
 \PlusBreak{}
 i \pi {}
\biggl[
  {1 \over 3 y^3}   \biggl(
    -\li2(-x)
    +{x \over 8} (6 x^2 + 19 x + 17) X^2
 \PlusBreak{\null + i \pi \bigglB {1 \over 3 y^3} \bigglP }
     {\pi^2 \over 48} (48 x^3 + 135 x^2 + 126 x + 47)
         \biggr)  + {7 x + 3 \over 12 y} Y^2
 \PlusBreak{\null + i \pi \bigglB }
   {1 \over 12 y^2} (10 x^2 + 15 x + 8) X Y
%
  + {x \over 18 y^3} (5 x^2 - 11 x - 7) X
 \PlusBreak{\null + i \pi \bigglB }
   {1 \over 18 y^2} (5 x^2 + 12 x + 1) Y
%
  - {587 x + 407  \over 216 y^2 }    
\biggr]
\,, \label{EE42}\\[1pt plus 4pt]
%
F_4^{[2]} & = &
{1 \over y}   \biggl(  
{1 \over 24} (X^2 + Y^2) + {1 \over 36} X Y  
- {5 \over 54} (X + Y) - {23 \over 216} \pi^2 
\biggr)
 \PlusBreak{}
  i {\pi \over 9 y} \biggl( (X + Y) - {5\over 3 } \biggr)  %
\,, \label{FF42}\\[1pt plus 4pt]
%
G_4^{[2]} & = & 
- G_4^{[1]} 
\,. \label{GG42}
\end{eqnarray}

\noindent
For $h=4$ in \eqn{hel4}
and color factor $\trc^{[3]}$ in \eqn{TraceBasisqggq}:
\begin{eqnarray}
H_4^{[3]} & = & 
{1 \over 2 y^3}  
\biggl[
  -{1 \over 2} (16 x^3 + 47 x^2 - 23 x - 10) \li4(-x)
 \MinusBreak{ {1 \over 2 y^3} \bigglB }
   (13 x^3 + 31 x^2 + 56 x + 5) \li4 \biggl( -{x \over y} \biggr)
  - (8 x^3 + 32 x^2 + 7 x + 16) \li4(-y)
 \MinusBreak{ {1 \over 2 y^3} \bigglB }
   (10 x^3 + 25 x^2 + 20 x + 1) X \li3(-y)
  - (x^3 - x^2 + 28 x + 2) X \li3(-x)
 \PlusBreak{ {1 \over 2 y^3} \bigglB }
   (8 x^3 + 15 x^2 + 39 x + 1) Y \li3(-x)
  - (5 x^3 + 9 x^2 + 3 x - 3) X Y \li2(-x)
 \MinusBreak{ {1 \over 2 y^3} \bigglB }
   {1 \over 4} (2 x + 3) (3 x^2 + 5 x + 4) X^2 \li2(-x)
  - {x \over 24} (4 x^2 - 11 x + 11) X^4
 \PlusBreak{ {1 \over 2 y^3} \bigglB }
   {\pi^2 \over 6} (25 x^3 + 53 x^2 + 70 x + 11) \li2(-x)
  \MinusBreak{ {1 \over 2 y^3} \bigglB }
   {1 \over 4} (13 x^3 + 21 x^2 + 3 x - 13) X^2 Y^2
  - {1 \over 6} (9 x^3 + 38 x^2 + 49 x + 22) X^3 Y
 \PlusBreak{ {1 \over 2 y^3} \bigglB }
   {1 \over 6} (34 x^3 + 86 x^2 + 103 x + 18) X Y^3
  - {1 \over 24} (10 x^3 + 22 x^2 + 47 x + 2) Y^4
 \PlusBreak{ {1 \over 2 y^3} \bigglB }
   {\pi^2 \over 12} (x + 3) (14 x^2 - 13 x + 4) X^2
  + {\pi^2 \over 3} (4 x^3 + 8 x^2 + 4 x - 5) X Y
  \MinusBreak{ {1 \over 2 y^3} \bigglB }
   {\pi^2 \over 12} (7 x^3 + 9 x^2 + 30 x - 5) Y^2
  - (9 x^3 + 18 x^2 + 42 x + 2) \zeta_3 Y
  \PlusBreak{ {1 \over 2 y^3} \bigglB }
   (9 x^3 + 14 x^2 + 34 x + 1) \zeta_3 X
 +  {\pi^4 \over 360} (10 x^3 + 334 x^2 - 121 x + 184)
\biggr]
 \PlusBreak{}
  {1 \over 4 y} \biggl(
    2 (15 x + 13) Y \li3(-y)
  + (9 x + 5) Y^2 \li2(-x) 
  +2  (1-y) (X^2 - Y^2) \pi^2
           \biggr)
%
 \PlusBreak{}
 {1 \over 12 y^3}
\biggl[
    (53 x^3 + 352 x^2 + 392 x + 124) ( \li3(-x) - X \li2(-x) )
 \MinusBreak{\null -  {1 \over 12 y^3} \bigglB }
    (19 x^3 + 43 x^2 + 47 x + 19) X^2 Y
 \PlusBreak{\null - {1 \over 12 y^3} \bigglB }
    {x \over 3} (68 x^2 - 7 x - 35) X^3
  - (52 x^2 + 83 x + 62) \zeta_3
 \MinusBreak{\null - {1 \over 12 y^3} \bigglB }
   {\pi^2 \over 12} (585 x^3 + 2038 x^2 + 2147 x + 780) X
\biggr]
 \PlusBreak{}
 {1 \over 12 y^2}  
\biggl[
  - (27 x^2 + 263 x + 62) ( \li3(-y) + Y \li2(-x) )
 \MinusBreak{\null -  {1 \over 12 y^2} \bigglB }
    (4 x^2 + 132 x + 41) X Y^2
   - {\pi^2 \over 12} (309 x^2 - 646 x - 307) Y
\biggr]
 \PlusBreak{}
 {11 \over 18} Y^3
%
- {x \over 72 y^3} (108 x^3 + 949 x^2 + 1193 x + 487) X^2
 \MinusBreak{}
 {1 \over 12 y^2} (36 x^3 + 185 x^2 + 108 x + 10) X Y
 \PlusBreak{}
 {\pi^2 \over 72 y^2} (108 x^3 + 450 x^2 + 363 x + 250)
+{1 \over 108 y} (211 x + 76) Y
 \MinusBreak{}
 {1 \over 72 y} (108 x^2 + 161 x - 208) Y^2
%
+ {x \over 108 y^2} (211 x + 121) X
%
\PlusBreak{}
i \pi {}
\biggl[
   {1 \over 2 y^3} \biggl(
      (7 x^3 + 16 x^2 + 11 x - 1) \li3(-x)
 \PlusBreak{\null + i \pi \bigglB {1 \over 2 y^3} \bigglP}
      (5 x^3 + 18 x^2 + 21 x + 12) \li3(-y)
 \MinusBreak{\null + i \pi \bigglB {1 \over 2 y^3} \bigglP}
      {1 \over 2} (16 x^3 + 37 x^2 + 29 x + 6) X \li2(-x)
 \MinusBreak{\null + i \pi \bigglB {1 \over 2 y^3} \bigglP}
      2 (y - 1) (2 x^2 + 3 x + 2) Y \li2(-x)
 \MinusBreak{\null + i \pi \bigglB {1 \over 2 y^3} \bigglP}
      {\pi^2 \over 3} (8 x^3 + 28 x^2 + 32 x + 13) Y        
 \PlusBreak{\null + i \pi \bigglB {1 \over 2 y^3} \bigglP}
      {1 \over 2} (9 x^3 + 36 x^2 + 45 x + 22) X Y^2        
 \MinusBreak{\null + i \pi \bigglB {1 \over 2 y^3} \bigglP}
      {1 \over 6} (6 x^3 + 9 x^2 + 9 x + 8) X^3
     -(4 x^2 + 8 x + 1) \zeta_3
        \biggr)
 \MinusBreak{+\null  i \pi \bigglB }
   {\pi^2 \over 4 y^2} (2 x^2 + 3 x + 2) X    
  -{1 \over y} \biggl(
      {1 \over 4} (7 x + 11) X^2 Y
     +{1 \over 6} Y^3
         \biggr)
%
 \MinusBreak{\null + i \pi \bigglB}
    {1 \over 12 y^3}  \biggl(
      (26 x^3 + 62 x^2 + 67 x + 62) \li2(-x)
 \MinusBreak{\null +  i \pi \bigglB {1 \over 12 y^3}  \bigglP}
      {1 \over 2} (151 x^3 + 252 x^2 + 228 x + 86) X^2
 \PlusBreak{ \null + i \pi \bigglB {1 \over 12 y^3}  \bigglP}
      (57 x^3 + 104 x^2 + 73 x + 18) X Y
 \MinusBreak{\null + i \pi \bigglB {1 \over 12 y^3}  \bigglP}
      {\pi^2 \over 12} (44 x^3 + 17 x^2 - 80 x + 9)
            \biggr)
  -{21 x + 8 \over 8 y} Y^2
%
 \MinusBreak{\null +  i \pi \bigglB }
   {1 \over 36 y^3} (286 x^3 + 314 x^2 + 133 x - 30) X
 \MinusBreak{\null +  i \pi \bigglB }
   {1 \over 36 y^2} (286 x^2 + 371 x + 238) Y
%
  -{83 x + 38 \over 54 y^2}
\biggr]
\,. \hskip 5 cm \label{HH43}\\[1pt plus 4pt]
%
I_4^{[3]} & = & 
{1 \over 2 y^3}   
\biggl[
  -3 (x - y) x \li4(-y)
  -{1 \over 2} (8 x^3 - 9 x^2 - 3 x - 2) \li4(-x)
 \MinusBreak{ {1 \over 2 y^3} \bigglB }
   (5 x^3 + 5 x^2 + 4 x + 1) \li4 \biggl( -{x \over y} \biggr)
  - (4 x^3 + x^2 + 2 x + 1) X \li3(-y)
 \MinusBreak{ {1 \over 2 y^3} \bigglB }
   (x^3 + 5 x^2 + 4 x + 2) X \li3(-x)
  + (2 x^3 + 6 x^2 + 3 x + 1) \zeta_3 X
 \PlusBreak{ {1 \over 2 y^3} \bigglB }
   (x - y) (1 - x y) Y{} (\li3(-x) - \zeta_3)
  + {1 \over 4} (2 x^3 + x^2 + 5 x + 4) X^2 \li2(-x)
 \MinusBreak{ {1 \over 2 y^3} \bigglB }
   (x^3 - x^2 + x + 1) X Y \li2(-x)
 \PlusBreak{ {1 \over 2 y^3} \bigglB }
   {\pi^2 \over 6} (x^3 - x^2 - 2 x - 1) \li2(-x)
  +{1 \over 24} (3 x - 1) (x - y) x X^4
 \MinusBreak{ {1 \over 2 y^3} \bigglB }
   {1 \over 24} (6 x^3 + 8 x^2 + 7 x + 2) Y^4
  -{1 \over 3} (x^3 + 3 x^2 - 1) X^3 Y
 \MinusBreak{ {1 \over 2 y^3} \bigglB }
   {1 \over 4} (4 x^3 - 5 x^2 + 2 x + 3) X^2 Y^2
  + {1 \over 6} (7 x^3 + 12 x^2 + 12 x + 4) X Y^3
 \MinusBreak{ {1 \over 2 y^3} \bigglB }
   {\pi^2 \over 12} (9 x^2 + 8 x + 1) x X^2
  +{\pi^2 \over 3} (4 x^2 - 2 x - 1) x X Y
 \MinusBreak{ {1 \over 2 y^3} \bigglB }
   {\pi^2 \over 12} (4 x^3 + 5 x^2 + 7 x + 3) Y^2
  + {\pi^4 \over 360} (62 x^2 + 140 x + 79) x 
\biggr]
 \PlusBreak{}
{3 x + 1 \over 2 y} Y \li3(-y)
-{1 \over 4} Y^2 \li2(-x)
%
 \PlusBreak{}
 {x \over 4 y^3}   
\biggl[
    3 (5 x^2 + 4 x + 2) (\li3(-x) - X \li2(-x))
 \MinusBreak{\null + {x \over 4 y^3} \bigglB }
    {2 \over 3} (8 x^2 + 7 x + 2) X^3
   +{1 \over 3} (6 x^2 + 11 x + 1) X^2 Y
 \MinusBreak{\null + {x \over 4 y^3} \bigglB }
    {\pi^2 \over 6} (22 x^2+28 x+11) X
   +(2 x - 7) \zeta_3
\biggr]
 \MinusBreak{}
 {x \over 4 y^2}   
\biggl[
    (9 x + 7) (\li3(-y) + Y \li2(-x))
 \PlusBreak{\null + {x \over 4 y^2} \bigglB }
    {1 \over 3} (13 x + 10) X Y^2
   -2 (x-y) \pi^2 Y
\biggr]
%
 \MinusBreak{}
 {x \over 24 y^3} (6 x^3 - 3 x^2 - 49 x - 37) X^2
 - {1 \over 12 y^2} (6 x^3 - 15 x^2 - 38 x - 24) X Y
 \PlusBreak{}
 {\pi^2 \over 8 y^2} (2 x^3 - 9 x - 8) 
-{1 \over 8 x y} (2 x^3 - 9 x^2 - 12 x - 2) Y^2
%
 \MinusBreak{}
 {x \over 12 y^2} (47 x + 41) X
-{47 x + 46 \over 12 y} Y
%
 \PlusBreak{}
 i \pi {}
\biggl[
   {1 \over 2 y^3}  \biggl(
      (3 + 6 x - x^2) x \li3(-y)
     +(x^3 - 2 x^2 - x - 1) \li3(-x)
 \PlusBreak{\null + i \pi \bigglB {1 \over 2 y^3}  \bigglP }
      2 (x - y) x Y \li2(-x)
     +{1 \over 2} (3 x^2 + 3 x + 2) X \li2(-x)
 \PlusBreak{\null + i \pi \bigglB {1 \over 2 y^3}  \bigglP }
      {x \over 6} (x - 1) (3 x + 2) X^3
     +{x \over 2} (x^2 + 10 x + 5) X Y^2
 \MinusBreak{\null + i \pi \bigglB {1 \over 2 y^3}  \bigglP }
      {\pi^2 \over 6} (3 x^2 + 10 x + 5) x Y
     +3 \zeta_3 x^2
              \biggr)
 \MinusBreak{\null + i \pi \bigglB }
   {1 \over 4 y} \biggl(
      x Y^3
     +(x - 1) X^2 Y
     +\pi^2 x X
          \biggr)
%
 \PlusBreak{\null + i \pi \bigglB }
  {x \over 4 y^3} \biggl(
    -(6 x^2 - 4 x - 1) \li2(-x)
    -{1 \over 6} (39 x^2 + 26 x + 4) X^2
 \PlusBreak{\null +  i \pi \bigglB {x \over 4 y^3} \bigglP }
     {1 \over 3} (11 x^2 + 20 x + 1) X Y
    +{\pi^2 \over 6} (2 x^2 - 8 x - 1) 
        \biggr)
 \MinusBreak{\null +  i \pi \bigglB }
   {x \over 24 y} Y^2
%
  -{1 \over 12 y^3} (6 x^3 + 4 x^2 + 25 x + 24) X
 \MinusBreak{\null + i \pi \bigglB }
   {1 \over 12 x y^2} (6 x^3 + 25 x^2 + 18 x + 6) Y
%
  + {26 x + 23 \over 6 y^2}
\biggr]
\,. \label{II43}\\[1pt plus 4pt]
%
J_4^{[3]} & = &
{x{}(x - 1) \over y^3}
\biggl[
      \li4(-x) + \li4(-y) - \li4 \biggl( -{x \over y} \biggr)
  -(\li3(-x) - \zeta_3) X
 \PlusBreak{ {x (x - 1) \over y^3} \bigglB }
   (\li3(-x) - \zeta_3) Y
  + {\pi^2 \over 6} \li2(-x)
  -{1 \over 24} X^4
  +{1 \over 6} X Y^3
 \MinusBreak{ {x (x - 1) \over y^3} \bigglB }
   {1 \over 24} Y^4
  -{\pi^2 \over 6} X^2
  -{\pi^2 \over 12} Y^2
  -{7 \pi^4 \over 120}
\biggr]
%
 \PlusBreak{}
 {1 \over 72 y^3} \biggl(
    x{} (34 + 11 x - 20 x^2) X^3
   +3 (3 x + 2) (x^2 - x - 3) X^2 Y
      \biggr)
 \PlusBreak{}
  {1 \over 3 y^2}   \biggl[
    (x^2 + 9 x + 2) ( \li3(-x) - X \li2(-x) )
 \MinusBreak{\null +  {1 \over 3 y^2} \bigglP }
    (x^2 - 6 x - 1) ( \li3(-y) + Y \li2(-x) )
 \MinusBreak{\null + {1 \over 3 y^2} \bigglP }
    {1 \over 8} (x - 2) (9 x + 1) X Y^2
   +{\pi^2 \over 24} (41 x^2 - 22 x - 15) Y
    \biggr]
 \PlusBreak{}
  {\pi^2 \over 36 y} (20 x + 33) X
+ {1 \over 3 y} \zeta_3
- {1 \over 9} Y^3
%
 + {x \over 12 y^3} (6 x^3+44 x^2+66 x+25) X^2
 \PlusBreak{}
  {1 \over 72 y^2}   \biggl[
        (72 x^3 + 303 x^2 + 259 x + 22) X Y
    -    (36 x^3 + 126 x^2 + 139 x + 73) \pi^2 
             \biggr]
 \PlusBreak{}
  {1 \over 24 x y} (12 x^3 + 13 x^2 - 5 x + 6) Y^2
%
 -{x \over 72 y^2} (19 x - 17) X
-{19 x - 20 \over 72 y} Y
%
%
 \PlusBreak{}
 i \pi {}
\biggl[
   {1 \over 24 y^3}   \biggl(
      2 (8 x^3 + 11 x^2 - 2 x - 4) X Y
      -(21 x^3 + 30 x^2 + 21 x + 14) X^2
               \biggr)
 \PlusBreak{\null + i \pi \bigglB }
     {1 \over 3 y}   \biggl(
     {1 \over 8} (13 x + 10) Y^2
   + (x - y) \li2(-x)
   - {\pi^2 \over 24} (x - 25) 
          \biggr)
%
 \PlusBreak{\null + i \pi \bigglB }
   {1 \over 72 y^3} (153 x^3 + 230 x^2 + 19 x - 22) X
 \PlusBreak{\null + i \pi \bigglB }
   {1 \over 72 x y^2} (153 x^3 + 211 x^2 + 16 x - 36) Y
%
  +{4 x - 5 \over 18 y^2} 
\biggr]
\,, \label{JJ43}\\[1pt plus 4pt]
%
K_4^{[3]} & = & 
{x{}(x - 1) \over 6 y^3}   
\biggl[
    12 \biggl( \li4(-x) + \li4(-y) - \li4 \biggl( -{x \over y}
\biggr) \biggr)
 \PlusBreak{ {x (x - 1) \over 6 y^3} \bigglB }
    12 (Y - X) (\li3(-x) - \zeta_3)
   +2 \pi^2 \li2(-x)
   -{1 \over 2} X^4
 \PlusBreak{ {x (x - 1) \over 6 y^3} \bigglB }
    2 X Y^3
   -{1 \over 2} Y^4
   -\pi^2 (2 X^2 + Y^2)
   -{7 \pi^4 \over 10} 
\biggr]
%
 \PlusBreak{}
 {1 \over 3 y^3} 
\biggl[
   -(x^3 + 15 x^2 + 15 x + 2) (\li3(-x) - X \li2(-x))
 \MinusBreak{\null - {1 \over 3 y^3} \bigglB } 
    {x \over 24} (2 x^2 - 30 x - 27) X^3
   +{1 \over 8} (5 x^3 + 20 x^2 + 22 x + 8) X^2 Y
 \MinusBreak{\null - {1 \over 3 y^3} \bigglB } 
    {\pi^2 \over 24} x{} (x^2 - 90 x - 87) X
   +\zeta_3
\biggr]
 \PlusBreak{}
  {1 \over 3 y^2}   
\biggl[
     (x^2 + 14 x + 1) Y {} \biggl( \li2(-x) - {\pi^2 \over 3} \biggr)
 \PlusBreak{\null + {1 \over 3 y^2} \bigglB }
     {1 \over 8} (7 x^2 + 63 x + 8) X Y^2
    +(x^2 + 14 x + 1) \li3(-y)
\biggr]
%
 \PlusBreak{}
 {x \over 8 y^3} (6 x^3 + 23 x^2 + 32 x + 13) X^2
 \PlusBreak{}
  {x \over y^2}   \biggl(
    {1 \over 8} (12 x^2 + 35 x + 21) X Y
   -{\pi^2 \over 36} (27 x^2 + 81 x + 28) 
             \biggr)
 \PlusBreak{}
 {1 \over 4 y} (3 x^2 + 6 x - 1) Y^2
%
-{x{} (29 x + 9) \over 24 y^2} X
-{29 x + 8 \over 24 y} Y
%
%
%
 \PlusBreak{}
 i \pi {}
\biggl[
  {1 \over 3 y^3}  \biggl(
     \li2(-x)
    -{x \over 8} (x^2 + 10 x + 11) X^2
    + {1 \over 4} (2 x^3 + 10 x^2 + 11 x + 4) X Y
 \MinusBreak{\null + i \pi \bigglB {1 \over 3 y^3}  \bigglP }
     {\pi^2 \over 24} (9 x^3 + 30 x^2 + 33 x + 16) 
          \biggr)
  - {3 x + 4 \over 24 y} Y^2
%
 \MinusBreak{\null + i \pi \bigglB }
    {x \over 8 y^3} (x^2 - 8 x - 5) X
  - {1 \over 8 y^2} (x^2 - x - 4) Y
%
  +{7 x + 2 \over 6 y^2}  
\biggr]
\,, \label{KK43}\\[1pt plus 4pt]
%
L_4^{[3]} & = & 
{1 \over 6 y}   \biggl(
   {1 \over 12} (4 + 3 x) X Y
  -{5 \over 12} x X^2
  +{5 \over 9} x X
  -{\pi^2 \over 2}
        \biggr)
  -{1 \over 36} Y^2
  +{5 \over 54} Y
 \PlusBreak{}
  i {\pi \over 72 y}   
\biggl(
  -(7 x - 4) X
  +(7 x + 8) Y
  -{20 \over 3}
\biggr)
\,. \label{LL43}
\end{eqnarray}


\noindent
For $h=5$ in \eqn{hel5}
and color factor $\trc^{[1]}$ in \eqn{TraceBasisqggq}:
\begin{eqnarray}
A_5^{[1]} & = & 
{1 \over 4 y} \biggl[
   -(20 x^3 + 17 x - 2 + 36 x^2) \li4(-x)
 \PlusBreak{ {1 \over 4 y} \bigglB }
   2 (18 x^2 + 10 x^3 - 1 + 8 x) X \li3(-x)
   +{1 - y \over 2} X^2 \li2(-x)
 \MinusBreak{ {1 \over 4 y} \bigglB }
    {1 \over 12} (10 x^3 + 18 x^2 + 11 x + 6) X^4
 \PlusBreak{ {1 \over 4 y} \bigglB }
    {\pi^2 \over 3} (5 x^3+9 x^2+7 x+7) X^2
   +(5 x+7) \zeta_3 X
 \PlusBreak{ {1 \over 4 y} \bigglB }
    {\pi^4 \over 720} (1280 x^3 + 2304 x^2 + 1105 x + 57) 
\biggr]
 \PlusBreak{}
  x{}(4 + 5 x) \biggl[ 
    \li4(-y)
   -\li4 \left( -{x \over y} \right)
   -X \li3(-y)
   -{1 \over 24} Y^4
 \PlusBreak{ x(4 + 5 x) \bigglB }
    {1 \over 6} Y^3 X
   -{1 \over 4} X^2 Y^2
   -{\pi^2 \over 12} Y^2
       \biggr]
 \PlusBreak{}
    {1 \over 12} \biggl[
    (10 x^2 + 8 x + 2) \pi^2 \li2(-x)
   - (10 x^2 + 8 x + 1) Y X{} (X^2 - 2  \pi^2) 
\biggr]
%
 \PlusBreak{}
 {1 \over y} \biggl[
    {1 \over 6} (30 x^2 + 28 x - 11) 
          \biggl( -\li3(-x) + X \li2(-x) + {1 \over 2} X^2 Y \biggr)
 \MinusBreak{\null - {1 \over y} \bigglB }
    {1 \over 36} (21 x + 30 x^2 - 25) X^3
   +{1 \over 48} (240 x^2 + 107 x - 205) \zeta_3
 \PlusBreak{\null + {1 \over y} \bigglB }
    {\pi^2 \over 288} (240 x^2 + 159 x - 265) X
\biggr]
%
 \PlusBreak{}
 {1 \over 288 y^2} (721 x^2 + 1449 x + 368) X^2
-{\pi^2 \over 12} (10 x - 3) 
%
 \MinusBreak{}
 {1 \over 96 y} (167 + 407 x) X
%
+ {1 \over 16} x - {30377 \over 3456}
%
 \PlusBreak{}
 i \pi {} \biggl[
    {1 \over 2 y}  \biggl(
            (10 x^3 + 18 x^2 + 8 x - 1) \li3(-x)
        +{1 - y \over 2} X \li2(-x)
 \MinusBreak{\null + i \pi \bigglB {1 \over 2 y}  \bigglP }
         {1 \over 6} (10 x^3 + 18 x^2 + 11 x + 6) X^3
 \MinusBreak{\null + i \pi \bigglB {1 \over 2 y}  \bigglP }
         {\pi^2 \over 3} (5 x^3 - 1 + 4 x + 9 x^2) X
        -(x - y) \zeta_3 
       \biggr)
 \MinusBreak{\null + i \pi \bigglB }
    x{} (4 + 5 x) \li3(-y)
   -{1 \over 4} (10 x^2 + 8 x + 1) X^2 Y
%
 \PlusBreak{\null + i \pi \bigglB }
  {1 \over 6 y}  \biggl(
       (30 x^2 + 28 x - 11) ( \li2(-x) + X Y )
 \MinusBreak{\null + i \pi \bigglB {1 \over 6 y}  \bigglP }
      {1 \over 2} (30 x^2 + 21 x - 25) X^2
     -{\pi^2 \over 24} (120 x^2 + 83 x - 73) 
   \biggr)
  -{9\over 4} \zeta_3   
%
 \PlusBreak{\null + i \pi \bigglB }
   {1 \over 144 y^2} (721 x^2 + 1449 x + 368) X
  + {11\over 288} \pi^2  
%
+  {5 \over 2 y} + {407\over 96}
\biggr]
\,, \label{AA51}\\[1pt plus 4pt]
%
B_5^{[1]} & = &
{1 \over 2 y}  \biggl[
  -(7 x^3 - 6 x^2 - 11 x - 5) \li4(-x)
  -x{} (7 x^2 - 6 x - 14) \li4(-y)
 \MinusBreak{ {1 \over 2 y}  \bigglB }
   (5 x^3 + 12 x^2 - 4) \li4 \left( -{x \over y} \right)
  -2 (3 x^2 - 3 x - 2) y Y \li3(-y)
 \PlusBreak{ {1 \over 2 y}  \bigglB }
   (6 x^3 + 12 x^2 + x - 4) Y \li3(-x)
 \PlusBreak{ {1 \over 2 y}  \bigglB }
   (x^3 - 12 x^2 - 7 x - 1) X \li3(-x)
  +(x^3 + 5 x + 4) X \li3(-y)
 \PlusBreak{ {1 \over 2 y}  \bigglB }
   {1 \over 2} (6 x^2 + 3 x + 1) X^2 \li2(-x)
  -(3 x^2 + 3 x + 2) Y^2 \li2(-x)
 \PlusBreak{ {1 \over 2 y}  \bigglB }
   {\pi^2 \over 6} (5 x^3 + 12 x^2 - 18 x - 14) \li2(-x)
  -4 y \li2(-x) X Y
 \MinusBreak{ {1 \over 2 y}  \bigglB }
   {1 \over 24} (11 x^3 + 8 x + 24 x^2 - 2) Y^4
  +{1 \over 24} x{}(2 - x^2) X^4
 \MinusBreak{ {1 \over 2 y}  \bigglB }
   {4 - x \over 4} (x^2 - 2 x - 2) X^2 Y^2
  +{1 \over 6} (x^3 + 6 x^2 + 5 x + 1) Y X^3
 \PlusBreak{ {1 \over 2 y}  \bigglB }
   {y^2 \over 6} (11 x - 10) Y^3 X
  +(4 + 5 x) Y \zeta_3
  -{1 \over 2} (9 + 11 x) X \zeta_3
 \PlusBreak{ {1 \over 2 y}  \bigglB }
   {\pi^2 \over 12} (x^3 - 9 x^2 -14 x +12) X^2         
  +{\pi^2 \over 12} (x^3 - 3 x^2 + 12 x + 14) Y^2       
 \MinusBreak{ {1 \over 2 y}  \bigglB }
   {\pi^2 \over 6} (2 x^3 - 9 x^2 + 15 x + 12) Y X
 \PlusBreak{ {1 \over 2 y}  \bigglB }
   {\pi^4 \over 1440} (488 x^3 - 1152 x^2 - 987 x - 343)   
\biggl]
%
 \PlusBreak{}
{1 \over 4 y}  \biggl[
   x{}(7 x^2 + 16 x + 21) (\li3(-y) + Y \li2(-x))
 \PlusBreak{\null + {1 \over 4 y}  \bigglB }
    x{}(7 x^2 + 2 x + 1) (\li3(-x) - X \li2(-x))
 \MinusBreak{\null + {1 \over 4 y}  \bigglB }
   {x \over 18} (3 x^2 + 18 x + 29) X^3
  -{y \over 3} (6 x^2 + 15 x + 10) Y^3
 \PlusBreak{\null + {1 \over 4 y}  \bigglB }
   {2 x \over 3} (4 + 9 x) Y X^2
  +{1 \over 6} (3 x^3 - 12 x^2 + 21 x - 40) X Y^2
 \MinusBreak{\null + {1 \over 4 y}  \bigglB }
   {1 \over 36} (360 x^2 + 133 x - 443) \zeta_3
  -{\pi^2 \over 3} (x^3 + x^2 + 26 x - 10) Y
 \MinusBreak{\null + {1 \over 4 y}  \bigglB }
   {\pi^2 \over 72} (96 x^3 + 72 x^2 - 541 x - 477) X 
\biggr]
%
 \PlusBreak{}
 {1 \over 144 y^2} (99 x^4 + 63 x^3 - 98 x^2 + 154 x + 144) X^2
 \PlusBreak{}
 {x \over 72 y} (99 x^2 + 73 + 72 x) X Y
+{1 \over 144 x} (99 x^3 + 81 x^2 + 173 x + 27) Y^2
 \MinusBreak{}
 {\pi^2 \over 288 y} (198 x^3 + 108 x^2 - 145 x - 455) 
%
 \PlusBreak{}
 {1 \over 216 y} (81 x^2 + 647 x + 782) X
-{3 \over 8} (1 - x) Y
%
+ {1 \over 8} x
+ {19139 \over 2592}
%
 \PlusBreak{}
  i \pi {}  \biggl[
 {1 \over 2 y} \biggl(
   (7 x^3 - 5 - 6 x) \li3(-x)
  -2 x{} (3 x + 1) Y \li2(-x)
 \MinusBreak{\null + i \pi  \bigglB {1 \over 2 y} \bigglP }
    x{} (5 - 7 x^2) \li3(-y)
  +(6 x^2 + 7 x + 5) X \li2(-x)
 \MinusBreak{\null + i \pi  \bigglB {1 \over 2 y} \bigglP }
   {x \over 6} (6 x^2 + 6 x + 1) Y^3
  -{x \over 6} (x^2 - 6 x - 8) X^3
 \PlusBreak{\null + i \pi  \bigglB {1 \over 2 y} \bigglP }
   (3 x^3 - x + 1)  X Y^2                                     
  + {1 \over 2} (x^3+ 2 x + 1) Y X^2                          
 \MinusBreak{\null + i \pi  \bigglB {1 \over 2 y} \bigglP }
   {\pi^2 \over 6} (6 x^3 - 6 x^2 - 15 x - 8)  Y              
 \MinusBreak{\null + i \pi  \bigglB {1 \over 2 y} \bigglP }
   {\pi^2 \over 6} (x^3 - 12 x^2  - 5 x - 6)  X               
   +{1 \over 2} y \zeta_3                                     
         \biggr)
%
 \PlusBreak{\null + i \pi  \bigglB }
 {1 \over 2 y}  \biggl( 
    \li2(-x) x{} (7 x + 10)
   +{x \over 6} (3 x - 1) (3 x + 5) X^2
 \PlusBreak{\null +  i \pi  \bigglB {1 \over 2 y} \bigglP }
    {1 \over 6} (9 x^3 + 33 x^2 + 54 x + 10) Y^2
 \MinusBreak{\null + i \pi  \bigglB {1 \over 2 y} \bigglP }
    {1 \over 3} (9 x^3 + 12 x^2 + 13 x + 20) X Y
 \PlusBreak{\null + i \pi  \bigglB {1 \over 2 y} \bigglP }
    {\pi^2 \over 48} (72 x^3 + 112 x^2 + 111 x + 79)         
   \biggr)
%
 \MinusBreak{\null + i \pi  \bigglB }
 {1 \over 8 y^2} (27 x^2 + 12 x^3 - 16 - 9 x) X
 \MinusBreak{\null + i \pi  \bigglB }
 {1 \over 72 y x} (181 x^2 + 108 x^3 + 200 x + 27) Y
%
+ {1 \over y} - {647 \over 216}                              
\biggr]
\,, \label{BB51}\\[1pt plus 4pt]
C_5^{[1]} & = &
 {x \over 2 y} \biggl[
   {1 \over 24} (x^2 - 2) \biggl(
                  -24 \li4(-y) + 24 \li4 \left( -{x \over y} \right)
                  - 4 \pi^2 \li2(-x)
 \PlusBreak{\null + {x \over 2 y} \bigglB {1 \over 24} (x^2 - 2) }
                    Y^4 - 4 Y^3 X +2 \pi^2 Y^2
                \biggr)
 \PlusBreak{\null + {x \over 2 y} \bigglB }
    {1 \over 2} (7 - 2 x^2) \li4(-x)
   +(\li3(-x) - \zeta_3) Y
 \MinusBreak{\null + {x \over 2 y} \bigglB }
    y{} (x - 1) X \li3(-y)
   +(x^2 - 3) X \li3(-x)
   +{1 \over 4} \li2(-x) X^2
 \MinusBreak{\null + {x \over 2 y} \bigglB }
    {1 \over 24} (1 + x^2) X^4
   +{\pi^2 \over 12} x^2 X^2 
\biggr]
 \PlusBreak{}
 {x \over 24} (1 - x) X^2 Y{} (3 Y + 2 X)
+{\pi^2 \over 6} x{}(x - 1) Y X 
 \MinusBreak{}
 {1 \over 2 y} (x + 3) X \zeta_3
+{\pi^4 \over 720 y} (32 x^3 - 48 x + 11)
%
 \PlusBreak{}
 {x \over 12} (x + 3)  \biggl(
  - 3 \li3(-y)
   -3 Y \li2(-x)
   -{3 \over 2} X Y^2
   +\pi^2 Y 
       \biggr)
 \PlusBreak{}
 {1 \over 4 y}   \biggl[
    x{}(7 + x^2 + 2 x) ( \li3(-x) - X \li2(-x) )
 \MinusBreak{\null + {1 \over 4 y} \bigglB }
    {x \over 6} (x^2 + 6 x - 1) X^3
   + x{}(x - 2) Y X^2
 \PlusBreak{\null + {1 \over 4 y} \bigglB }
    {1 \over 2} (4 x^2 + 7 x + 15) \zeta_3
   -{\pi^2 \over 6} (2 x^3 + 6 x^2 + 7 x - 3) X
        \biggr]
%
 \MinusBreak{}
 {x \over 32 y^2} (2 x^3 + 10 x^2 + 21 + 21 x) X^2
+{x \over 8} (x + 3) X Y
 \MinusBreak{}
 {1 \over 16 x} (x^3 + 3 x^2 + 3 x - 3) Y^2
-{\pi^2 \over 96} (6 x^2 + 30 x - 29) 
%
 \PlusBreak{}
 {3 \over 8} (x - 1) Y
+{1 \over 32 y} (12 x^2 + 25 x - 3) X
%
+ {x \over 16} 
+ {255 \over 128}
%
%
 \PlusBreak{}
  i \pi {} \biggl[
  {1 \over 2 y} \biggl( 
  - x{}(2 - x^2) \li3(-x)
  +{x \over 2} \li2(-x) X
  - (2 x +3) \zeta_3 
 \MinusBreak{\null - i \pi  \bigglB {1 \over 2 y} \bigglP } 
   {x^3 \over 6} X^3
  +{\pi^2 \over 6} x{}(2 - x^2) X
          \biggr)
 + {x \over 4} (1 - x) \Bigl(2 \li3(-y) + Y X^2 \Bigr)
%
 \PlusBreak{\null -  i \pi  \bigglB }
  {x{}(x - 2) \over 2 y}  \biggl(
     \li2(-x)
    +X Y
    -{1 \over 2} X^2
    -{\pi^2 \over 6}
          \biggr)
   -{1 \over 8} \pi^2             
%
 \MinusBreak{\null - i \pi  \bigglB }
 {x \over 16 y^2} (7 x + 15) X
-{3 (x - 1) \over 8 x} Y
%
-{1 \over 2 y} - {25 \over 32}   
 \biggr]
\,, \label{CC51}\\[1pt plus 4pt]
%
D_5^{[1]} & = &
x{} (x - y)  \biggl[ 
   - \li4(-x) - \li4(-y) + \li4 \left( -{x \over y} \right) 
 \PlusBreak{ x (x - y)  \bigglB }
    (\li3(-x) + \li3(-y)) X
   -{\pi^2 \over 6} \li2(-x)
    -{1 \over 6} Y^3 X
 \PlusBreak{ x (x - y)  \bigglB }
     {1 \over 4} X^2 Y^2
    +{\pi^2 \over 12} Y^2
    +{1 \over 24} Y^4
    +{\pi^2 \over 12} X^2
    -{1 \over 24} X^4
 \PlusBreak{ x (x - y)  \bigglB }
     {1 \over 6} Y X^3
    -{\pi^2 \over 3} Y X  
    +{4 \over 45} \pi^4
\biggr]
%
 \MinusBreak{}
 {y \over 3} \li3(-x) 
-{7 \over 3} x \li3(-x) 
-{1 \over 6} (1 - 6 x) Y X^2
 \MinusBreak{}
 {1 \over 3} (1 - 6 x) X \li2(-x)
+{1 \over 72 y} (24 x^2 + 15 x - 14) X^3
 \MinusBreak{}
 {\pi^2 \over 144 y} (48 x^2 + 21 x - 47) X
+{65 \over 24} \zeta_3 x
+ {17 \over 24} \zeta_3 y
%
+ {\pi^2 \over 27} (4 + 9 x) 
 \MinusBreak{}
 {1 \over 72 y^2} (70 x^2 + 141 x + 35) X^2
%
+{1 \over 216 y} (163 x - 53) X
%
+ {863 \over 864}
%
 \PlusBreak{}
 i \pi {} \biggl[
   x{} (x - y)  \biggl(
      (\li3(-x) + \li3(-y))
     +{1 \over 2} Y X^2
     -{1 \over 6} X^3
     -{\pi^2 \over 6} X
            \biggr)
%
 \MinusBreak{\null + i \pi  \bigglB }
   {1 \over 3} (1 - 6 x) \li2(-x) 
  +{1 \over 24 y} (24 x^2 + 15 x - 14) X^2
 \PlusBreak{\null + i \pi  \bigglB }
   {1 \over 3} (6 x - 1) X Y
  -{\pi^2 \over 48} (16 x - 3)   
%
 \MinusBreak{ \null + i \pi  \bigglB }
   {1 \over 36 y^2} (141 x + 70 x^2 + 35) X
%
  -{1 \over y}
  -{163 \over 216}      
\biggr]
\,, \label{DD51}\\[1pt plus 4pt]
%
E_5^{[1]} & = & 
x{} (x - y)  \biggl[
     2 \biggl( \li4(-x) + \li4(-y) - \li4 \left( -{x \over y} \right) \biggr)
 \MinusBreak{ x (x - y)  \bigglB }
     2 X{} (\li3(-x) + \li3(-y))
    +{\pi^2 \over 3} \li2(-x)
    +{1 \over 3} Y^3 X
 \MinusBreak{ x (x - y)  \bigglB }
     {1 \over 12} Y^4
    -{1 \over 3} Y X^3
    -{1 \over 2} X^2 Y^2
    +{1 \over 12} X^4
    -{\pi^2 \over 6} Y^2
 \PlusBreak{ x (x - y)  \bigglB }
     {2 \over 3} \pi^2 Y X 
    -{\pi^2 \over 6} X^2 
    -{8 \over 45} \pi^4
\biggr]
%
 \PlusBreak{}
 {1 \over 3 y} \biggl[
    (x^3 - 9 x^2 - 9 x + 2) ( \li3(-x) - X \li2(-x) )
 \PlusBreak{\null -  {1 \over 3 y} \bigglB } 
    {x \over 24} (8 x^2 - 24 x - 19) X^3
   + {1 \over 4} (4 x^3 + 15 x^2 + 15 x + 6) X Y^2
 \MinusBreak{\null -  {1 \over 3 y} \bigglB }
    {1 \over 8} (8 x^3 - 21 x^2 - 23 x + 8) Y X^2
  -  {\pi^2 \over 24} (8 x^3 + 33 x^2 + 21 x + 14) Y
 \PlusBreak{ \null - {1 \over 3 y} \bigglB }
    {\pi^2 \over 48} (8 x^3 + 120 x^2 + 73 x - 27) X 
  +  {1 \over 24} (288 x^2 + 287 x - 25) \zeta_3
\biggr]
 \MinusBreak{}
 {y^2 \over 3} (\li3(-y) + Y \li2(-x) )
+{1 \over 24} (3 x + 2) Y^3
%
 \MinusBreak{}
 {x \over 72 y^2} (18 x^3 + 24 x^2 - 79 x - 139) X^2
 - {1 \over 36} (9 x^2 - 18 x + 1) Y^2
 \MinusBreak{}
 {1 \over 144 y}  \biggl[
      8 x{} (9 x^2 - 3 x - 7) X Y
      -(36 x^3 + 39 x - 33 + 68 x^2) \pi^2
         \biggr]
%
 \MinusBreak{}
 {1 \over 216 y} (191 + 335 x - 180 x^2) X
 + {1 \over 6} (5 x-4) Y
%
-{4085 \over 2592}
 \PlusBreak{}
  i \pi {} \biggl[
{x \over 3}  (x - y) \Bigl(X^3 - 3 Y X^2 - 6 \li3(-x) - 6 \li3(-y) 
                                                        + \pi^2 X \Bigr)
%
 \PlusBreak{\null + i \pi   \bigglB }
 {1 \over 3 y}  \biggl(
    (12 x^2 + 12 x - 1) \li2(-x)
   +{x \over 8} (4 x^2 - 39 x - 32) X^2
 \MinusBreak{\null +  i \pi   \bigglB {1 \over 3 y}  \bigglP }
    {1 \over 8} (4 x^2 + x + 1) (y - 1) Y^2
   -{x \over 4} (4 x^2 - 39 x - 41) X Y
 \PlusBreak{\null +  i \pi   \bigglB {1 \over 3 y}  \bigglP }
    {\pi^2 \over 48} (24 x^3 - 42 x^2 - 45 x + 17)           
   \biggr)
%
 -{x \over 36 y^2} (12 x^2 - 59 x - 125) X
 \MinusBreak{\null + i \pi   \bigglB }
{1 \over 18 y} (6 x^2 + 10 x - 1) Y
%
 + {3\over 2 y} + {371 \over 216}                            
\biggr]
\,, \label{EE51}\\[1pt plus 4pt]
%
F_5^{[1]} & = &
F_4^{[1]}
\,, \label{FF51}\\[1pt plus 4pt]
%
G_5^{[1]} & = & 
G_4^{[1]}
\,.  \hskip 5.5 cm \label{GG51}
\end{eqnarray}


\noindent
For $h=5$ in \eqn{hel5}
and color factor $\trc^{[2]}$ in \eqn{TraceBasisqggq}:
\begin{eqnarray}
A_5^{[2]} & = & 
{1 \over y} 
\biggl[
   {1 \over 48} (1 - y) (1 - x y) \biggl(
      6 X^2 \li2(-x) - 12 X Y \li2(-x)  
 \PlusBreak{ {1 \over y} \bigglB {1 \over 48} (1 - y) (1 - x y) \bigglP }  
      6 Y^2 \li2(-x) + 3 X^4 +2 Y^4
    +  9 X^2 Y^2 + \pi^2 Y^2  \biggr)
 \MinusBreak{ {1 \over y} \bigglB }
   (3 x^3 + 9 x^2 + 9 x + 2) 
     \biggl( {1 \over 4} \li4 \biggl( -{x \over y} \biggr)
    -{\pi^2 \over 24} \li2(-x) \biggr)
 \MinusBreak{ {1 \over y} \bigglB }
   {1 \over 12} (3 x^3 + 9 x^2 + 9 x + 5) X^3 Y
  - {1 \over 24} X {} (x^3 + 3 x^2 + 3 x + 4) (Y^3 - \pi^2 X)
 \MinusBreak{ {1 \over y} \bigglB }
  {\pi^2 \over 6} X Y
  - {5 \over 4} \zeta_3 X
  -\zeta_3 Y
  -{\pi^4 \over 320} (4 x^3 + 12 x^2 + 12 x - 7) 
\biggr]
 \PlusBreak{}
 {y^2 \over 2} (X - Y) (\li3(-x) + \li3(-y))
%
 \PlusBreak{}
 {1 \over 24 y} (6 x^2 + 15 x + 44) 
      ( \li3(-x) + \li3(-y) 
            + (Y - X) \li2(-x) )
 \MinusBreak{}
 {1 \over y} 
\biggl[
    {1 \over 72} (23 x^3 + 60 x^2 + 42 x + 50) X^3
   - {1 \over 48} (2 x^3 - 18 x^2 - 63 x - 32) X^2 Y
 \MinusBreak{\null + {1 \over y} \bigglB }
    {1 \over 24} (21 x^3 + 78 x^2 + 105 x + 82) X Y^2
 \PlusBreak{\null + {1 \over y} \bigglB }
    {1 \over 144} (86 x^3 + 282 x^2 + 327 x + 208) Y^3
 \PlusBreak{ \null +{1 \over y} \bigglB }
    {\pi^2 \over 288} (92 x^3 + 288 x^2 + 288 x + 335) X
 \PlusBreak{\null + {1 \over y} \bigglB }
    {\pi^2 \over 144} (86 x^3 + 276 x^2 + 312 x + 267) Y
   + {3 \over 16} (4 x^2 + 10 x - 13) \zeta_3
\biggr]
%
 \PlusBreak{}
 {1 \over 288 y} 
\biggl[
    (475 x^3 + 798 x^2 + 186 x + 368) X^2
 \MinusBreak{\null + {1 \over 288 y} \bigglB }
    2 (475 x^3 + 1062 x^2 + 846 x + 280) X Y
 \PlusBreak{\null + {1 \over 288 y} \bigglB }
    (475 x^3 + 1326 x^2 + 1506 x + 1160) Y^2
 \PlusBreak{\null + {1 \over 288 y} \bigglB }
    \pi^2 (475 x^3 + 982 x^2 + 622 x - 528) 
\biggr]
%
 \PlusBreak{}
   {1 \over y}  
\biggl[
   {1 \over 288} (950 x^2 + 1433 x + 501) X
 \MinusBreak{\null + {1 \over y} \bigglB }
   {1 \over 864} (2850 x^2 + 5091 x - 32) Y
\biggr]
%
-{36077 \over 3456 y} - {475 \over 288}
%
 \PlusBreak{}
  i \pi {}
 \biggl[ 
      - {9 \over 4 y} \zeta_3
%
   - 11 {(1 - y) \over 12 y} (1 - x y) \Bigl((X - Y)^2 + \pi^2\Bigr)
   + {11 \over 288 y} \pi^2 
%
 \PlusBreak{\null + i \pi \bigglB }
  {11 \over 36 y} \Bigl(
   (6 x^2 + 15 x - 2) (Y-X)
    +22 Y   
        \Bigr)
%
+ {2327 \over 864 y}
+ {11 \over 12} 
\biggr]
\,, \label{AA52}\\[1pt plus 4pt]
%
B_5^{[2]} & = & 
{1 \over 2 y} 
\biggl[
  -{1 \over 2} (y - 1) (4 x^2 - 5 x - 5) \li4 \biggl( -{x \over y} \biggr)
  -(12 x + 13) x \li4(-y)
 \MinusBreak{ {1 \over 2 y} \bigglB }
   (12 x^2 + 9 x + 4) \li4(-x)
  +(x^3 + 9 x^2 + x - 1) X \li3(-x)
 \MinusBreak{ {1 \over 2 y} \bigglB }
   (x^3 - 9 x^2 - 5 - 14 x) Y \li3(-y)
  + (x - 5)  y^2  X \li3(-y)
 \MinusBreak{ {1 \over 2 y} \bigglB }
   (x^3 + 3 x^2 - 5 - 4 x) Y \li3(-x)
   -{1 \over 4} (9 x^2 + x - 2) X^2 \li2(-x)
 \MinusBreak{ {1 \over 2 y} \bigglB }
   {\pi^2 \over 12} (4 x^3 + 3 x^2 - 47 x - 26) \li2(-x)
  + {1 \over 2} (y - 1) (3 x + 5) X Y \li2(-x)
 \PlusBreak{ {1 \over 2 y} \bigglB }
   {5 \over 4} (3 x^2 + 3 x + 2) Y^2 \li2(-x)
  +{x \over 24} (x^2 - 3 x - 3) X^4
 \MinusBreak{ {1 \over 2 y} \bigglB }
   {1 \over 6} (x^3 + 3 x^2 + 2 x - 1) X^3 Y
  +  {1 \over 8} (6 x^3 - 3 x^2 - 29 x - 22) X^2 Y^2
 \MinusBreak{ {1 \over 2 y} \bigglB }
   {1 \over 12} (10 x^3 - 21 x^2 - 47 x - 28) X Y^3
 \PlusBreak{ {1 \over 2 y} \bigglB }
   {1 \over 24} (5 x^3 + 6 x^2 - 2 x - 4) Y^4
  -{\pi^2 \over 2} (x^3 + x^2 - 7 x - 4) X Y
 \PlusBreak{ {1 \over 2 y} \bigglB }
   {\pi^2 \over 12} (2 x^3 + 6 x^2 + 3 x - 12) X^2      
  +{1 \over 2} (12 x + 11) \zeta_3 X
 \PlusBreak{ {1 \over 2 y} \bigglB }
   {\pi^2 \over 24} (4 x^3 - 3 x^2 - 37 x - 34) Y^2     
  -(6 x + 5) \zeta_3 Y
 \PlusBreak{ {1 \over 2 y} \bigglB }
   {\pi^4 \over 1440} (68 x^3 + 1404 x^2 + 744 x + 343) 
\biggr]
%
 \PlusBreak{}
{1 \over 4 y}   
\biggl[
  -x {} (3 x^2 + 2 x - 4) (\li3(-x) - X \li2(-x) )
 \MinusBreak{\null + {1 \over y 4} \bigglB }  
   x {} (3 x^2 + 2 x + 8) (\li3(-y) + Y \li2(-x) )
 \MinusBreak{\null + {1 \over 4 y } \bigglB } 
   {x \over 18} (41 x^2 + 24 x + 15) X^3
   + {x \over 6} (19 x^2 - 6 x - 4) X^2 Y
 \PlusBreak{\null + {1 \over 4 y } \bigglB } 
    {1 \over 6} (3 x^3 + 36 x^2 + 9 x + 40) X Y^2
  - {1 \over 18} (43 x^3 + 78 x^2 + 120 x + 60) Y^3
 \MinusBreak{\null + {1 \over 4 y} \bigglB } 
   {\pi^2 \over 72} (128 x^3 + 72 x^2 + 552 x + 477) X
 \MinusBreak{\null + {1 \over 4 y} \bigglB } 
   {\pi^2 \over 18} (25 x^3 + 54 x^2 - 75 x + 60) Y
  + {1 \over 36} (144 x^2 + 36 x - 443) \zeta_3
\biggr]
%
 \PlusBreak{}
{1 \over 72 y}  
\biggl[
   {1 \over 2} (359 x^3 + 186 x^2 + 186 x + 144) X^2
 \MinusBreak{\null + {1 \over 72 y} \bigglB }
   (359 x^3 + 372 x^2 + 265 x + 144) X Y
 \PlusBreak{\null + {1 \over 72 y} \bigglB}
   {\pi^2 \over 4} (718 x^3 + 784 x^2 + 382 x - 167) 
\biggr]
 \PlusBreak{}
 {1 \over 144 x y} (359 x^4 + 558 x^3 + 344 x^2 + 308 x - 27) Y^2
%
 \PlusBreak{}
 {1 \over 216 y} (915 x^2 + 402 x - 782) X
 \MinusBreak{}
 {1 \over 72 y} (305 x^2 + 227 x + 213) Y
%
+ {14135 \over 2592 y} - {139 \over 72}
%
 \PlusBreak{}
 i \pi {}
\biggl[ 
  {1 \over y}  \biggl(
     {1 \over 2} (6 x^2 + 5 x + 4) \li3(-x)
    +{x \over 2} (6 x + 5) \li3(-y)
 \PlusBreak{\null + i \pi \bigglB  {1 \over y} \bigglP } 
     (3 x + 1) x Y \li2(-x)
    -(3 x^2 + 3 x + 2) X \li2(-x)
 \PlusBreak{\null +  i \pi \bigglB  {1 \over y} \bigglP } 
     {1 \over 4} x {} (x^2-2) X^2 Y                         
    +{x^3 \over 12} Y^3
    +{x y \over 12} (x + 5) X^3
 \MinusBreak{\null +  i \pi \bigglB  {1 \over y} \bigglP } 
     {1 \over 4} (x^3 - 6 x^2 - 3 x + 2) X Y^2              
   -  {\pi^2 \over 12} (x^3 + 12 x^2 + 5 x + 6) X           
 \PlusBreak{\null + i \pi \bigglB  {1 \over y} \bigglP }
     {\pi^2 \over 12} (x^3 - 12 x^2 - 16 x - 8) Y           
   + {1\over 4} \zeta_3                                     
       \biggr)
%
 \PlusBreak{\null + i \pi \bigglB }
  {1 \over 24 y}  \biggl(
    -72 x \li2(-x)
    -x {} (31 x^2 + 36 x + 7) X^2
 \PlusBreak{\null + i \pi \bigglB {1 \over 24 y}  \bigglP }
     2 (31 x^3 + 36 x^2 + 29 x + 40) X Y
    - (31 x^3 + 36 x^2 + 87 x + 20) Y^2
 \MinusBreak{\null + i \pi \bigglB {1 \over 24 y}  \bigglP }
     {\pi^2\over 4} (124 x^3 + 128 x^2 + 116 x + 79)       
        \biggr)
%
   -{x \over 72 y} (186 x + 79) X
 \PlusBreak{\null + i \pi \bigglB }
   {1 \over 72 x y} (186 x^3 + 79 x^2 + 164 x - 27) Y
%
  -{571 \over 108 y} + {31 \over 24}                     
\biggr]
\,, \hskip 3.8 cm \label{BB52}\\[1pt plus 4pt]
%
C_5^{[2]} & = & 
{1 \over 2 y}  
\biggl[
  -{x \over 2} (5 x^2 + 12 x + 2) \li4 \biggl( -{x \over y} \biggr)
 \MinusBreak{ {1 \over 2 y} \bigglB }
   x {} (6 x^2 + 6 x + 1) ( \li4(-x) + \li4(-y) )
 \PlusBreak{ {1 \over 2 y} \bigglB }
   X {} (2 x^3 \li3(-x) + 2 x^3 \li3(-y) + 3 \zeta_3)
 \PlusBreak{ {1 \over 2 y} \bigglB }
   (4 x^2 + 6 x + 1) x Y {} (\li3(-x) + \li3(-y))
 \MinusBreak{ {1 \over 2 y} \bigglB }
   {x^3 \over 4} (X - Y)^2 \li2(-x)
  -{x \over 24} (9 x^2 + 12 x + 2) Y^4
 \MinusBreak{ {1 \over 2 y} \bigglB }
   {x^3 \over 24} X^2 (-8 X Y - 9 Y^2 + 2 X^2)
 \PlusBreak{ {1 \over 2 y} \bigglB }
   {\pi^2 \over 12} (5 x^2 + 12 x + 2) x \li2(-x)
  +{x \over 12} (15 x^2 + 24 x + 4) X Y^3
 \PlusBreak{ {1 \over 2 y} \bigglB }
   {\pi^2 \over 24} x^3 (2 X^2 - 12 X Y + 3 Y^2)
  + {\pi^4 \over 360} (116 x^3 + 90 x^2 + 15 x - 11) 
\biggr]
%
 \PlusBreak{}
{1 \over 8 y}  
\biggl[
   3 x {} (2 x^2 - 2 x + 1) ( \li3(-x) - X \li2(-x) )
 \PlusBreak{\null + {1 \over 8 y} \bigglB }
    x {} (2 x^2 + 6 x + 5) \Bigl(3 \li2(-x) Y + 3 \li3(-y) - \pi^2 Y \Bigr)
 \PlusBreak{\null + {1 \over 8 y} \bigglB }
   {x \over 3} (6 x^2 - 2 x + 1) X^3
  -{3 \over 2} (4 x^2 - 2 x + 1) x X^2 Y
 \PlusBreak{\null +  {1 \over 8 y} \bigglB }
   2 (3 x^2 + 4 x + 4) x X Y^2
  +{1 \over 6} (22 x + 13) x Y^3
 \MinusBreak{\null + {1 \over 8 y} \bigglB }
   {y \over 3} (3 x^2 + x - 3) \pi^2 X
  -(10 x^2 + 19 x + 15) \zeta_3
\biggr]
%
 \PlusBreak{}
 {1 \over 16 y}  
\biggl[
    {x \over 2} (11 x^2 + 22 x - 6) X^2
  -x {} (11 x^2 + 10 x + 4) X Y
 \PlusBreak{\null + {1 \over 16 y} \bigglB }
   {\pi^2 \over 6} (33 x^3 - 26 x^2 + 29) 
\biggr]
%
 \PlusBreak{}
 {1 \over 32 y} (46 x^2 - 17 x + 3) X
-{1 \over 32 y} (46 x^2 - 5 x - 12) Y
 \PlusBreak{}
   {1 \over 32 x y} (11 x^4 - 2 x^3 - 10 x^2 - 8 x - 6) Y^2
 +{187 \over 128 y} - {17 \over 32}
%
 \PlusBreak{}
 i \pi {}
\biggl[
   {1 \over y}  \biggl(
      {x \over 2} (6 x^2 + 6 x + 1) (\li3(-x) + \li3(-y))
 \MinusBreak{\null + i \pi \bigglB {1 \over y}  \bigglP }
      {x^3 \over 12} X^3
     +{x^3 \over 4} X^2 Y
     -{\pi^2 \over 12} x^3 X
     +{3\over 2} \zeta_3          
         \biggr)
 \MinusBreak{\null + i \pi \bigglB }
  {x \over 12} (5 x + 1) Y {} (3 X Y - Y^2 - \pi^2)
%
 \PlusBreak{\null + i \pi \bigglB }
   {1 \over y}  \biggl(
      {3 \over 2} (x - y) x \li2(-x)
     +{\pi^2 \over 24}\,  {} (9 x^3 - 2 x^2 + x - 3)  
  \PlusBreak{\null + i \pi \bigglB {1 \over y}  \bigglP }
      {x \over 8}{} (3 x^2 - 2 x + 1) X {} (X - 2 Y)
         \biggr)
   -{x \over 8} (3 x + 7) Y^2
%
 \PlusBreak{\null + i \pi \bigglB }
  {x \over 8 y} (6 x - 5) X
  +{1 \over 8 x} (6 x^2 + x + 3) Y
%
  + {27 \over 32 y} + {3 \over 8}       
\biggr]
\,, \label{CC52}\\[1pt plus 4pt]
%
D_5^{[2]} & = &
%
{1 \over 3 y}   
\biggl[
  -\li3(-x)
  -\li3(-y)
  +X \li2(-x)
  -Y \li2(-x)
 \PlusBreak{ {1 \over 3 y} \bigglB }
   {1 \over 24} (5 x^3 + 15 x^2 + 15 x + 14) X^3
  -{1 \over 8} (y - 1) (x y - 1) X^2 Y
 \PlusBreak{ {1 \over 3 y} \bigglB }
   {7 \over 24} (y - 1) (x y - 1) Y^3
  -{1 \over 8} (3 x^3 + 9 x^2 + 9 x + 10) X Y^2
 \PlusBreak{ {1 \over 3 y} \bigglB }
   {\pi^2 \over 48} (10 x^3 + 30 x^2 + 30 x + 37) X
 \PlusBreak{ {1 \over 3 y} \bigglB }
   {7 \over 24} (x^3 + 3 x^2 + 3 x + 3) \pi^2 Y
  -{9 \over 8} \zeta_3
\biggr]
%
 \PlusBreak{}
{1 \over 72 y}   
\biggl[ 
     -(37 x^3 + 60 x^2 + 15 x + 35) X^2
   + 2 (37 x^3 + 72 x^2 + 45 x + 9) X Y
 \MinusBreak{\null + {1 \over 72 y} \bigglB }
    (37 x^3+84 x^2 + 75 x + 71) Y^2
   - {\pi^2 \over 3} (111 x^3 + 210 x^2 + 120 x - 191) 
\biggr]
%
 \MinusBreak{}
 {1 \over 216 y} (222 x^2 + 294 x - 53) X
  + {1 \over 36 y} (37 x^2 + 55 x + 28) Y
%
+{1307 \over 864 y} + {37 \over 72}
%
%
 \PlusBreak{}
i \pi {}
\biggl[
 {1 - y \over 6 y} (1 - x y) \Bigl((X - Y)^2 + \pi^2 \Bigr)
  - {1 \over 144 y} \pi^2                
%
 \PlusBreak{\null + i \pi \bigglB }
    {1 \over 18 y}   \biggl(
     (6 x^2 + 15 x - 13) X
    -(6 x^2 + 15 x + 31) Y
          \biggr)
%
  +{185\over 216 y} - {1 \over 6}      
\biggr]
\,, \label{DD52}\\[1pt plus 4pt]
%
E_5^{[2]} & = & 
x {} (x - y) 
\biggl[
   2 \li4(-x)
  +2 \li4(-y)
  -2 \li4 \biggl( -{x \over y} \biggr)
 \MinusBreak{ x (x - y) \bigglB }
   2 X {} ( \li3(-x) + \li3(-y) )
  +{\pi^2 \over 3} \li2(-x)
 \PlusBreak{ x (x - y) \bigglB }
   {1 \over 12} X^4
  -{1 \over 3} X^3 Y
  -{1 \over 2} X^2 Y^2
  +{1 \over 3} X Y^3
 \MinusBreak{ x (x - y) \bigglB }
   {1 \over 12} Y^4
  -{\pi^2 \over 6} (X^2 - 4 X Y + Y^2)
  -{8 \over 45} \pi^4
\biggr]
%
 \PlusBreak{}
  {1 \over 3 y}   
\biggl[
   (x^3 - 9 x^2 - 9 x + 2) ( \li3(-x) - X \li2(-x) )
 \PlusBreak{\null + {1 \over 3 y} \bigglB }
   {1 \over 24}  (7 x^3 + 9 x^2 + 15 x + 6) Y^3
     +  {1 \over 24} x {} (13 x^2 - 24 x - 24) X^3 
 \MinusBreak{\null + {1 \over 3 y} \bigglB }
   {1 \over 8} (9 x^3 - 27 x^2 - 25 x + 8) X^2 Y
  - {1 \over 8} (y - 1) (5 x^2 + 8 x + 2) X Y^2
 \PlusBreak{\null + {1 \over 3 y} \bigglB }
   {\pi^2 \over 16} (6 x^3 + 40 x^2 + 48 x + 9) X
 \PlusBreak{\null + {1 \over 3 y} \bigglB }
   {\pi^2 \over 24} (y - 1) (x^2 + 13 x + 1) Y
  +{1 \over 24} (288 x^2 + 288 x - 23) \zeta_3
\biggr]
 \MinusBreak{}
 {y^2 \over 3} ( \li3( - y) + Y \li2(-x) )
%
+{x \over 72 y^2} (x^3 - 11 x^2 + 57 x + 123) X^2
 \PlusBreak{}
  {1 \over 144 y}  \biggl[
    4 (x^2 + 12 x + 31) x X Y
   -2 (x^3 + 36 x^2 + 65 x + 38) Y^2
 \MinusBreak{\null + {1 \over 144 y}  \bigglB} 
    (2 x^3 - 60 x^2 - 36 x - 33) \pi^2
             \biggr]
%
-{1 \over 36 y} (11 x^2 + 8 x - 36) Y
 \PlusBreak{}
 {1 \over 216 y} (66 x^2 - 96 x + 191) X
%
-{3401 \over 2592 y} + {19 \over 72}
%
 \PlusBreak{}
 i \pi {}
\biggl[
  x {} (x - y)  \biggl( 
   -2 ( \li3(-x) + \li3(-y) )
   +{1 \over 3} X^3
   -X^2 Y
   +{\pi^2 \over 3} X
      \biggr)
%
 \PlusBreak{\null + i \pi \bigglB}
   {1 \over y}  
\biggl(
     {1 \over 3} (12 x^2 + 12 x - 1) \li2(-x)
    +{x \over 24} (x - 5) (8 x + 7) X^2
 \MinusBreak{null + \! i \pi \bigglB {1 \over y} \bigglP }
     {1 \over 12} (8 x^3 - 33 x^2 - 31 x + 8) X Y
 \PlusBreak{null + \! i \pi \bigglB {1 \over y} \bigglP }
     {1 \over 24} (8 x^3 + 15 x^2 + 21 x + 6) Y^2
    + {\pi^2 \over 144} (48 x^3 -6 x^2 + 47 - 6 x)           
\biggr)
%
 \MinusBreak{\null + i \pi \bigglB }
   {x \over 18 y^2} (12 x^2 - 7 x - 46) X
  -{1 \over 18 y} (12 x^2 + 17 x + 19) Y
%
 \PlusBreak{\null + i \pi \bigglB }
  {551 \over 216 y}               
 + {2 \over 3}
\biggr]
\,, \label{EE52}\\[1pt plus 4pt]
%
F_5^{[2]} & = &
F_4^{[2]}
\,, \label{FF52}\\[1pt plus 4pt]
%
G_5^{[2]} & = & 
G_4^{[2]}
\,. \label{GG52}
\end{eqnarray}


\noindent
For $h=5$ in \eqn{hel5}
and color factor $\trc^{[3]}$ in \eqn{TraceBasisqggq}:
\begin{eqnarray}
H_5^{[3]} & = & 
{1 \over 2 y}  
\biggl[
  -(33 x^3 + 66 x^2 + 41 x - 5) ( \li4(-x) + \li4(-y) )
 \MinusBreak{ {1 \over 2 y} \bigglB }
   21 \li4(-y)
  +{1 \over 2} (44 x^3 + 63 x^2 - 7 x - 10) \li4 \biggl( -{x \over y}
\biggr)
 \PlusBreak{ {1 \over 2 y} \bigglB }
   (28 x^3 + 51 x^2 + 22 x - 2) X \li3(-x)
 \PlusBreak{ {1 \over 2 y} \bigglB }
   (24 x^3 + 39 x^2 + 5 x - 1) X \li3(-y)
 \PlusBreak{ {1 \over 2 y} \bigglB }
   (3 x^3 + 9 x^2 + 14 x + 1) Y \li3(-x)
 \PlusBreak{ {1 \over 2 y} \bigglB }
   (7 x^3 + 21 x^2 + 29 x + 13) Y \li3(-y)
 \MinusBreak{ {1 \over 2 y} \bigglB }
   {1 \over 4} (x + 3) (2 x^2 + 3 x + 4) X^2 \li2(-x)
 \MinusBreak{ {1 \over 2 y} \bigglB }
   {1 \over 2} (2 x^3 + 3 x^2 + 11 x - 6) X Y \li2(-x)
 \PlusBreak{ {1 \over 2 y} \bigglB }
   {1 \over 4} (6 x^3 + 15 x^2 + 27 x + 10) Y^2 \li2(-x)
 \MinusBreak{ {1 \over 2 y} \bigglB }
   {\pi^2 \over 12} (44 x^3 + 63 x^2 - 31 x - 22) \li2(-x)
   - {x \over 24} (26 x^2 + 33 x + 11) X^4
 \PlusBreak{ {1 \over 2 y} \bigglB }
   {1 \over 6} (24 x^3 + 27 x^2 - 6 x - 22) X^3 Y
   + {1 \over 8} (50 x^3 + 105 x^2 + 29 x + 26) X^2 Y^2
 \MinusBreak{ {1 \over 2 y} \bigglB }
   {1 \over 12} (30 x^3 + 39 x^2 - 59 x - 36) X Y^3
   +{1 \over 12} (11 x^3 + 18 x^2 + 2 x - 1) Y^4
 \PlusBreak{ {1 \over 2 y} \bigglB }
   {\pi^2 \over 6} (14 x^3 + 27 x^2 + 26 x + 18) X^2        
  - {\pi^2 \over 6} (52 x^3 + 93 x^2 + 33 x + 10) X Y
 \MinusBreak{ {1 \over 2 y} \bigglB }
   {\pi^2 \over 24} y {} (50 x^2 + 43 x - 38) Y^2              
  + (9 x + 1) \zeta_3 X
  -(9 x + 2) \zeta_3 Y
 \PlusBreak{ {1 \over 2 y} \bigglB }
   {\pi^4 \over 360} (985 x^3 + 1875 x^2 + 900 x + 184) 
\biggr]
%
 \PlusBreak{}
{1 \over 12 y}  
\biggl[
   (31 x^3 - 60 x^2 - 20 x + 124) ( \li3(-x) - X \li2(-x) )
 \PlusBreak{\null + {1 \over 12 y} \bigglB }
   (31 x^3 + 114 x^2 + 119 x + 62) ( \li3(-y) + Y \li2(-x) )
 \PlusBreak{ \null +{1 \over 12 y} \bigglB }
   {x \over 3} (40 x^2 - 63 x - 35) X^3
  -{y \over 3} (3 x^2 - 30 x - 22) Y^3
 \MinusBreak{\null + {1 \over 12 y} \bigglB }
   (36 x^3 - 24 x^2 - 14 x + 19) X^2 Y
 \PlusBreak{\null + {1 \over 12 y} \bigglB }
   (32 x^3 + 102 x^2 + 101 x + 41) X Y^2
 \PlusBreak{ \null +{1 \over 12 y} \bigglB }
   {\pi^2 \over 12} (98 x^3 + 192 x^2 - 275 x - 780) X
 \MinusBreak{\null + {1 \over 12 y} \bigglB }
   {\pi^2 \over 12} (112 x^3 + 588 x^2 + 257 x + 307) Y
   + (132 x^2 + 123 x - 62) \zeta_3
\biggr]
%
 \PlusBreak{}
 {x \over 72 y^2} (135 x^3 + 24 x^2 + 268 x + 487) X^2
 \PlusBreak{}
 {1 \over 72 y}  \biggl[
    6 (45 x^3 + 14 x^2 + x + 10) X Y
   - (135 x^3 - 34 x^2 - 178 x + 250) \pi^2
      \biggr]
 \PlusBreak{}
 {1 \over 72} (135 x^2 + 60 x - 172) Y^2
%
 +{x \over 108 y} (90 x - 121) X
+{1 \over 54} (45 x - 38) Y
%
%
 \PlusBreak{}
 i \pi {}
\biggl[
   {1 \over 2 y}  \biggl(
      (31 x^3 + 60 x^2 + 34 x + 12) \li3(-y)
 \PlusBreak{ \null + i \pi \bigglB {1 \over 2 y}  \bigglP }
      (31 x^3 + 60 x^2 + 36 x - 1) \li3(-x)
 \MinusBreak{\null + i \pi \bigglB {1 \over 2 y}  \bigglP }
      (2 x^3 + 6 x^2 + 12 x + 3) X \li2(-x)
 \PlusBreak{\null + i \pi \bigglB {1 \over 2 y}  \bigglP }
      2 (1 - y) (2 - x y) Y \li2(-x)
 \MinusBreak{\null + i \pi \bigglB {1 \over 2 y}  \bigglP }
      {1 \over 3} (14 x^3 + 24 x^2 + 13 x + 4) X^3
 \PlusBreak{\null + i \pi \bigglB {1 \over 2 y}  \bigglP }
      {1 \over 2} (26 x^3 + 42 x^2 + 9 x - 11) X^2 Y        
 \PlusBreak{\null + i \pi \bigglB {1 \over 2 y}  \bigglP }
      {1 \over 2} (5 x^3 + 18 x^2 + 22 x + 22) X Y^2        
 \MinusBreak{\null + i \pi \bigglB {1 \over 2 y}  \bigglP }
      {1 \over 6} (3 x^3 + 12 x^2 + 9 x + 2) Y^3
     - (5 x - 1) y^2 \pi^2 X                                 
 \MinusBreak{\null + i \pi \bigglB {1 \over 2 y}  \bigglP }
      {1 \over 6} (1 - y) (7 x^2 + 10 x + 13) \pi^2 Y       
     -\zeta_3
           \biggr)
%
 \PlusBreak{\null + i \pi \bigglB }
   {1 \over y}   \biggl(
     {1 \over 12} (174 x^2 + 139 x - 62) \li2(-x)
 \PlusBreak{\null + i \pi \bigglB {1 \over 2y}  \bigglP }
     {1 \over 24} (39 x^3 - 138 x^2 - 62 x + 86) X^2
 \MinusBreak{\null +i \pi \bigglB {1 \over 2 y}  \bigglP }
     {1 \over 4} (13 x^3 - 46 x^2 - 37 x + 6) X Y
 \PlusBreak{\null + i \pi \bigglB {1 \over 2 y}  \bigglP }
     {1 \over 8} (13 x^3 + 12 x^2 - 7 x - 8) Y^2
 \PlusBreak{\null + i \pi \bigglB {1 \over 2 y}  \bigglP }
     {\pi^2 \over 144} (234 x^3 - 180 x^2 - 136 x + 9) 
         \biggr)
%
 \MinusBreak{\null + i \pi \bigglB }
 {1 \over 36 y^2} (153 x^3 - 223 x^2 - 454 x + 30) X
 \MinusBreak{\null + i \pi \bigglB }
 {1 \over 36 y} Y{} (153 x^2 - 115 x - 202)
%
+{211 \over 108 y} + {5 \over 4}
\biggr]
\,, \label{HH53}\\[1pt plus 4pt]
%
I_5^{[3]} & = &
{1 \over y}  
\biggl[
   -{3 \over 2} (x^2 - 1) x \li4(-y)
  -{1 \over 2} (3 x^3 + x - 1) \li4(-x)
 \MinusBreak{ {1 \over y} \bigglB }
   {1 \over 4} (16 x^3 + 9 x^2 + 3 x + 2) \li4 \biggl( -{x \over y}
         \biggr)
  - {1 \over 2} (2 x^3 + 3 x^2 + 2 x + 2) X \li3(-x)
 \MinusBreak{ {1 \over y} \bigglB }
   {1 \over 2} (6 x^3 + 3 x^2 + x + 1) X \li3(-y)
  + {1 \over 2} (3 x^3 + 3 x^2 + 2 x + 1) Y \li3(-x)
 \PlusBreak{ {1 \over y} \bigglB }
   {1 \over 2} (7 x^3 + 3 x^2 - x + 1) Y \li3(-y)
  - {1 \over 8} (2 x^3 - 3 x^2 - 7 x - 4) X^2 \li2(-x)
 \PlusBreak{ {1 \over y} \bigglB }
   {\pi^2 \over 24} (16 x^3 + 9 x^2 - 5 x - 2) \li2(-x)
 \PlusBreak{ {1 \over y} \bigglB }
   {1 \over 48} (x - 1) (4 x + 1) x X^4
  -{1 \over 12} (x - y) (3 x^2 - 3 x - 2) X^3 Y
 \MinusBreak{ {1 \over y} \bigglB }
   {1 \over 16} (10 x^3 + 15 x^2 + 13 x + 6) X^2 Y^2
  + {1 \over 24} (30 x^3 + 21 x^2 + 5 x + 8) X Y^3
 \MinusBreak{ {1 \over y} \bigglB }
   {1 \over 24} (4 x^3 + 3 x^2 + 2 x + 1) Y^4
  -{\pi^2 \over 24} (2 x^2 + 3 x + 10) x X^2
 \PlusBreak{ {1 \over y} \bigglB }
   {\pi^2 \over 6} (4 x^2 + 3 x + 3) x X Y
  -{\pi^2 \over 48} (10 x^3 + 9 x^2 + 7 x + 6) Y^2
 \PlusBreak{ {1 \over y} \bigglB }
   {1 \over 2} (x - y) \zeta_3 X
  -{1 \over 2} (2 x + 1) \zeta_3 Y
  +{\pi^4 \over 720} (25 x^2 - 27 x + 10) x
\biggr]
 \MinusBreak{}
   {1 \over 8} (6 x^2 - 3 x+2) Y^2 \li2(-x)
  +{1 \over 4} (2 x^2 + x + 2) X Y \li2(-x)
%
 \PlusBreak{}
  {1 \over y}  
\biggl[
  -{x \over 4} (9 x^2 + 2 x - 1) ( \li3(-y) + \li2(-x) Y )
 \MinusBreak{\null - {1 \over y} \bigglB }
   {3 \over 4} (3 x^2 + 2) x{} (\li3(-x) - X \li2(-x))
 \MinusBreak{\null - {1 \over y} \bigglB }
   {x \over 6} (3 x^2 - 3 x + 2) X^3
  -{1 \over 12} (26 x + 3) x^2 X Y^2
 \PlusBreak{\null - {1 \over y} \bigglB }
   {1 \over 12} (22 x^2 - 9 x + 11) x X^2 Y
  -{\pi^2 \over 24} (3 x^2 - 6 x - 5) x X
 \PlusBreak{\null - {1 \over y} \bigglB }
   {\pi^2 \over 24} (20 x^2 + 4 x - 1) x Y
  -{x \over 4} (16 x + 1) \zeta_3 
\biggr]
 \PlusBreak{}
 {x y \over 12} Y^3
%
+{x \over 24 y^2} (3 x^3 - 16 x^2 - 62 x - 37) X^2
 \PlusBreak{}
 {1 \over 12 y} (3 x^3 - 26 x^2 - 47 x - 24) X Y
 \MinusBreak{}
 {\pi^2 \over 8 y} (x^3 - 10 x^2 - 12 x - 8)
-{x \over 12 y} (6 x - 41) X
 \PlusBreak{}
 {1 \over 8 x} (x^3 - 12 x^2 - 6 x - 2) Y^2
%
-{1 \over 6} (3 x - 23) Y
%
%
 \PlusBreak{}
 i \pi {}
\biggl[
    {1 \over 2 y} \biggl(
      (x - 1) (x^2 + x + 1) \li3(-x)
     +(x^2 - 2) x \li3(-y)
 \MinusBreak{\null -  i \pi \bigglB {1 \over 2 y} \bigglP }
      (2 x^3 - 2 x - 1) X \li2(-x)
     -2 (1 - x^2) x \li2(-x) Y
 \MinusBreak{ \null - i \pi \bigglB {1 \over 2 y} \bigglP }
      {1 \over 2} x^3 Y^3
     +{x \over 6} (2 x^2 + 1) X^3
     +{x \over 2} (5 x^2 - 4) X Y^2
 \MinusBreak{ \null - i \pi \bigglB {1 \over 2 y} \bigglP }
      {x - 1 \over 2} (x-y)^2 X^2 Y
     -{\pi^2 \over 2} x X
     -{\pi^2 \over 6} x{} (7 x^2 - 4) Y
          \biggr)
%
 \PlusBreak{\null + i \pi \bigglB }
   {1 \over 4 y}  \biggl(
    -(2 x - 7) x \li2(-x)
    -{x \over 6} (19 x^2 - 18 x + 20) X^2
 \MinusBreak{\null -  i \pi \bigglB {1 \over 4 y} \bigglP }
     {x \over 6} (19 x^2 - 12 x - 3) Y^2
    -{\pi^2 \over 6} (19 x^2 - 6 x + 6) x
 \PlusBreak{\null -  i \pi \bigglB {1 \over 4 y} \bigglP }
     {x \over 3} (19 x^2 - 18 x + 19) X Y
            \biggr)
%
 \PlusBreak{\null - i \pi \bigglB }
   {1 \over 12 y^2} (7 x^3 + 11 x^2 + 34 x + 24) X
 \PlusBreak{\null -  i \pi \bigglB }
   {1 \over 12 x y} (7 x^3 + 7 x^2 + 6) Y
%
  -{47 \over 12 y}
  -{1 \over 12}         
\biggr]
\,, \label{II53}\\[1pt plus 4pt]
%
J_5^{[3]} & = &
x{} (x - y)  
\biggl[
  -\li4(-x)
  -\li4(-y)
  +\li4 \biggl( -{x \over y} \biggr)
  + X{} (\li3(-x) + \li3(-y))
 \MinusBreak{ x (x - y) \bigglB }
  {\pi^2 \over 6} \li2(-x)
  -{1 \over 24} X^4
  + {1 \over 6} X^3 Y
  +{1 \over 4} X^2 Y^2
  -{1 \over 6} X Y^3
  +{1 \over 24} Y^4
 \PlusBreak{ x (x - y) \bigglB }
  {\pi^2 \over 12} X^2
  - {\pi^2 \over 3} X Y
  +{\pi^2 \over 12} Y^2
  +{4 \over 45} \pi^4
\biggr]
%
 \PlusBreak{}
  {1 \over 3 y} 
\biggl[
   (6 x^2 + 5 x - 2) ( \li3(-x) - X \li2(-x) )
 \PlusBreak{\null + {1 \over 3 y} \bigglB }
   (x - 1) ( \li3(-y) + Y \li2(-x) )
 \MinusBreak{\null + {1 \over 3 y} \bigglB }
   (6 x^2 + 6 x - 1) \zeta_3
  +{x \over 24} (3 x^2 + 57 x + 34) X^3
 \MinusBreak{\null + {1 \over 3 y} \bigglB }
   {y \over 6} (3 x + 2) Y^3
  -{1 \over 8} (2 x^3 + 42 x^2 + 37 x + 6) X^2 Y
 \PlusBreak{\null + {1 \over 3 y} \bigglB }
   {1 \over 8} (x^3 + 3 x^2 + 11 x - 2) X Y^2
  +{\pi^2 \over 24} (4 x - 3) (3 x - 5) Y
 \PlusBreak{\null + {1 \over 3 y} \bigglB }
   {\pi^2 \over 24} (3 x^3 + 9 x^2 + 46 x + 66) X
\biggr]
%
 \PlusBreak{}
 {x \over 12 y^2} (3 x^3 + 13 x^2 - 9 x - 25) X^2
 \PlusBreak{}
 {1 \over 72 y} \biggl[
   (36 x^3 + 126 x^2 + 39 x - 22) X Y
  -  (18 x^3 + 90 x^2 + 72 x - 73) \pi^2
        \biggr]
 \PlusBreak{}
 {1 \over 24 x} (6 x^3 + 16 x^2 + 23 x - 6) Y^2
%
 -{x \over 72 y} (36 x + 17) X
-{1 \over 18} (9 x + 5) Y
%
%
 \PlusBreak{}
 i \pi {}
\biggl[
    x {} (x - y)  \biggl(
     (\li3(-x) + \li3(-y))  
     -{1 \over 6} X^3
     +{1 \over 2} X^2 Y
     -{\pi^2 \over 6} X
           \biggr)
%
 \MinusBreak{\null - i \pi \bigglB }
 {1 \over 3 y}  
\biggl(
   (6 x^2 + 4 x - 1) \li2(-x)
  - {1 \over 8} (x^3 + 39 x^2 + 17 x - 14) X^2
 \PlusBreak{ \null - i \pi \bigglB {1 \over 3 y} \bigglP }
   {1 \over 4} (x^3 + 39 x^2 + 30 x + 4) X Y
  - {1 \over 8} (x^3 + 15 x^2 + 27 x + 10) Y^2
 \MinusBreak{\null -  i \pi \bigglB {1 \over 3 y} \bigglP }
   {\pi^2 \over 24} (3 x^3 + 69 x^2 + 65 x + 25) 
\biggr)
%
 \MinusBreak{\null - i \pi \bigglB }
 {1 \over 72 y^2} (6 x^3 + 273 x^2 + 317 x - 22) X
 \MinusBreak{ \null - i \pi \bigglB }
 {1 \over 72 x y} (6 x^3 + 195 x^2 + 124 x - 36) Y
%
-{19 \over 72 y} 
-{13 \over 24}
\biggr]
\,, \label{JJ53}\\[1pt plus 4pt]
%
K_5^{[3]} & = &
x{} (x - y) 
\biggl[
    2 \biggl( -\li4(-x) - \li4(-y) + \li4 \biggl( -{x \over y}
\biggr) \biggr)
   -{\pi^2 \over 3} \li2(-x)
 \PlusBreak{ x (x - y) \bigglB }
    2 (\li3(-x) + \li3(-y)) X
   -{1 \over 12} (X^4 - Y^4)
 \PlusBreak{ x (x - y) \bigglB }
    {1 \over 3} X Y{} (X^2 - Y^2)
   +{1 \over 2} X^2 Y^2
   + {\pi^2 \over 6} (X  - Y)^2
   -{\pi^2 \over 3} X Y
    +{8 \over 45} \pi^4  
\biggr]
%
 \PlusBreak{}
 {1 \over 3 y} 
\biggl[
  -(x^3 - 9 x^2 - 9 x + 2) ( \li3(-x) - X \li2(-x) )
 \MinusBreak{\null + {1 \over 3 y} \bigglB }
   {x \over 24} (5 x^2 - 24 x - 27) X^3
   +{1 \over 8} (6 x^3 - 24 x^2 - 25 x + 8) X^2 Y
 \PlusBreak{\null + {1 \over 3 y} \bigglB }
   {1 \over 8} (y - 1) (7 x^2 + 10 x + 4) X Y^2
 \MinusBreak{\null + {1 \over 3 y} \bigglB }
   {\pi^2 \over 24} x{} (x^2 + 60 x + 60) X
  -(12 x^2 + 12 x - 1) \zeta_3
\biggr]
 \PlusBreak{}
  {y^2 \over 3}  \biggl(
    \li3(-y)
   +\li2(-x) Y
   -{\pi^2 \over 3} Y \biggr)
%
+{1 \over 4} (x^2 - 2 x - 1) Y^2
 \PlusBreak{}
 {x \over 8 y^2} (2 x^3 + 2 x^2 - 7 x - 13) X^2
+{x \over 8 y} (4 x^2 - 2 x - 5) X Y
 \PlusBreak{}
 {\pi^2 \over 36} x{} (9 x + 8)
%
-{x \over 24 y} (20 x - 9) X
-{1 \over 6} (5 x - 2) Y
%
%
 \PlusBreak{}
 i \pi {} \biggl[
   x{} (x - y)  \biggl(
      2 (\li3(-x) + \li3(-y))
     -{1 \over 3} X^3
     +X^2 Y
     -{\pi^2 \over 3} X
   \biggr)
%
 \PlusBreak{ \null + i \pi \bigglB }
   {1 \over y}  
\biggl(
     -{1 \over 3} (12 x^2 + 12 x - 1) \li2(-x)
 \PlusBreak{\null +  i \pi \bigglB {1 \over y} \bigglP }
      {1 \over 12} (3 x^3 - 36 x^2 - 37 x + 4) X Y
     - {1 \over 24} (3 x^3 + 12 x^2 + 12 x + 4) Y^2
 \MinusBreak{\null +  i \pi \bigglB {1 \over y} \bigglP }
      {x \over 24} (3 x^2 - 36 x - 38) X^2
     -{\pi^2 \over 72} (3 x^2 - 4) (3 x - 4)
\biggr)
%
 \PlusBreak{\null + i \pi \bigglB }
   {x \over 8 y^2} (2 x^2 - 7 x - 21) X
  +{1 \over 8 y} (2 x^2 + 7 x + 4) Y
%
  - {29 \over 24 y} - {7 \over 8}
\biggr]
\,, \hskip 2.7 cm \label{KK53}\\[1pt plus 4pt]
%
L_5^{[3]} & = &
L_4^{[3]}
\,. \label{LL53}
\end{eqnarray}


\section{Auxiliary functions for two-loop scheme shifts}
\label{OrdepsdeltaRemainderAppendix}

In this appendix we present auxiliary functions appearing in
\eqns{Twoloopc12}{Twoloopc3} for the shift in the two-loop amplitudes
under scheme changes.  These functions correspond to the
$\delta_R$-dependent parts of the $\Ord(\e)$ terms in the one-loop 
amplitude remainders.  They are given by,
\begin{eqnarray}
 M^{(1),[1]\,\e,\delta_R}_1 &=&
          N {} \biggl( {1\over 4} + {x \over 6} \biggr) + {1 \over 4 N}
 \,,  \label{M11_e_deltaR} \\[1pt plus 4pt]
 M^{(1),[2]\,\e,\delta_R}_1 &=& 
N { x {} (2 x -1) \over 12 y}  - {x\over 4 y N} 
 \,,  \label{M12_e_deltaR} \\ [1pt plus 4pt]
 M^{(1),[3]\,\e,\delta_R}_1 &=& 
0 
\,, \label{M13_e_deltaR} \\[1pt plus 4pt]
M^{(1),[1]\,\e,\delta_R}_2  &=&
 {1 \over 4 y^3} \biggl[ N  \Bigl(
  x X^2 + y {} (1 - x)  X + y^2 (2 + 3 x) 
+  i \pi x {} (2 X -  y {} (3+x) )  
            \Bigr)
 \PlusBreak{ }
 {1\over N} \Bigl(
                   -x^2 X^2
                   - y {} (1 + 3 x) X
                   - y^2 (2 + x) 
  \MinusBreak{ {1 \over 4 y^3} \bigglB  {1\over N} \BiglP }
                    i \pi {} x {} ( 2 x {} X +  y {} (1-x))  
                             \Bigr)
         \biggr] 
 \,,  \label{M21_e_deltaR} \\[1pt plus 4pt]
M^{(1),[2]\,\e,\delta_R}_2  &=&  
{x \over 4 y} \biggl(
          - N {} ( 3  + i \pi ) 
            + {1 \over  N} (1 + i \pi )  \biggr)
 \,, \label{M22_e_deltaR} \\[1pt plus 4pt]
M^{(1),[3]\,\e,\delta_R}_2  &=& 
 - {x \over 4 y^2} \Bigl( X^2 + 2 i {} \pi {} X \Bigr)
\,,  \label{M23_e_deltaR}  \\[1pt plus 4pt]
 M^{(1),[1]\,\e,\delta_R}_3 &=&
 N {} \biggl( {1\over 4} + {1 \over 6 x} \biggr) 
          + {1 \over 4 N}
 \,, \label{M31_e_deltaR} \\ [1pt plus 4pt]
 M^{(1),[2]\,\e,\delta_R}_3 &=& 
 N {} { (2   - x) \over 12 x y}  - {1\over 4 y N} 
 \,, \label{M32_e_deltaR} \\ [1pt plus 4pt]
 M^{(1),[3]\,\e,\delta_R}_3 &=&
 0 
\,, \label{M33_e_deltaR} \\ [1pt plus 4pt]
 M^{(1),[1]\,\e,\delta_R}_4 &=& 
{1 \over 4 y^3} \biggl[
 N {} \Bigl( x^2 X^2
      + y {} (1 + 3 x) X 
      + y^2 (3 + 2 x) 
 +  i \pi {} ( 2 x^2 X + x y {} (1 - x) )
       \Bigr)
 \MinusBreak{ }
 {1 \over N} \Bigl(
  x X^2 
 - y {} (x-1) X 
  +(2 x + 1) y^2
  + i \pi {} (2 x X - x y {} (x+3) )
        \Bigr)  \biggr]
\,, \label{M41_e_deltaR} \\ [1pt plus 4pt]
M^{(1),[2]\,\e,\delta_R}_4 &=& 
{1\over 4 y}  \biggl( N {} (X-3) - {1\over N} (X- 1) \biggr)
\,, \label{M42_e_deltaR} \\ [1pt plus 4pt]
M^{(1),[3]\,\e,\delta_R}_4 &=& 
- {x \over 4 y^2} \Bigl( X^2 +  2 i {} \pi {} X \Bigr)
\,, \label{M43_e_deltaR} \\[1pt plus 4pt]
M^{(1),[1]\,\e,\delta_R}_5 &=& 
      {1\over 4} N {} (X - 3) - {1\over 4 N} (X - 1)
\,, \label{M51_e_deltaR} \\[1pt plus 4pt]
M^{(1),[2]\,\e,\delta_R}_5 &=& 
 {N \over 4} \biggl[
   x^2 \Bigl((X - Y )^2 + \pi^2 \Bigr)
   + X {}(2 x - 1)
   + {x \over y} (2 x + 1) Y
   - {x+3 \over y} \biggr]
 \PlusBreak{ } 
 {1\over 4 N} 
   \biggl[
    x {} (1 + x) X^2
   -2 x {}(1 + x) X Y
   + \pi^2 x {} (1 + x) 
   + X {}(2 x + 1)
 \PlusBreak{\null + {1\over 4 N}  \bigglB } 
    x {} (1 + x) Y^2
   + {x\over y} (2 x + 3) Y
   + {1 - x \over y} \biggr]
\,, \label{M52_e_deltaR} \\ [1pt plus 4pt]
M^{(1),[3]\,\e,\delta_R}_5 &=& 
- {x\over 4} \Bigl((X-Y)^2 + \pi^2 \Bigr)
\,. \label{M53_e_deltaR}
\end{eqnarray}
%



\begin{thebibliography}{99}

\bibitem{BRY}
Z.~Bern, J.S.~Rozowsky and B.~Yan,
``Two-loop four-gluon amplitudes in $N=4$ super-Yang-Mills,''
Phys.\ Lett.\ B {\bf 401}, 273 (1997)
[arXiv:hep-ph/9702424];\\
%
Z.~Bern, L.J.~Dixon, D.C.~Dunbar, M.~Perelstein and J.S.~Rozowsky,
``On the relationship between Yang-Mills theory and gravity and its
implication for ultraviolet divergences,''
Nucl.\ Phys.\ B {\bf 530}, 401 (1998)
[arXiv:hep-th/9802162].

\bibitem{AllPlusTwo}
Z.~Bern, L.J.~Dixon and D.A.~Kosower,
``A two-loop four-gluon helicity amplitude in QCD,''
JHEP {\bf 0001}, 027 (2000)
[arXiv:hep-ph/0001001].

\bibitem{BhabhaTwoLoop}
Z.~Bern, L.J.~Dixon and A.~Ghinculov,
``Two-loop correction to Bhabha scattering,''
Phys.\ Rev.\ D {\bf 63}, 053007 (2001)
[arXiv:hep-ph/0010075].

\bibitem{GOTYqqqq}
C.~Anastasiou, E.W.N.~Glover, C.~Oleari and M.E.~Tejeda-Yeomans,
``Two-loop QCD corrections to $q\bar{q} \to q' \bar{q}'$,''
Nucl.\ Phys.\ B {\bf 601}, 318 (2001)
[hep-ph/0010212];\\
``Two-loop QCD corrections to $q \bar{q} \to q \bar{q}$,''
Nucl.\ Phys.\ B {\bf 601}, 341 (2001)
[hep-ph/0011094].

\bibitem{GOTYqqgg}
C.~Anastasiou, E.W.N.~Glover, C.~Oleari and M.E.~Tejeda-Yeomans,
``Two-loop QCD corrections to massless quark-gluon scattering,''
Nucl.\ Phys.\ B {\bf 605}, 486 (2001)
[hep-ph/0101304].

\bibitem{GOTYgggg}
E.W.N.~Glover, C.~Oleari and M.E.~Tejeda-Yeomans,
``Two-loop QCD corrections to gluon-gluon scattering,''
Nucl.\ Phys.\ B {\bf 605}, 467 (2001) [arXiv:hep-ph/0102201].

\bibitem{gggamgamPaper}
Z.~Bern, A.~De Freitas and L.J.~Dixon,
``Two-loop amplitudes for gluon fusion into two photons,''
JHEP {\bf 0109}, 037 (2001)
[arXiv:hep-ph/0109078].

\bibitem{PhotonPaper}
Z.~Bern, A.~De Freitas, L.J.~Dixon, A.~Ghinculov and H.L.~Wong,
``QCD and QED corrections to light-by-light scattering,''
JHEP {\bf 0111}, 031 (2001)
[arXiv:hep-ph/0109079].

\bibitem{BDDgggg}
Z.~Bern, A.~De Freitas and L.~Dixon,
``Two-loop helicity amplitudes for gluon-gluon scattering in QCD and
  supersymmetric Yang-Mills theory,''
JHEP {\bf 0203}, 018 (2002)
[arXiv:hep-ph/0201161].

\bibitem{TwoLoopee3Jets}
L.W.~Garland, T.~Gehrmann, E.W.N.~Glover, A.~Koukoutsakis and E.~Remiddi,
``The two-loop QCD matrix element for $e^+ e^- \to 3$ jets,''
Nucl.\ Phys.\ B {\bf 627}, 107 (2002)
[arXiv:hep-ph/0112081].

\bibitem{TwoLoopee3JetsHel}
L.W.~Garland, T.~Gehrmann, E.W.N.~Glover, A.~Koukoutsakis and E.~Remiddi,
``Two-loop QCD helicity amplitudes for $e^+ e^- \to 3$ jets,''
Nucl.\ Phys.\ B {\bf 642}, 227 (2002)
[arXiv:hep-ph/0206067];\\
%
%
S.~Moch, P.~Uwer and S.~Weinzierl,
``Two-loop amplitudes for $e^+ e^- \to q \qb g$: The $\Nf$-contribution,''
Acta Phys.\ Polon.\ B {\bf 33}, 2921 (2002)
[arXiv:hep-ph/0207167].

\bibitem{IBP}
F.V.~Tkachov,
``A theorem on analytical calculability of four-loop renormalization 
group functions,''
Phys.\ Lett.\ B {\bf 100}, 65 (1981); \\
K.G.~Chetyrkin and F.V.~Tkachov,
``Integration by parts: the algorithm to calculate beta functions 
in 4 loops,'' 
Nucl.\ Phys.\ B {\bf 192}, 159 (1981).

\bibitem{PBScalar}
V.A.~Smirnov,
``Analytical result for dimensionally regularized massless on-shell  
double box,''
Phys.\ Lett.\  {\bf B460}, 397 (1999)
[arXiv:hep-ph/9905323].

\bibitem{NPBScalar}
J.B.~Tausk,
``Non-planar massless two-loop Feynman diagrams with four on-shell legs,''
Phys.\ Lett.\  {\bf B469}, 225 (1999)
[arXiv:hep-ph/9909506].

\bibitem{Lorentz}
T.~Gehrmann and E.~Remiddi,
``Differential equations for two-loop four-point functions,''
Nucl.\ Phys.\ B {\bf 580}, 485 (2000)
[arXiv:hep-ph/9912329].

\bibitem{NPBReduction}
C.~Anastasiou, T.~Gehrmann, C.~Oleari, E.~Remiddi and J.B.~Tausk,
``The tensor reduction and master integrals of the two-loop massless 
crossed box with light-like legs,''
Nucl.\ Phys.\  {\bf B580}, 577 (2000)
[arXiv:hep-ph/0003261].

\bibitem{PBReduction}
V.A.~Smirnov and O.L.~Veretin,
``Analytical results for dimensionally regularized massless on-shell  
double boxes with arbitrary indices and numerators,''
Nucl.\ Phys.\  {\bf B566}, 469 (2000)
[arXiv:hep-ph/9907385].

\bibitem{IntegralsAGO}
C.~Anastasiou, E.W.N.~Glover and C.~Oleari,
``Application of the negative-dimension approach to massless scalar box  
integrals,''
Nucl.\ Phys.\  {\bf B565}, 445 (2000)
[arXiv:hep-ph/9907523]; \\
C.~Anastasiou, E.W.N.~Glover and C.~Oleari,
``The two-loop scalar and tensor pentabox graph with light-like legs,''
Nucl.\ Phys.\  {\bf B575}, 416 (2000),
err. ibid.\  {\bf B585}, 763 (2000)
[arXiv:hep-ph/9912251].

\bibitem{Catani}
S.~Catani,
``The singular behaviour of QCD amplitudes at two-loop order,''
Phys.\ Lett.\  {\bf B427}, 161 (1998)
[arXiv:hep-ph/9802439].

\bibitem{GloverReview}
E.W.N.~Glover,
``Progress in NNLO calculations for scattering processes,''
arXiv:hep-ph/0211412.

\bibitem{Aversa}
F.~Aversa, P.~Chiappetta, M.~Greco and J.P.~Guillet,
``Higher order corrections to QCD jets,''
Phys.\ Lett.\ B {\bf 210}, 225 (1988);
``Jet production in hadronic collisions to $\Ord(\alpha_s^3)$,''
Z.\ Phys.\ C {\bf 46}, 253 (1990).

\bibitem{EKS}
S.D.~Ellis, Z.~Kunszt and D.E.~Soper,
``The one-jet inclusive cross section at $\Ord(\alpha_s^3)$: 
quarks and gluons,''
Phys.\ Rev.\ Lett.\  {\bf 64}, 2121 (1990);
``Two jet production in hadron collisions at $\Ord(\alpha_s^3)$ in QCD,''
Phys.\ Rev.\ Lett.\  {\bf 69}, 1496 (1992).

\bibitem{JETRAD}
W.T.~Giele, E.W.N.~Glover and D.A.~Kosower,
``Higher order corrections to jet cross sections in hadron colliders,''
Nucl.\ Phys.\ B {\bf 403}, 633 (1993)
[arXiv:hep-ph/9302225].

\bibitem{MRSTNNLO}
A.D.~Martin, R.G.~Roberts, W.J.~Stirling and R.S.~Thorne,
``Estimating the effect of NNLO contributions on global parton analyses,''
Eur.\ Phys.\ J.\ C {\bf 18}, 117 (2000)
[arXiv:hep-ph/0007099];
``NNLO global parton analysis,''
Phys.\ Lett.\ B {\bf 531}, 216 (2002)
[arXiv:hep-ph/0201127].

\bibitem{NNLOPDFApprox}
W.L.~van Neerven and A.~Vogt,
``NNLO evolution of deep-inelastic structure functions: 
 The non-singlet case,''
Nucl.\ Phys.\ B {\bf 568}, 263 (2000)
[arXiv:hep-ph/9907472];\\
``NNLO evolution of deep-inelastic structure functions: 
 The singlet case,''
Nucl.\ Phys.\ B {\bf 588}, 345 (2000)
[arXiv:hep-ph/0006154];\\
``Improved approximations for the three-loop splitting functions in QCD,''
Phys.\ Lett.\ B {\bf 490}, 111 (2000)
[arXiv:hep-ph/0007362];\\
S.~Moch, J.A.M.~Vermaseren and A.~Vogt,
``Non-singlet structure functions at three loops: Fermionic  contributions,''
Nucl.\ Phys.\ B {\bf 646}, 181 (2002)
[arXiv:hep-ph/0209100].

\bibitem{PDFuncertainty}
S.~Alekhin,
``Extraction of parton distributions and $\alpha_s$ from DIS data within 
 the Bayesian treatment of systematic errors,''
Eur.\ Phys.\ J.\ C {\bf 10}, 395 (1999)
[arXiv:hep-ph/9611213];\\
W.T.~Giele, S.A.~Keller and D.A.~Kosower,
``Parton distributions with errors,'' in
{\it La Thuile 1999, Results and perspectives in particle physics};\\
D.~Stump {\it et al.},
``Uncertainties of predictions from parton distribution functions. I: 
 The Lagrange multiplier method,''
Phys.\ Rev.\ D {\bf 65}, 014012 (2002)
[arXiv:hep-ph/0101051];
``Uncertainties of predictions from parton distribution functions. II:  
 The Hessian method,''
Phys.\ Rev.\ D {\bf 65}, 014013 (2002)
[arXiv:hep-ph/0101032];\\
%
W.T.~Giele, S.A.~Keller and D.A.~Kosower,
``Parton distribution function uncertainties,''
arXiv:hep-ph/0104052;\\
A.D.~Martin, R.G.~Roberts, W.J.~Stirling and R.S.~Thorne,
``Uncertainties of predictions from parton distributions. I: 
  Experimental  errors,''
arXiv:hep-ph/0211080.

\bibitem{Kidonakis}
N.~Kidonakis,
``Resummation for heavy quark and jet cross sections,''
Int.\ J.\ Mod.\ Phys.\ A {\bf 15}, 1245 (2000)
[arXiv:hep-ph/9902484];\\
N.~Kidonakis and J.F.~Owens,
``Effects of higher-order threshold corrections in high-$E_T$ jet  
production,''
Phys.\ Rev.\ D {\bf 63}, 054019 (2001)
[arXiv:hep-ph/0007268].

\bibitem{TreeSixPoint}
F.A.~Berends and W.~Giele,
``The six gluon process as an example of Weyl-Van Der Waerden spinor 
calculus,''
Nucl.\ Phys.\ B {\bf 294}, 700 (1987);\\
M.L.~Mangano, S.~Parke and Z.~Xu,
``Duality and multi-gluon scattering,''
Nucl.\ Phys.\ B {\bf 298}, 653 (1988).

\bibitem{MPReview}
M.L.~Mangano and S.J.~Parke,
``Multiparton amplitudes in gauge theories,''
Phys.\ Rept.\ {\bf 200}, 301 (1991).

\bibitem{OneloopFivePoint}
Z.~Bern, L.J.~Dixon and D.A.~Kosower,
``One-loop corrections to five-gluon amplitudes,''
Phys.\ Rev.\ Lett.\  {\bf 70}, 2677 (1993)
[arXiv:hep-ph/9302280];\\
Z.~Kunszt, A.~Signer and Z.~Tr\'ocs\'anyi,
``One-loop radiative corrections to the helicity amplitudes 
of QCD processes involving four quarks and one gluon,''
Phys.\ Lett.\ B {\bf 336}, 529 (1994)
[arXiv:hep-ph/9405386].

\bibitem{Oneloopqqggg}
Z.~Bern, L.J.~Dixon and D.A.~Kosower,
``One-loop corrections to two-quark three-gluon amplitudes,''
Nucl.\ Phys.\ B {\bf 437}, 259 (1995)
[arXiv:hep-ph/9409393].

\bibitem{SpinorHelicity}
F.A.~Berends, R.~Kleiss, P.~De Causmaecker, R.~Gastmans and T.T.~Wu,
``Single bremsstrahlung processes in gauge theories,''
Phys.\ Lett.\ B {\bf 103}, 124 (1981);\\
%
P.~De Causmaecker, R.~Gastmans, W.~Troost and T.T.~Wu,
``Helicity amplitudes for massless QED,''
Phys.\ Lett.\ B {\bf 105}, 215 (1981);\\
%
Z.~Xu, D.~Zhang and L.~Chang,
``Helicity amplitudes for multiple bremsstrahlung in 
massless nonabelian gauge theories,''
Nucl.\ Phys.\ B {\bf 291}, 392 (1987).

\bibitem{SWI}
M.T.~Grisaru, H.N.~Pendleton and P.~van Nieuwenhuizen,
``Supergravity and the S matrix,''
Phys.\ Rev.\ D {\bf 15}, 996 (1977); \\
M.T.~Grisaru and H.N.~Pendleton,
``Some properties of scattering amplitudes in supersymmetric theories,''
Nucl.\ Phys.\ B {\bf 124}, 81 (1977); \\
S.J.~Parke and T.R.~Taylor,
``Perturbative QCD utilizing extended supersymmetry,''
Phys.\ Lett.\ B {\bf 157}, 81 (1985),
err. ibid.\  {\bf 174B}, 465 (1985).

\bibitem{TwoLoopSUSY}
Z.~Bern, A.~De Freitas, L.~Dixon and H.L.~Wong,
``Supersymmetric regularization, two-loop QCD amplitudes and 
 coupling  shifts,''
Phys.\ Rev.\ D {\bf 66}, 085002 (2002)
[arXiv:hep-ph/0202271].

\bibitem{Neq4}
Z.~Bern, L.J.~Dixon, D.C.~Dunbar and D.A.~Kosower,
``One-loop $n$-point gauge theory amplitudes, unitarity 
and collinear limits,''
Nucl.\ Phys.\ {\bf B425}, 217 (1994)
[arXiv:hep-ph/9403226].

\bibitem{LoopReview}
Z.~Bern, L.J.~Dixon and D.A.~Kosower,
``Progress in one-loop QCD computations,''
Ann.\ Rev.\ Nucl.\ Part.\ Sci.\ {\bf 46}, 109 (1996)
[arXiv:hep-ph/9602280].

\bibitem{BFKL}
E.A.~Kuraev, L.N.~Lipatov and V.S.~Fadin,
``Multi-reggeon processes in the Yang-Mills theory,''
Sov.\ Phys.\ JETP {\bf 44}, 443 (1976)
[Zh.\ Eksp.\ Teor.\ Fiz.\  {\bf 71}, 840 (1976)].

\bibitem{TwoloopBFKL}
V.~Del Duca and E.W.N.~Glover,
``The high energy limit of QCD at two loops,''
JHEP {\bf 0110}, 035 (2001)
[arXiv:hep-ph/0109028];\\
%
A.V.~Bogdan, V.~Del Duca, V.S.~Fadin and E.W.N.~Glover,
``The quark Regge trajectory at two loops,''
JHEP {\bf 0203}, 032 (2002)
[arXiv:hep-ph/0201240].

\bibitem{StermanIR}
G.~Sterman and M.E.~Tejeda-Yeomans,
``Multi-loop amplitudes and resummation,''
Phys.\ Lett.\ B {\bf 552}, 48 (2003)
[arXiv:hep-ph/0210130].

\bibitem{SofferVirey}
J.~Soffer and J.M.~Virey,
``Testing various polarized parton distributions at RHIC,''
Nucl.\ Phys.\ B {\bf 509}, 297 (1998)
[arXiv:hep-ph/9706229].

\bibitem{dFFSV}
D.~de Florian, S.~Frixione, A.~Signer and W.~Vogelsang,
``Next-to-leading order jet cross sections in polarized hadronic collisions,''
Nucl.\ Phys.\ B {\bf 539}, 455 (1999)
[arXiv:hep-ph/9808262].

\bibitem{GTYqqgg}
E.W.N.~Glover and M.E.~Tejeda-Yeomans, ``Two-loop QCD helicity amplitudes
for massless quark-massless gauge boson scattering,''
arXiv:hep-ph/0304169.

\bibitem{CDR}
J.C.~Collins, {\it Renormalization: an introduction to renormalization group,
and the operator-product expansion}, Cambridge Monographs on Mathematical
Physics (Cambridge Univ. Press, 1984).

\bibitem{EllisSexton}
R.K.~Ellis and J.C.~Sexton,
``QCD Radiative corrections to parton-parton scattering,''
Nucl.\ Phys.\ B {\bf 269}, 445 (1986).

\bibitem{HV}
G.~'t Hooft and M.J.G.~Veltman,
``Regularization and renormalization of gauge fields,''
Nucl.\ Phys.\ {\bf B44}, 189 (1972).

\bibitem{BKgggg}
Z.~Bern and D.A.~Kosower,
``The computation of loop amplitudes in gauge theories,''
Nucl.\ Phys.\ {\bf B379}, 451 (1992).

\bibitem{DR}
W.~Siegel,
``Supersymmetric dimensional regularization via dimensional reduction,''
Phys.\ Lett.\  {\bf B84}, 193 (1979);\\
%
D.M.~Capper, D.R.T.~Jones and P.~van Nieuwenhuizen,
``Regularization by dimensional reduction of supersymmetric and 
nonsupersymmetric gauge theories,''
Nucl.\ Phys.\  {\bf B167}, 479 (1980);\\
%
I.~Jack, D.R.T.~Jones and K.L.~Roberts,
``Equivalence of dimensional reduction and dimensional regularization,''
Z.\ Phys.\  {\bf C63}, 151 (1994)
[arXiv:hep-ph/9401349].

\bibitem{BGMvdB}
T.~Binoth, E.W.N.~Glover, P.~Marquard and J.J.~van der Bij,
``Two-loop corrections to light-by-light scattering in supersymmetric  QED,''
JHEP {\bf 0205}, 060 (2002)
[arXiv:hep-ph/0202266].

\bibitem{CataniGrazziniSoft}
S.~Catani and M.~Grazzini,
``The soft-gluon current at one-loop order,''
Nucl.\ Phys.\ B {\bf 591}, 435 (2000)
[arXiv:hep-ph/0007142].

\bibitem{KSTfourparton}
Z.~Kunszt, A.~Signer and Z.~Tr\'ocs\'anyi, 
``One-loop helicity amplitudes for all 2 $\to$ 2 processes in QCD 
and $N=1$ supersymmetric Yang-Mills theory,''
Nucl.\ Phys.\ {\bf B411}, 397 (1994)
[arXiv:hep-ph/9305239].

\bibitem{CST}
S.~Catani, M.H.~Seymour and Z.~Tr\'ocs\'anyi,
``Regularization scheme independence and unitarity in QCD cross sections,''
Phys.\ Rev.\ {\bf D55}, 6819 (1997)
[arXiv:hep-ph/9610553].

\bibitem{AGTYPrivate}
C.~Anastasiou, E.W.N.~Glover and M.E.~Tejeda-Yeomans,
private communications.

\bibitem{AGTYphotons}
C.~Anastasiou, E.W.N.~Glover and M.E.~Tejeda-Yeomans,
``Two-loop QED and QCD corrections to massless fermion boson scattering,''
Nucl.\ Phys.\ B {\bf 629}, 255 (2002)
[arXiv:hep-ph/0201274].

\bibitem{QGRAF}
P.~Nogueira,
``Automatic Feynman graph generation,''
J.\ Comput.\ Phys.\  {\bf 105}, 279 (1993).

\bibitem{FORM}
J.A.M.~Vermaseren,
``New features of FORM,''
arXiv:math-ph/0010025.

\bibitem{BernMorgan}
Z.~Bern and A.~G.~Morgan,
``Massive loop amplitudes from unitarity,''
Nucl.\ Phys.\ B {\bf 467}, 479 (1996)
[arXiv:hep-ph/9511336].

\bibitem{NielsenIds}
K.S.~K\"olbig, J.A.~Mignaco and E.~Remiddi, 
``On Nielsen's generalized polylogarithms and their numerical
calculation,'' 
B.I.T. {\bf 10}, 38 (1970).

\bibitem{Lewin}
L.~Lewin, {\it Dilogarithms and associated functions} (Macdonald, 1958).

\bibitem{OneloopSplit}
Z.~Bern and G.~Chalmers,
``Factorization in one-loop gauge theory,''
Nucl.\ Phys.\ B {\bf 447}, 465 (1995)
[arXiv:hep-ph/9503236];\\
%
Z.~Bern, V.~Del Duca and C.R.~Schmidt,
``The infrared behavior of one-loop gluon amplitudes at  
next-to-next-to-leading order,'' Phys.\ Lett.\ B {\bf 445}, 168 (1998)
[arXiv:hep-ph/9810409];\\
%
D.A.~Kosower and P.~Uwer,
``One-loop splitting amplitudes in gauge theory,''
Nucl.\ Phys.\ B {\bf 563}, 477 (1999)
[arXiv:hep-ph/9903515];\\
%
Z.~Bern, V.~Del Duca, W.B.~Kilgore and C.R.~Schmidt,
``The infrared behavior of one-loop QCD amplitudes at 
next-to-next-to-leading order,''
Phys.\ Rev.\ D {\bf 60}, 116001 (1999)
[arXiv:hep-ph/9903516].

\bibitem{GunionKunszt}
J.F.~Gunion and Z.~Kunszt,
``Improved analytic techniques for tree graph calculations and the 
$ggq\bar{q}\ell\bar{\ell}$ subprocess,''
Phys.\ Lett.\ B {\bf 161}, 333 (1985).

\bibitem{OneloopColor}
Z.~Bern and D.A.~Kosower,
``Color decomposition of one-loop amplitudes in gauge theories,''
Nucl.\ Phys.\ B {\bf 362}, 389 (1991).

\bibitem{Higgs}
Z.~Bern, L.~Dixon and C.~Schmidt,
``Isolating a light Higgs boson from the di-photon background at the LHC,''
Phys.\ Rev.\ D {\bf 66}, 074018 (2002)
[arXiv:hep-ph/0206194];\\
%
L.~Dixon and M.S.~Siu,
``Resonance-continuum interference in the di-photon Higgs signal at the  LHC,''
arXiv:hep-ph/0302233.

\bibitem{NNLOPS}
D.A.~Kosower,
``Multiple singular emission in gauge theories,''
Phys.\ Rev.\ D {\bf 67}, 116003 (2003)
[arXiv:hep-ph/0212097];\\
S.~Weinzierl,
``Subtraction terms at NNLO,''
JHEP {\bf 0303}, 062 (2003)
[arXiv:hep-ph/0302180].

\end{thebibliography}
\end{document}